\documentclass[aps,amsmath,amssymb,twocolumn,prx]{revtex4-2}

\usepackage{graphicx}
\usepackage{dcolumn}
\usepackage{bm}
\usepackage{color}
\usepackage[normalem]{ulem}
\usepackage{amsmath}

\usepackage{dutchcal}

\begin{document}

\title{Full Counting Statistics of Yu-Shiba-Rusinov Bound States}

\author{David Christian Ohnmacht}
\email{david.ohnmacht@uni-konstanz.de}

\author{Wolfgang Belzig}
\affiliation{Fachbereich Physik, Universit{\"a}t Konstanz, D-78457 Konstanz, Germany}

\author{Juan Carlos Cuevas}
\affiliation{Departamento de F\'{i}sica Te\'{o}rica de la Materia Condensada and 
Condensed Matter Physics Center (IFIMAC), Universidad Aut\'{o}noma de Madrid, E-28049 Madrid, Spain}

\date{\today}

\begin{abstract}
With the help of scanning tunneling microscopy (STM) it has become possible to address single magnetic 
impurities on superconducting surfaces and to investigate the peculiar properties of the in-gap states 
known as Yu-Shiba-Rusinov (YSR) states. These systems are an ideal playground to investigate multiple
aspects of superconducting bound states, such as the occurrence of quantum phase transitions or the
interplay between Andreev transport physics and the spin degree of freedom, with profound implications
for disparate topics like Majorana modes or Andreev spin qubits. However, until very recently YSR states 
were only investigated with conventional tunneling spectroscopy, missing the crucial information contained 
in other transport properties such as shot noise. In this work we adapt the concept of full counting 
statistics (FCS) to provide the deepest insight thus far into the spin-dependent transport in these
hybrid atomic-scale systems. We illustrate the power of FCS by analyzing different situations in which
YSR states show up including single-impurity junctions with a normal and a superconducting STM tip, 
as well as double-impurity systems where one can probe the tunneling between individual YSR states [Nat.\ 
Phys.\ {\bf 16}, 1227 (2020)]. The FCS concept allows us to unambiguously identify every tunneling process 
that plays a role in these situations and to classify them according to the charge transferred in them. 
Moreover, FCS provides all the relevant transport properties, including current, shot noise and all the 
cumulants of the current distribution. In particular, our approach is able to reproduce the experimental 
results recently reported on the shot noise of a single-impurity junction with a normal STM tip 
[Phys.\ Rev.\ Lett.\ {\bf 128}, 247001 (2022)]. We also predict the signatures of resonant (and non-resonant) 
multiple Andreev reflections in the shot noise and Fano factor of single-impurity junctions with two 
superconducting electrodes and show that the FCS approach allows us to understand conductance features
that have been incorrectly interpreted in the literature. In the case of double-impurity junctions we 
show that the direct tunneling between YSR states is characterized by a strong reduction of the Fano factor
that reaches a minimum value of $7/32$, a new fundamental result in quantum transport. The FCS approach 
presented here can be naturally extended to investigate the spin-dependent superconducting transport in a 
variety of situations, such as atomic spin chains on surfaces or superconductor-semiconductor nanowire junctions, 
and it is also suitable to analyze multi-terminal superconducting junctions, irradiated contacts, and many other 
basic situations.
\end{abstract}

\maketitle

\section{Introduction} \label{sec:intro}

Yu-Shiba-Rusinov (YSR) bound states are one of the most fundamental consequences of the interplay between magnetism
and superconductivity at the atomic scale \cite{Yu1965,Shiba1968,Rusinov1969}. They appear when a magnetic impurity 
is coupled to one or several superconducting leads, and they have been extensively investigated in the context of 
STM experiments and impurities on superconducting surfaces 
\cite{Yazdani1997,Ji2008,Franke2011,Bauer2013,Menard2015,Ruby2015,Hatter2015,Ruby2016,Randeria2016,Choi2017,
Cornils2017,Hatter2017,Farinacci2018,Brand2018,Malavolti2018,Kezilebieke2019,Senkpiel2019,Schneider2019,
Liebhaber2020,Huang2020a,Huang2020b,Odobesko2020,Huang2021,Karan2022,Homberg2022,Li2022,Trahms2023,Trivini2023}, for 
recent reviews see Refs.~\cite{Heinrich2018,Choi2019}. The appeal of these hybrid junctions is manifold. On the one 
hand, they enable the investigation of a fundamental quantum phase transition
\cite{Franke2011,Bauer2013,Malavolti2018,Farinacci2018,Karan2022}. They also provide the chance to study states that
are the precursors of Majorana modes, which might potentially appear when magnetic impurities are combined to form 
magnetic chains \cite{Nadj-Perge2014,Ruby2015b,Kezilebieke2018,Ruby2017,Ruby2018,Kamlapure2021,Schneider2022}. 
On the other hand, these systems are also a new kind of Andreev spin qubit, a topic that is rapidly growing in 
the community of superconducting qubits \cite{Tosi2019,Hays2021,Matute2022,Pita2022}. More important for the topic 
of this work is the fact that magnetic impurities on superconducting surfaces offer the possibility of exploring the 
role of the spin degree of freedom in situations where the electronic transport is dominated by Andreev reflections, 
which is a topic of great interest for the field of superconducting spintronics \cite{Linder2015,Eschrig2015}.

Until very recently, YSR states had been only investigated by the means of standard tunneling spectroscopy, i.e.,
via current or conductance measurements. This has already allowed us to elucidate many basic properties of these
superconducting bound states. However, as it is well-known in mesoscopic physics, the analysis of other transport
properties, such as shot noise, may provide very valuable information that is out of scope of conventional tunneling 
spectroscopy \cite{Blanter2000,Nazarov2003}. An experimental breakthrough illustrating this idea was recently reported 
in which the shot noise through a magnetic impurity featuring YSR states was measured \cite{Thupakula2022}. These
experiments revealed the power of shot noise measurements by accessing scales like the intrinsic YSR lifetimes, which 
are very difficult to obtain via conductance measurements. These experiments call for more comprehensive theories able 
to provide knowledge about the YSR states and their transport properties beyond the conventional calculations of the 
charge current. In this regard, the concept of full counting statistics (FCS), introduced some time ago in the context 
of quantum transport in mesoscopic systems \cite{Levitov1993,Levitov1996}, is clearly the most powerful tool that we 
have at our disposal to reach this goal. This concept has already provided unprecedented insight into the physics of
superconducting junctions \cite{Belzig2001a,Belzig2001b,Belzig2003,Cuevas2003,Cuevas2004}, which includes not only the
current, shot noise and all possible cumulants of the current distribution, but also the possibility to unambiguously
identify the contribution of every possible tunneling process and the charge transferred in it. Unfortunately, the 
concept of FCS has not been used in the context of the superconducting hybrid junctions featuring YSR states mainly 
due to technical difficulties implementing the spin-dependent scattering that occurs in these systems.

In this work we fill this void by providing a systematic study of the FCS in a variety of junctions featuring
YSR bound states. Those cases include single-impurity junctions with one and two superconducting leads, and two-impurity
systems in which the tunneling between individual YSR states has been recently reported for the first time 
\cite{Huang2020b,Huang2021}. In all these cases we show how the concept of FCS allows us to identify all the 
relevant tunneling processes and to classify them according to the charge transferred in them, providing so
a very deep insight into the physics of YSR states. Moreover, from the knowledge of the FCS in these situations 
we obtain all the relevant transport properties: current, shot noise, and all the cumulants of the current 
distribution. Among the main result of this work we can highlight the full analysis of the shot noise through 
a single-magnetic impurity coupled to a normal and to a superconducting lead in excellent agreement with very 
recent experimental results \cite{Thupakula2022}. This analysis reveals the possibility to access energy and
time scales that are usually out of the scope of conductance measurements, and it provides a very fresh insight
into the nature of a resonant Andreev reflection. We also present very concrete predictions for the shot noise 
and Fano factor in the case in which a magnetic impurity is coupled to two superconducting leads for arbitrary 
junction transparency. In particular, we elucidate the signatures of resonant and YSR-mediated multiple Andreev 
reflections (MARs) in both the current and shot noise and amend some misinterpretations related to these processes 
that have been reported in the literature. In the case of two-impurity junctions, we demonstrate that the direct 
quasiparticle tunneling between YSR states provides an unique signature in the noise and Fano factor. In particular, 
this tunneling can be identified by a strong reduction of the Fano factor at the resonant voltage at which the two
states align and it can reach a minimum value of $7/32$. We quantitatively explain this result in terms of the 
quasiparticle tunneling between two sharp levels, which constitutes an extreme example of quantum tunneling. 
It is also worth stressing that the FCS approach presented here can be readily adapted to analyze the charge 
transport properties in a plethora of superconducting nanoscale junctions including magnetic atomic chains, 
superconductor-semiconductor nanowire hybrid junctions, multi-terminal Josephson junctions, irradiated 
junctions, just to mention a few.

The rest of this paper is organized as follows. First, in Sec.~\ref{sec:fcs} we remind the basics of FCS in 
the context of electronic quantum transport. Then, Sec.~\ref{Sec:action} presents the details on how the FCS 
can be obtained in practice in all the systems analyzed in this work with the help of a powerful Keldysh action.
Section~\ref{sec:NS-scatt-matrix} shows how this action can be used in combination with a mean-field 
model for the YSR states to describe the FCS in single-impurity junctions. The results of this combination 
are discussed in detail in Sec.~\ref{Sec:NS} for the case in which an impurity is coupled to a 
superconducting lead and a normal reservoir. In particular, we focus on the analysis of the resonant
Andreev reflection that can take place in the presence of YSR states and we present a detailed discussion 
on very recent shot noise experiments \cite{Thupakula2022}. This analysis is extended to the case 
of single-magnetic impurities with two superconducting electrodes in Sec.~\ref{Sec:SS}. In this case, we focus 
on the prediction of the signatures of MARs in the noise and Fano factor to guide future experiments. 
Section~\ref{Sec:2imp} deals with the case of two-impurity junctions in close connection with recent 
experiments \cite{Huang2020b,Huang2021}, and shows how the tunneling between individual YSR states is 
revealed in the current fluctuations. Finally, we present an outlook and our main conclusions in 
Sec.~\ref{sec:conclusions}. Some of the technical details related to the Keldysh action and the corresponding 
Green's functions are discussed in Appendices \ref{App:GFs} and \ref{App:action}.

\section{Full Counting Statistics: A Reminder} \label{sec:fcs}

The electronic transport in a quantum device can be viewed as a stochastic process which can be completely 
characterized by a probability distribution. Most experiments focus on the measurement of the electrical 
current, i.e. the average of that distribution. However, it has been shown in numerous systems that 
nonequilibrium current fluctuations (shot noise), i.e., the second cumulant of the current distribution, 
contain very valuable information out the scope of current measurements such as the charge of the carriers 
or the distribution of conduction channels \cite{Blanter2000,Nazarov2003}. Ideally, one would like to have 
access to all the cumulants of the current distribution to completely characterize the electronic transport. 
This idea was developed in the context of quantum optics and led to the introduction of the concept of photon 
counting statistics \cite{Scully1997}, which turned out to be key to characterize quantum states of the 
light such as those realized in a laser. In the 1990s, Levitov and Lesovik adapted this concept to 
mesoscopic electron transport \cite{Levitov1993,Levitov1996}, in which the electrons passing a certain 
conductor are counted. Later on, Nazarov and coworkers showed that these ideas could be combined with the 
powerful Keldysh-Green's function approach \cite{Nazarov1999,Nazarov2003}, which enabled the analysis of the 
electron counting statistics in numerous situations including those that involve superconducting electrodes
\cite{Belzig2001a,Belzig2001b,Belzig2003}. In this section, we briefly remind the reader of the basics of 
full counting statistics in the context of quantum electronic devices, and in Sec.~\ref{Sec:action} we 
shall address how it can be computed in the situations in which we are interested in this work, namely 
superconducting hybrid junctions featuring YSR states.   

Since the charge transfer in any junction is fundamentally discrete, one can aim at counting those charges. 
In a measuring time $t_0$, there is a certain likelihood for $N$ particles to cross the junction, which is 
described by a set of probabilities $\{P_{t_0}(N) \}_{N }$, the so-called \emph{full counting 
statistics} (FCS), which contains all the information concerning the possible transport processes. In practice, 
the FCS can be obtained from the so-called cumulant generating function (CGF) $\mathcal{A}_{t_0}(\chi)$, which 
is given by
\begin{equation}\label{Eq:CGF}
    \mathcal{A}_{t_0}(\chi)= \ln \left( \sum_{N} P_{t_0}(N) e^{\imath N \chi}\right),
\end{equation}
where $\chi$ is the so-called \emph{counting field}. Notice that because $\sum_{N }P_{t_0}(N)=1$, 
the CGF is normalized: $\mathcal{A}_{t_0}(0) = 0$. The counting field $\chi$ is the conjugate variable to the 
charge number $N$ and it is used as an auxiliary variable. By performing derivatives of the CGF with respect 
to the counting field, one obtains the cumulants $C_n$ as follows
\begin{equation}
    C_n = (-\imath)^n \frac{\partial^n \mathcal{A}_{t_0}(\chi)}{\partial \chi^n} \Big |_{\chi = 0} .
\end{equation}
Thus, for instance, the first cumulant reads
\begin{equation} \label{eq-C1}
    C_1 = \sum_{N } NP_{t_0}(N) = \langle N \rangle ,
\end{equation}
which is the expectation value of the charge number. Physically, it corresponds to the current $I$ averaged 
over the time interval $t_0$, $I = eC_1/t_0$. The second cumulant is given by
\begin{equation} \label{eq-C2}
    C_2 = \sum_{N} N^2 P_{t_0}(N) - \left(\sum_{N } N P_{t_0}(N) \right)^2 = 
    \langle N^2 \rangle - \langle N \rangle^2 ,
\end{equation}
which describes the variance. This second cumulant can be related to the zero-frequency shot noise $S$ via
$S=2e^2C_2/t_0$. An important measurable quantity is the so-called \emph{Fano factor}, which is given by
\begin{equation}
    F^{\ast} = \frac{S}{2e|I|} = \frac{C_2}{|C_1|} ,
\end{equation}
which is easily accessible in the framework of FCS. The Fano factor is a measure of the effective charge of 
the carriers in a system when the junction transmission is low and tunneling events are uncorrelated (Poissonian 
limit). In superconducting (SC) junctions, the Fano factor can be super-Poissonian ($F^{\ast}>1$), indicating 
that the transferred charge is larger than one due to the occurrence of Andreev reflections. In contrast, 
resonant tunneling might result in a so-called sub-Poissonian ($F^{\ast} < 1$) Fano factor, where the 
interpretation of the Fano factor as an effective charge breaks down \cite{Blanter2000}.

In the simple case of a two-terminal device featuring a single conduction channel characterized by an 
energy-independent transmission coefficient $\tau$, the CGF at zero temperature reads (ignoring spin)
\begin{equation}
    \mathcal{A}_{t_0}(\chi) = \frac{e V t_0}{h} \ln \left( 1+ \tau (e^{\imath \chi}-1) \right) ,
\end{equation}
where $V$ is the bias voltage between the two terminals. The interpretation of this result is that the 
transport is dominated by single-electron tunneling with a probability $P_1 = \tau$. This interpretation 
becomes apparent when considering the corresponding full counting statistics, which is given by
\begin{equation}
    P_{t_0}(N) = \binom{M}{N} \tau^N (1 - \tau)^{M-N} .
\end{equation}
This is a binomial distribution where we define the number of attempts $M = \lceil eVt_0/h \rceil $, 
where $\lceil \ \rceil$ describes the next highest integer. This is justified by the fact that $t_0$ is chosen 
to be sufficiently large. Hence, it is evident that the transport can be described as $N$ particles being 
transmitted in individual tunneling processes transferring a single electron charge with a probability $\tau$. 

In the case of junctions containing SCs, there are additional multi-particle tunneling processes, namely 
MARs, whose probabilities can also be obtained from the knowledge of the CGF. As we shall show below, in
the case where there is no spin-flip scattering the CGF of a superconducting contact can be expressed as
\cite{Cuevas2003,Cuevas2004}  
\begin{equation}\label{Eq:CGF_p}
    \mathcal{A}_{t_0}(\chi) = \frac{t_0}{2h} \sum_{\sigma}\int d E \ln \left[ 
    \sum^{\infty}_{n=-\infty} P_n^\sigma(E,V) e^{\imath n \chi} \right] ,
\end{equation}
where $P_n^\sigma(E,V)$ is the spin-, energy- and bias-dependent probability of a transport process 
transferring $n$ electron charges with spin $\sigma$. Thus, the evaluation of the CGF in the different 
physical situations that we shall address in this work will allow us to classify the different tunneling 
processes that can take place according to the charge they transfer, something that cannot be objectively
done with any other theoretical method. Moreover, from the knowledge of those probabilities we can readily 
obtain the different transport quantities: current, noise, and higher-order cumulants. In the most 
general case of spin-flip, like in the example of Section~\ref{Sec:2imp}, the main difference will be the
impossibility to spin-resolve the tunneling probabilities (see discussion below).

\section{Keldysh Action} \label{Sec:action}

Our goal is to obtain the probabilities of the different tunneling processes $P_n(E,V)$ from the knowledge 
of the CGF, like in Eq.~(\ref{Eq:CGF_p}). Here, we shall make use of a remarkable result obtained by Snyman 
and Nazarov \cite{Snyman2008,Nazarov2015} in which the CGF of an arbitrary mesoscopic/nanoscopic device can be 
expressed in terms of two basic ingredientes: (i) the Green's functions of the electron reservoirs or leads, 
which can be superconducting, and (ii) the scattering matrix of the device in the normal state. Technically 
speaking, these researchers showed that the CGF of any type of junction (ignoring inelastic interactions) 
can be expressed as (in this section we drop the subindex $t_0$ and omit the prefactor $t_0/h$)
\begin{equation}\label{action}
    \mathcal{A}(\chi) = \frac{1}{2}\mathrm{Tr} \ln \left[ \underbrace{ \frac{\hat{1} + \hat{G}(\chi)}{2} + 
    \hat{S}\frac{\hat{1}-\hat{G}(\chi)}{2}}_{\hat{Q}(\chi)} \right] - \frac{1}{2}\mathrm{Tr} \ln \hat{Q}(0),
\end{equation}
where $\hat{G}(\chi)$ contains the information of the reservoir Green's functions (GFs) and $\hat{S}$ is 
the normal-state scattering matrix whose structure will be explained in what follows. First, $\hat{G}(\chi)$ 
is a GF in the lead-time-Keldysh-spin-Nambu space, denoted by the symbol $(\ \hat{} \ )$. In this work, we 
focus on two-terminal settings for which this GF adopts the generic form
\begin{equation}
    \hat{G}(\chi) = \mathrm{diag}(G_{\mathrm{L}}(\chi), G_{\mathrm{R}}),
\end{equation}
which is a block-diagonal matrix with the two time-Keldysh-spin-Nambu GFs as entries on the diagonal. 
The counting field, similarly to the voltage, can be gauged away from the right onto the left lead. In what 
follows the right terminal will always be superconducting $G_{\mathrm{R}} = G_{\mathrm{SC}}$, while the left 
terminal can be either normal or superconducting. The GFs are in fact infinite matrices in time space, which 
means that the element $(t,t^\prime)$ of the GF $G_{\mathrm{L,R}}(\chi)$ is the Keldysh GF 
$\check{g}_{\rm L,R}(\chi,t,t^\prime)$. In other words, the lead GF is an infinitely large matrix in time space 
whose entries are $16 \times 16$-matrices in lead-Keldysh-spin-Nambu space. For a detailed explanation of the 
structure of the GFs we refer to Appendix \ref{App:GFs}. Equivalently, the normal state scattering matrix 
$\hat{S}$ is also an infinite matrix in time space whose $(t,t^\prime)$ entry is a $16 \times 16$-matrix in 
lead-Keldysh-spin-Nambu space, which we denote as $\tilde{s}(t,t^\prime)$, where the symbol $(\ \tilde{}\ )$ indicates 
that the scattering matrix is expressed in lead-Keldysh-spin-Nambu space. However, the scattering matrix 
only depends on the relative time $t_{\mathrm{rel}}=t-t^\prime$, meaning that $\tilde{s}(t,t^\prime) = \tilde{s}
(t_{\mathrm{rel}})$. To elaborate on the structure of the scattering matrix, in a two-terminal 
setting it can be written in lead space as follows
\begin{equation}
    \tilde{s}(t_{\rm rel}) = \begin{pmatrix}
    \check{\mathfrak{r}}(t_{\mathrm{rel}}) & \check{\mathfrak{t}}^\prime (t_{\mathrm{rel}}) \\ 
    \check{\mathfrak{t}}(t_{\mathrm{rel}}) & \check{\mathfrak{r}}^\prime(t_{\mathrm{rel}})
    \end{pmatrix},
\end{equation}
where $\check{\mathfrak{t}}(t_{\mathrm{rel}})$ is the transmission matrix and $\check{\mathfrak{r}}(t_{\mathrm{rel}})$ 
the reflection matrix in Keldysh-spin-Nambu space. Upon a Fourier transformation the scattering matrix reads
\begin{equation}
     \tilde{s}(E) = \begin{pmatrix}
    \check{\mathfrak{r}}(E) & \check{\mathfrak{t}}^\prime(E) \\ \check{\mathfrak{t}}(E) & 
    \check{\mathfrak{r}}^\prime(E) \end{pmatrix}.
\end{equation}
The scattering matrix is unitary, meaning that $\tilde{s}(E)\tilde{s}^\dagger(E) = \tilde{1}$, where 
$\tilde{1}$ is the identity in lead-Keldysh-spin-Nambu space. If the transport preserves time-reversal 
symmetry, the scattering matrix is symmetric, meaning that $\tilde{s}(E) = \tilde{s}^T(E)$. The scattering 
matrix depends on the system under consideration and below we shall show how it can be obtained from models 
for the description of YSR states. 

Fourier transforming, we can write Eq.~(\ref{action}) in a Floquet representation as follows
\begin{align} \label{eq:action-v2}
    \mathcal{A}(\chi) =& \frac{1}{2}\mathrm{Tr} \ln \left[\underbrace{ \frac{\breve{1} + \breve{\cal G}(\chi,E)}{2} + 
    \breve{\cal S}(E)\frac{\breve{1}-\breve{\cal G}(\chi,E)}{2}}_{\breve{ \cal Q}(\chi,E)}\right] \nonumber \\
    &-\frac{1}{2}\mathrm{Tr} \ln \breve{ \cal Q}(0,E) ,
\end{align}
with the matrices $\breve{\cal G}(\chi,E)$ and $\breve{\cal S}(E)$ both expressed in 
Floquet-lead-Keldysh-spin-Nambu space which are the Floquet representations of the GF $\hat{G}(\chi)$ and the 
scattering matrix $\hat{S}$ respectively. In addition, we define the matrix $\breve{ \cal Q}(\chi,E)$. The trace 
that initially went over time space goes now about the Floquet space and Eq.~(\ref{eq:action-v2}) can be 
rewritten in terms of the Floquet energy $E \in [-eV,eV]$ as follows (see Appendix \ref{App:action} for more details)
\begin{equation}
    \mathcal{A}(\chi) = \frac{1}{2}\int_{-eV}^{eV} dE \, \mathrm{Tr} \left( \ln 
    \left[ \breve{\cal Q}(\chi,E)\right]-\ln 
    \left[ \breve{\cal Q}(0,E)\right] \right).
\end{equation}
The trace now goes over the matrix $ \breve{\cal Q}(\chi,E)$, which is infinite in 
the Floquet index with its entries being $16\times 16$-matrices in lead-Keldysh-spin-Nambu space. Note, that 
we can use that $\mathrm{Tr} \ln \breve{\cal Q}= \ln \det(\breve{\cal Q})$. In addition, in the case of single-impurity
junctions the matrix $\breve{\cal Q}(\chi,E)$ is a block-diagonal matrix in spin space and thus the determinant 
factorizes into two contributions, a spin $\Uparrow$ and $\Downarrow$ contribution, $\breve{\cal Q}^\Uparrow$ and 
$\breve{\cal Q}^\Downarrow$, which leads to the result
\begin{eqnarray}
    \mathcal{A}(\chi) & = & \frac{1}{2}\int_{-eV}^{eV} dE \, \left( \ln \det 
    \left[ \breve{\cal Q}(\chi,E)\right] - \ln \det \left[\breve{\cal Q}(0,E)\right] \right) \nonumber \\  & = &  \frac{1}{2}
    \int_{-eV}^{eV} dE \, \ln \frac{\det \breve{\cal Q}(\chi,E)}{\det \breve{\cal Q}(0,E)}  \nonumber \\
    & = & \frac{1}{2}\sum_{\sigma = \Uparrow,\Downarrow} \int_{-eV}^{eV} dE \, \ln 
    \frac{\det \breve{\cal Q}^\sigma(\chi,E)} {\det \breve{\cal Q}^\sigma(0,E)} \nonumber \\
    & = &\frac{1}{2}\sum_{\sigma = \Uparrow, \Downarrow} \int_{-eV}^{eV} dE \, 
    \ln \mathcal{P}^\sigma(\chi,E) , \label{CGF2}
\end{eqnarray}
where we have defined the charge- and spin-resolved counting polynomials $\mathcal{P}^\sigma(\chi,E)$. 
Note, that these counting polynomials follow from the determinant of infinitely large matrices, namely 
$\breve{\cal Q}^\sigma$. In addition, it holds that $\mathcal{P}^\sigma(0,E) = 1$. Therefore, it is evident 
that the counting polynomials take the following form
\begin{eqnarray}
    \mathcal{P}^\sigma(\chi,E) = \sum_{n=-\infty}^{\infty} P_n^\sigma(E)e^{\imath n \chi} ,
\end{eqnarray}
where we encounter the spin-, charge- and energy-resolved tunneling probabilities $P_n^\sigma(E)$ with 
the respective counting factor $e^{\imath n \chi}$. Thus, we obtain the action in the form of Eq.~(\ref{Eq:CGF_p}) 
by adding the prefactor $t_0/h$.

\section{Single-impurity YSR junctions: Model and scattering matrix} \label{sec:NS-scatt-matrix}

\begin{figure}[t]
\includegraphics[width=0.85\columnwidth,clip]{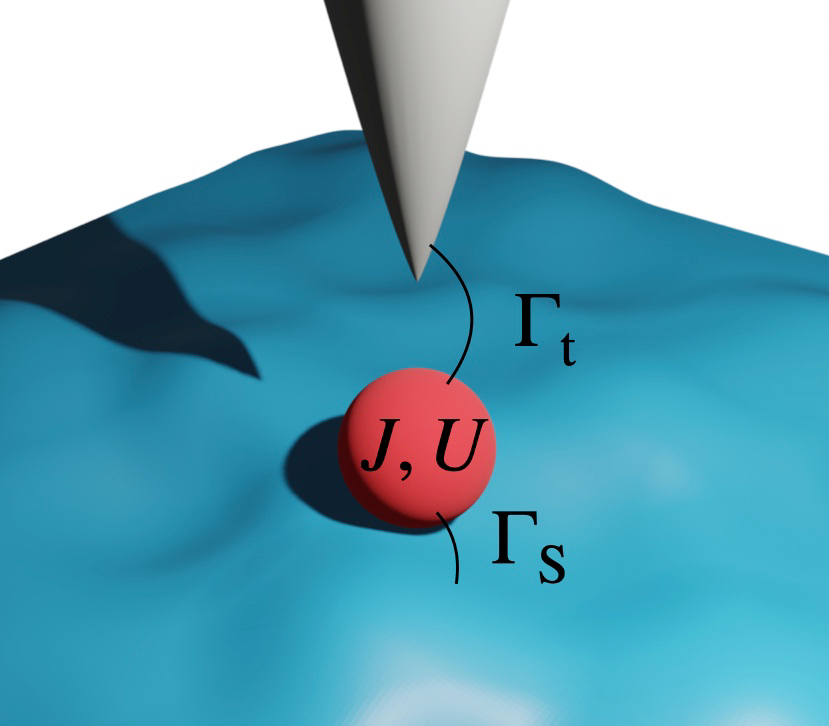}
\caption{Schematic representation of a magnetic impurity coupled to a superconducting substrate and 
to an STM tip that can be either normal or superconducting. The tunneling rates $\Gamma_{\rm t}$ 
and $\Gamma_{\rm S}$ describe the strength of the coupling of the impurity to the tip and substrate, 
respectively. The superconducting gaps of the electrodes are denoted by $\Delta_{\rm t}$ and 
$\Delta_{\rm S}$.}
\label{fig-system}
\end{figure}

The goal of this section is to use the concept of full counting statistics to describe the electronic transport 
properties of a junction featuring YSR states in which a single magnetic impurity is coupled to superconducting 
leads. As we showed in Sec.~\ref{Sec:action}, we need as an input a normal-state scattering matrix describing 
these types of junctions. To obtain it, we make use of a mean-field Anderson model with broken spin symmetry 
(see Ref.~\cite{Villas2020} and references therein for a discussion on its origin and range of 
applicability), and which has been very successful describing different transport 
characteristics \cite{Huang2020a,Karan2022}. This model, which is illustrated in Fig.~\ref{fig-system}, 
describes the experimentally relevant situation in which a magnetic impurity (an atom or a molecule) is coupled 
to a superconducting substrate (S) and to an STM tip (t), which can also be superconducting. The model 
used here is summarized in the following Hamiltonian
\begin{equation}
\label{eq-total-H}
H = H_{\rm t} + H_{\rm S} + H_{\rm imp} + H_\mathrm{hopping} .
\end{equation}
Here, $H_j$ (with $j={\rm t,S}$) is the BCS Hamiltonian of the lead $j$ given by
\begin{multline}
H_j = \sum_{{\boldsymbol k} \sigma} \xi_{{\boldsymbol k}j} 
c^{\dagger}_{{\boldsymbol k}j\sigma} c_{{\boldsymbol k}j\sigma} \\
+ \sum_{\boldsymbol k} \left( \Delta_j e^{i\varphi_j} 
c^{\dagger}_{{\boldsymbol k}j\uparrow} c^{\dagger}_{{-\boldsymbol k}j\downarrow} 
+ \Delta_j e^{-i\varphi_j} c_{{-\boldsymbol k}j\downarrow} c_{{\boldsymbol k}j\uparrow}
\right) ,
\end{multline}
where $c^{\dagger}_{{\boldsymbol k}j\sigma}$ and $c_{{\boldsymbol k}j\sigma}$ are the creation and annihilation 
operators, respectively, of an electron of momentum ${\boldsymbol k}$, energy $\xi_{{\boldsymbol k}j}$, and spin
$\sigma=\uparrow, \downarrow$ in lead $j$, $\Delta_j$ is the superconducting gap, and $\varphi_j$ is the 
corresponding superconducting phase. On the other hand, $H_{\rm imp}$ is the Hamiltonian of the magnetic impurity, 
which reads
\begin{equation}
H_{\rm imp}	= U(n_{\uparrow} + n_{\downarrow}) + J (n_{\uparrow} - n_{\downarrow}) .
\end{equation}
Here, $n_{\sigma} = d^{\dagger}_{\sigma} d_{\sigma}$ is the occupation number operator on the impurity, 
$U$ is the on-site energy, and  $J$ is the exchange energy that breaks the spin degeneracy on the impurity. 
Finally, $H_\mathrm{hopping}$ describes the coupling between the magnetic impurity and the leads and adopts 
the form
\begin{equation}
H_{\rm hopping} = \sum_{{\boldsymbol k},j,\sigma} t_j \left(d^{\dagger}_{\sigma}
c_{{\boldsymbol k}j\sigma} + c^{\dagger}_{{\boldsymbol k}j\sigma} d_{\sigma} 
\right) ,
\end{equation}
where $t_j$ describes the tunneling coupling between the impurity and the lead $j=\mathrm{t,S}$ and it is 
chosen to be real.

It is convenient to rewrite the previous Hamiltonian in terms of four-dimensional spinors that live in a 
space resulting from the direct product of the spin space and the Nambu (electron-hole) space. In the case 
of the leads, the relevant spinor is defined as
\begin{equation}
\label{eq-c-tilde}
\bar c^{\dagger}_{{\boldsymbol k}j} = \left( c^{\dagger}_{{\boldsymbol k}j\uparrow},
c_{{-\boldsymbol k}j\downarrow}, c^{\dagger}_{{\boldsymbol k}j\downarrow} , -c_{{-\boldsymbol k}j\uparrow} \right) ,
\end{equation}
while for the impurity states we define
\begin{equation}
\label{eq-d-tilde}
\bar d^{\dagger} = \left( d^{\dagger}_{\uparrow}, d_{\downarrow}, d^{\dagger}_{\downarrow}, -d_{\uparrow} \right) .
\end{equation}
Using the notation $\tau_i$ and $\sigma_i$ ($i=1,2,3$) for Pauli matrices in Nambu and spin space, respectively, 
and with $\tau_0$ and $\sigma_0$ as the unit matrices in those spaces, one can show that the Hamiltonian in 
Eq.~(\ref{eq-total-H}) can be cast into the form
\begin{subequations}
\begin{eqnarray}
H_j & = & \frac{1}{2} \sum_{\boldsymbol k} \bar c^{\dagger}_{{\boldsymbol k}j} 
\bar H_{{\boldsymbol k} j} \bar c_{{\boldsymbol k}j} , \\
H_{\rm imp} & = & \frac{1}{2} \bar d^{\dagger} \bar H_{\rm imp} \bar d ,	 \\
H_{\rm hopping} & = & \frac{1}{2} \sum_{{\boldsymbol k},j} \left\{
\bar c^{\dagger}_{{\boldsymbol k}j} \bar V_{j,{\rm imp}} \bar d +
\bar d^{\dagger} \bar V_{{\rm imp},j} \bar c_{{\boldsymbol k}j} \right\} ,
\end{eqnarray}
\end{subequations}
where
\begin{subequations}
\begin{eqnarray}
\bar H_{{\boldsymbol k} j} & = & \sigma_0 \otimes (\xi_{\boldsymbol k} \tau_3 + 
\Delta_j e^{i \varphi_j \tau_3} \tau_1 ), \\
\bar H_{\rm imp} & = & U (\sigma_0 \otimes \tau_3) + J (\sigma_3 \otimes \tau_0) , \\
\bar V_{j,{\rm imp}} & = & t_j(\sigma_0 \otimes \tau_3) = 
\bar V^{\dagger}_{{\rm imp},j} .
\end{eqnarray}
\end{subequations}

The $k$-dependent retarded and advanced GFs of the leads can be expressed in spin-Nambu space as a function 
of the energy as
\begin{equation}
    \bar{g}^{\rm r/a}_{\bm{k},j}(E) = \left( E \pm \imath \eta_j - \bar{H}_{\bm{k},j}\right)^{-1},
\end{equation}
where the Dynes' parameters $\eta_j$ are introduced. The $k$-dependence can be eliminated by summing over 
it and defining the spin-Nambu GF
\begin{eqnarray}
    \bar{g}^{\rm r/a}_j(E) & = & \sum_{\bm{k}} \bar{g}_{\bm{k},j}^{\rm r/a}(E) \\ & = & 
    \sigma_0 \otimes \frac{-1}{\sqrt{\Delta_j^2-(E\pm \imath \eta_j)^2}} 
    \begin{pmatrix} E\pm \imath \eta_j & \Delta_je^{\imath \phi_j} \\ \Delta_j e^{-\imath \phi_j} & 
    E\pm \imath \eta_j \end{pmatrix} , \nonumber
\end{eqnarray}
where $\phi_j$ is the superconducting phase of the order parameter of electrode $j$. In the case of a 
normal metal, the previous expression reduces to
\begin{equation}
    \bar{g}^{\rm r/a}_{\mathrm{N}} = \mp \imath \sigma_0 \otimes \tau_0. 
\end{equation}
Equivalently for the impurity, its GF is given by
\begin{equation}
    \bar{g}_{\mathrm{imp}}^{\rm r/a} = \left( E \pm \imath \eta_{\mathrm{imp}} -\bar{H}_{\mathrm{imp}}\right)^{-1},
\end{equation}
where $\eta_{\mathrm{imp}}$ is the regularization constant of the impurity GF. It will later be evident 
that it can be set to zero. In the following, we focus on the electron space of Nambu space. Namely, on the 
first and third component of the spinors in Eq.~(\ref{eq-c-tilde}) and Eq.~(\ref{eq-d-tilde}). Of special 
interest are the electron self-energies of the two leads which are given by
\begin{align}
    \bar{\Sigma}_{\rm t,e}^{\rm r/a} &= \bar{V}_{\rm imp, t,e} \, \bar{g}^{\rm r/a}_{\rm t,e} \, \bar{V}_{\rm t, imp,e} \\
    \bar{\Sigma}_{\rm S,e}^{\rm r/a} &= \bar{V}_{\rm imp, S,e} \, \bar{g}^{\rm r/a}_{\rm S,e} \, \bar{V}_{\rm S, imp,e},
\end{align}
where the index $( {}_{\rm e}  ) $ refers to extracting the first and third component of the respective quantity 
in spin-Nambu space. The dressed electron impurity retarded GF can be calculated using the self-energies as 
follows
\begin{equation}
    \bar{G}_{\mathrm{imp,e}}^{\rm r/a} = \left( (E\pm \imath \eta_{\mathrm{imp}}) \bar{1} - 
    \bar{H}_{\mathrm{imp,e}} -\bar{\Sigma}_{\mathrm{t,e}}^{\rm r/a}-\bar{\Sigma}_{\mathrm{S,e}}^{\rm r/a} \right)^{-1}.
\end{equation}
We define the electron coupling matrix in spin space with $\bar{\Gamma}_{t/S,e}$ by taking the imaginary 
part of the self-energy
\begin{equation}
    \bar{\Gamma}_{\rm t/S,e} = \Im \left( \bar{\Sigma}_{\rm t/S,e}^{\rm a} \right).
\end{equation}

Let us recall that in the usual regime in which the STM experiments are operated $\Gamma_{\rm t} \ll 
\Gamma_{\rm S}$, this model predicts the appearance of a pair of fully spin-polarized YSR bound states in 
the limit $J \gg \Delta_\mathrm{S}$, and they are inside the gap when also $\Gamma_\mathrm{S} \gg 
\Delta_\mathrm{S}$. In this case, the energy of the YSR states (measured with respect to the Fermi 
energy) is given by \cite{Villas2020}
\begin{equation} \label{eq-YSR1}
\epsilon_{\rm YSR} = \pm \Delta_{\rm S} \frac{J^2 - \Gamma^2_{\rm S} - U^2} {\sqrt{ \left[ 
\Gamma^2_{\rm S} + (J-U)^2 \right] \left[ \Gamma^2_{\rm S} + (J+U)^2 \right]}} ,
\end{equation}
In the electron-hole symmetric case $U=0$, the previous expression reduces to 
\begin{equation} \label{eq-YSR2}
\epsilon_{\rm YSR} = \pm \Delta_{\rm S} \frac{J^2 - \Gamma^2_{\rm S}} {J^2 + \Gamma^2_{\rm S}} . 
\end{equation}
The entries of the electron scattering matrix, namely the electron reflection matrices 
$\mathfrak{r}_{\rm e},\mathfrak{r}^\prime_{\rm e}$ and the transmission matrices $\mathfrak{t}_{\rm e},
\mathfrak{t}^\prime_{\rm e} $, can be computed using the Fisher-Lee relations following Ref.~\cite{Cuevas2017}
\begin{align}
   \bar{\mathfrak{r}}_{\rm e} &= \sigma_0-2\imath \left(\bar{\Gamma}_{\mathrm{t,e}}\right)^{1/2} 
   \bar{G}_{\mathrm{imp,e}}^{\rm r} (\bar{\Gamma}_{\rm t,e})^{1/2} \\
   \bar{\mathfrak{t}}^\prime_{\rm e} &= 2 \left(\bar{\Gamma}_{\mathrm{t,e}}\right)^{1/2} 
   \bar{G}_{\mathrm{imp,e}}^{\rm r} (\bar{\Gamma}_{\rm S,e})^{1/2} \\
    \bar{\mathfrak{t}}_{\rm e}&= 2 \left(\bar{\Gamma}_{\mathrm{S,e}}\right)^{1/2} 
    \bar{G}_{\mathrm{imp,e}}^{\rm r} (\bar{\Gamma}_{\rm t,e})^{1/2} \\
   \bar{\mathfrak{r}}^\prime_{\rm e} &= -\sigma_0 + 2\imath \left(\bar{\Gamma}_{\mathrm{S,e}}\right)^{1/2}
   \bar{G}_{\mathrm{imp,e}}^{\rm r} (\bar{\Gamma}_{\rm S,e})^{1/2}.
\end{align}
The hole-components of the scattering matrix follow from the electron-components with \cite{Nazarov2015}
\begin{equation}
    s_{\rm h}(E) = \sigma_2 s_{\rm e}(-E)^{\rm T} \sigma_2 .
\end{equation}
It is important to remark that in the action of Eq.~(\ref{action}), the scattering matrix is the 
normal-state one. Hence, to obtain the desired result, we have to evaluate the transmission matrix in the 
case in which both reservoirs are in the normal state, i.e., when the corresponding GFs are given by 
$\bar{g}_j^{\rm r/a} = \bar{g}_{\mathrm{N}}^{\rm r/a}$. Thus, for instance, it is straightforward to show that 
the transmission matrix $\bar{\mathfrak{t}}$ in the spin-Nambu basis $(\Psi_\uparrow^\dagger,
\Psi_\downarrow,\Psi_\downarrow^\dagger,-\Psi_\dagger)$ is given by the following diagonal matrix
\begin{widetext}
\begin{equation}
\bar{\mathfrak{t}} = \begin{pmatrix} t_{\rm e}^\Uparrow(E) & & &\\ & t_{\rm h}^\Uparrow(E) & & \\ & & t_{\rm e}^\Downarrow(E) & 
\\ & & & t_{\rm h}^\Downarrow(E)
\end{pmatrix} = \begin{pmatrix} \frac{2\sqrt{\Gamma_{\mathrm{t}} \Gamma_{\mathrm{S}}}}{E-U-J+\imath(\Gamma_{\mathrm{S}} +
\Gamma_{\mathrm{t}})} & & & \\ &\frac{2\sqrt{\Gamma_{\mathrm{t}} \Gamma_{\mathrm{S}}}}{-E-U+J+\imath(\Gamma_{\mathrm{S}} +
\Gamma_{\mathrm{t}})} & & \\ & & \frac{2\sqrt{\Gamma_{\mathrm{t}} \Gamma_{\mathrm{S}}}}{E-U+J+\imath(\Gamma_{\mathrm{S}} + 
\Gamma_{\mathrm{t}})}& \\ & & &\frac{2\sqrt{\Gamma_{\mathrm{t}} \Gamma_{\mathrm{S}}}}{-E-U-J+\imath(\Gamma_{\mathrm{S}} + 
\Gamma_{\mathrm{t}})} \end{pmatrix} ,
\end{equation}
\end{widetext}
where we have defined the tunneling rates $\Gamma_{\rm t/S} = \pi N_{0, \rm t/S} t^2_{\rm t/S}$,
where $N_{0, \rm t/S}$ corresponds to the normal density of state of the corresponding electrode. The 
tunneling rates describe the strength of the coupling between the impurity and the corresponding lead 
($j={\rm t,S}$). It is evident that the impurity regularization constant can be set to zero 
$\eta_{\rm imp}\approx 0$ as the substrate tunneling rate is always orders of magnitude larger than it.

Due to the unitary of the scattering matrix, it can be written in terms of the transmission matrix 
$\bar{\mathfrak{t}}$ as follows
\begin{equation}
  \tilde{s}(E) =\begin{pmatrix} \bar{\mathfrak{r}} & \bar{\mathfrak{t}}^\prime \\ \bar{\mathfrak{t}} & 
  \bar{\mathfrak{r}}^\prime \end{pmatrix}
  = \begin{pmatrix}1-\imath \bar{\mathfrak{t}}\sqrt{\frac{\Gamma_{\mathrm{t}}}
  {\Gamma_{\mathrm{S}}}} & \bar{\mathfrak{t}} \\ 
  \bar{\mathfrak{t}} & -\frac{\bar{\mathfrak{t}}}{\bar{\mathfrak{t}}^*}\left(1-\imath 
  \bar{\mathfrak{t}}\sqrt{\frac{\Gamma_{\mathrm{t}}} {\Gamma_{\mathrm{S}}}}\right)^{\ast} \end{pmatrix},
\end{equation}
which is a $8\times 8$-matrix in lead-spin-Nambu space. Note that the scattering matrix is proportional 
to the identity in Keldysh space in this case, thus its Keldysh structure is not included in the above formulas.

There are some interesting limiting cases to be discussed. First, in the case in which the energy of 
the impurity level is $U = 0$, the electron- and hole-transmission function are related via
\begin{equation}
    t_{\rm e}^\Uparrow(E) = -t_{\rm h}^\Uparrow(E)^*,
\end{equation}
and thus $|t_{\rm e}^\Uparrow|^2 = |t_{\rm h}^\Uparrow|^2$, so the transmission is the same for electrons and holes. 
Another interesting case is the high-coupling regime where $\Gamma_{\mathrm{L}}, \Gamma_{\mathrm{R}}\gg U,J$. 
In that case, the transmission at low energies adopts the form
\begin{equation}\label{highcoupling}
    t_{\rm e}^\Uparrow(E) = \frac{2\sqrt{\Gamma_{\mathrm{L}} \Gamma_{\mathrm{R}}}}{\imath (\Gamma_{\mathrm{R}} + 
    \Gamma_{\mathrm{L}})} = t_{\rm h}^\Uparrow(E),
\end{equation}
so the transmission matrix for electrons and holes are the same. In the case where $\Gamma_{\mathrm{t}} = 
\Gamma_{\mathrm{S}}$, it holds that $r_{\rm e}^\Uparrow = r_{\rm h}^\Uparrow \approx 0$ and the junction 
behaves as a single-channel highly transmissive point contact.

\section{Single-impurity YSR junctions: Normal conducting tip} \label{Sec:NS}

\begin{figure*}[t]
\includegraphics[width=0.9\textwidth,clip]{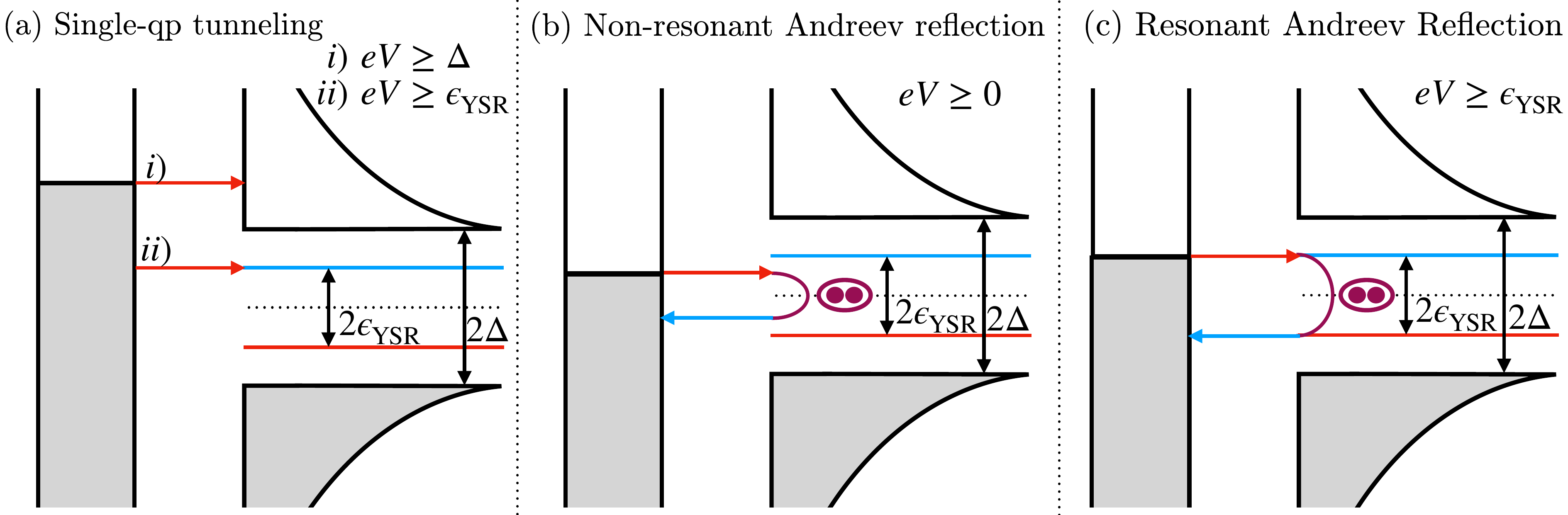}
\caption{Tunneling processes in the case of an impurity coupled to a normal and to a superconducting lead. 
In these diagrams the left density of states (DOS) corresponds to a normal STM tip and the right one to the 
impurity coupled to the superconducting substrate. The red lines correspond to electron-like quasiparticles 
and the blue ones to hole-like. In all cases, we indicate the threshold voltage at which the process starts 
to contribute to the transport. (a) Single-quasiparticle tunneling in which a quasiparticle tunnels either 
into the continuum DOS of the superconducting electrode (i) or resonantly into the excited YSR state inside 
the gap (ii). (b) Standard non-resonant Andreev reflection in which an electron is reflected as a hole. 
(c) A resonant Andreev reflection in which the electron that is retro-reflected impinges at the energy of 
a YSR state.}
\label{fig-NS-processes}
\end{figure*}

We now analyze the situation in which the STM tip is in the normal state, while the substrate is in the 
superconducting state. In this case the transport properties are a result of the competition between two 
tunneling processes, namely single-quasiparticle tunneling and an Andreev reflection, and the fact 
that both of them become resonant due to the presence of in-gap YSR states, see Fig.~\ref{fig-NS-processes}.
In this case, the GFs only depend on the relative time and we just have to integrate over all energies 
$E$ in the formula of the Keldysh action in Eq.~(\ref{action}). Moreover, in this single-impurity case 
the electrode GFs and the scattering matrix are block-diagonal in spin space. This allows us to carry out 
the calculation of the CGF analytically and the final result reads 
\begin{equation} \label{Action_NS}
    \mathcal{A}_{t_0}(\chi) = \frac{t_0}{2h}\sum_{\sigma = \Uparrow , \Downarrow} \int^{\infty}_{-\infty} dE \ln 
    \left[ \sum_{n=-2}^{2} P_n^\sigma(E,V) e^{\imath n\chi} \right] ,
\end{equation}
where $P_n^\sigma(E,V)$ corresponds to the probability of transferring $n$ charges across the junction 
for spin $\sigma$. The single-quasiparticle tunneling probabilities from tip to substrate ($n=1$) and 
from substrate to tip ($n=-1$) are given by
\begin{widetext}
\begin{eqnarray}
    P^\sigma_1(E,V) & = & \frac{\rho_{\mathrm{S}}}{D^\sigma} 
    \left[ (|t_{\rm e}^\sigma|^2-|t_{\rm e}^\sigma|^2|t_{\rm h}^\sigma|^2/2) f_1(1-f_{0}) +  
    (|t_{\rm h}^\sigma|^2-|t_{\rm e}^\sigma|^2 |t_{\rm h}^\sigma|^2/2) f_0(1-f_{-1}) \right] + \nonumber \\ & & 
    \frac{|t_{\rm e}^\sigma|^2|t_{\rm h}^\sigma|^2}{8D^\sigma} (1 + f^{\rm a}f^{\rm r} - g^{\rm a}g^{\rm r}) (2f_{0}-1)
    \left\{ (f_1+f_{-1}-1) \left[ 1 -(2f_{0}-1)(f_{1}+f_{-1}-1) \right] +  \right. \nonumber \\ & & 
    \hspace{5.8cm} \left. (2f_{0}-1) (1+f_{1}-f_{-1})(f_1-f_{-1}) \right\} , \label{eq-P1} \\ 
    P^\sigma_{-1}(E,V) & = & \frac{\rho_{\mathrm{S}}}{D^\sigma} 
    \left[ (|t_{\rm e}^\sigma|^2-|t_{\rm e}^\sigma|^2|t_{\rm h}^\sigma|^2/2) f_0(1-f_{1}) +  
    (|t_{\rm h}^\sigma|^2-|t_{\rm e}^\sigma|^2 |t_{\rm h}^\sigma|^2/2) f_{-1}(1-f_0) \right] + \nonumber \\ & & 
    \frac{|t_{\rm e}^\sigma|^2|t_h^\sigma|^2}{8D^\sigma} (1 + f^{\rm a}f^{\rm r} - g^{\rm a}g^{\rm r}) (2f_{0}-1)
    \left\{ (f_1+f_{-1}-1) \left[ 1 -(2f_{0}-1)(f_{1}+f_{-1}-1) \right] +  \right. \nonumber \\ & & 
    \hspace{5.8cm} \left. (2f_{0}-1) (1+f_{-1}-f_{1})(f_{-1}-f_{1}) \right\} , \label{eq-P2}
\end{eqnarray}
while the corresponding Andreev reflection probabilities from tip to substrate ($n=2$) and from substrate to tip 
($n=-2$) are given by
\begin{eqnarray}
P_2^\sigma(E,V) & = & \frac{|t_{\rm e}^\sigma|^2 |t_{\rm h}^\sigma|^2}{8 D^\sigma} f_1 (1-f_{-1})
 \left[ 1-f^{\rm a}f^{\rm r}-g^{\rm a}g^{\rm r}-(2f_{0}-1)^2(1+f^{\rm a}f^{\rm r}-g^{\rm a}g^{\rm r}) \right] , \\
P_{-2}^\sigma(E,V) & = & \frac{|t_{\rm e}^\sigma|^2 |t_{\rm h}^\sigma|^2}{8 D^\sigma} f_{-1} (1-f_1)
\left[ 1 - f^{\rm a}f^{\rm r} - g^{\rm a}g^{\rm r} -(2f_{0}-1)^2(1+f^{\rm a}f^{\rm r}-g^{\rm a}g^{\rm r}) \right]. 
\end{eqnarray}
\end{widetext}
In these expressions we have assumed that a bias voltage is applied to the normal metal, 
$g^{\rm r,a}(E) = -\imath (E\pm \imath \eta)/\sqrt{\Delta^2-(E\pm \imath \eta)^2}$ and 
$f^{\rm r,a}(E) = -\imath \Delta / \sqrt{\Delta^2-(E\pm \imath \eta)^2}$ are the GFs of the superconducting 
substrate, $\Delta$ being the gap of the SC electrode and $\eta$ the corresponding Dynes' parameter, 
$\rho_{\mathrm{S}} = (g^{\rm r}-g^{\rm a})/2$ is the substrate density of states (DOS), $f_n(E) = f(E+neV)$ with 
the Fermi function $f(E) = 1/(e^{E/k_{\rm B}T}+1)$, and the normalization factor $D^{\sigma}(E)$ is given by
\begin{equation}\label{ehtrans}
    D^\sigma = \left| 1 + \frac{1}{2} (1-r_{\rm e}^\sigma (r_{\rm h}^\sigma)^{\ast}) (g^r-1) \right|^2 .
\end{equation}
Let us say that these probabilities reduce to the known result for a NS junction with energy-independent
transmission and spin degeneracy \cite{Belzig2003}. Moreover, we have verified that they lead to the same
results for the current as in Ref.~\cite{Villas2020}. These probabilities have the expected structure. Thus,
for instance, the single-quasiparticle probabilities ($n=\pm 1$) are to a leading order proportional to 
the transmission coefficients (for electrons and holes) and to the DOS in the superconducting substrate. 
The Andreev reflection probabilities ($n=\pm 2$) are proportional to the product of the electron and 
hole transmission coefficients and to the Cooper pair density in the superconducting electrode (this 
will become more obvious in a moment). On the other hand, the denominators or normalization factors
$D^{\sigma}$ are responsible for higher-order terms in the transmission and they contain the information
of the energy of the bound states and their lifetimes. 

One can gain more insight into these expressions by considering the zero-temperature case where the above 
probabilities reduce to 
\begin{widetext}
\begin{eqnarray} \label{V>}
P_1^\sigma(E,V>0) & = & \frac{1}{{D^\sigma}} \left\{ \begin{array}{cc}
\rho_{\mathrm{S}} |t_{\rm h}^\sigma|^2+|t_{\rm h}^\sigma|^2|t_{\rm e}^\sigma|^2 
(1+f^{\rm a}f^{\rm r}-g^{\rm a}g^{\rm r}-2\rho_{\mathrm{S}}) & 
\mbox{if} \; E \in [0,eV] \\
 \rho_{\mathrm{S}} |t_{\rm e}^\sigma|^2+|t_{\rm e}^\sigma|^2|t_{\rm h}^\sigma|^2 
 (1+f^{\rm a}f^{\rm r}-g^{\rm a}g^{\rm r}-2\rho_{\mathrm{S}}) & 
 \mbox{if} \; E \in [-eV,0] \\
0 & \, \textrm{otherwise} \\ \end{array} \right. , \\
P_{2}^\sigma(E,V>0) & = & \frac{1}{4D^\sigma} \left\{ \begin{array}{cc}
-f^{\rm a}f^{\rm r}|t_{\rm e}^\sigma|^2 |t_{\rm h}^\sigma|^2 & \mbox{if} \;  
E \in [-eV,eV] \\ 0 & \mathrm{otherwise} \end{array} \right.,
\end{eqnarray}
where $-f^{\rm a}f^{\rm r}$ corresponds to the energy-dependent Cooper pair density. The other contributions are zero 
($P_{-1/-2}^\sigma = 0$) as there is no current flowing to the normal metal without thermal excitation. 
For negative voltages, we obtain
\begin{eqnarray}
P_{-1}^\sigma(E,V<0) & = & \frac{1}{D^\sigma} \left\{ \begin{array}{cc}
\rho_{\mathrm{S}} |t_{\rm h}^\sigma|^2+|t_{\rm h}^\sigma|^2|t_{\rm e}^\sigma|^2 
(1+f^{\rm a}f^{\rm r}-g^{\rm a}g^{\rm r}-2\rho_{\mathrm{S}}) & 
\mbox{if} \; E \in [-e|V|,0] \\
\rho_{\mathrm{S}} |t_{\rm e}^\sigma|^2+|t_{\rm e}^\sigma|^2|t_{\rm h}^\sigma|^2 
(1+f^{\rm a}f^{\rm r}-g^{\rm a}g^{\rm r}-2\rho_{\mathrm{S}}) & 
\mbox{if} \; E \in [0,e|V|] \\
0 & \, \textrm{otherwise} \\ \end{array} \right. \nonumber \\
P_{-2}^\sigma(E,V<0) & = & \frac{1}{4D^\sigma} \left\{ \begin{array}{ccc} 
 -f^{\rm a}f^{\rm r}|t_{\rm e}^\sigma|^2 |t_{\rm h}^\sigma|^2 & \mbox{if} \; 
 E \in [-e|V|,e|V|] \\ 0 & \mathrm{otherwise} \end{array} \right.,
\end{eqnarray}
\end{widetext}
where we see that now only the currents flowing from the substrate to the tip are nonzero and $P^\sigma_{1/2}=0$. 

From the knowledge of the probabilities $P_n^\sigma(E,V)$ we can easily compute all the cumulants of the 
current distribution. Here, we shall focus on the analysis of the current and the noise, which can
be obtained from the CGF of Eq.~(\ref{Action_NS}) using Eqs.~(\ref{eq-C1}) and (\ref{eq-C2}), and are given by
\begin{eqnarray} \label{current-NS}
    I(V) & = & \frac{e}{2h}\sum_{\sigma =\Uparrow,\Downarrow} \int^{\infty}_{-\infty} dE \sum^2_{n=-2} nP_n^\sigma(E,V), \\
    S(V) & = & \frac{e^2}{h} \sum_{\sigma = \Uparrow,\Downarrow} \int^{\infty}_{-\infty}  dE 
    \left\{ \sum^2_{n=-2} n^2 P_n^\sigma(E,V) - \right. \\ 
    & & \hspace{2.8cm} \left. \left(\sum^2_{n=-2} n P_n^\sigma(E,V)\right)^2 \right\} . \nonumber
\end{eqnarray}
\begin{figure}[t]
\includegraphics[width=1\columnwidth,clip]{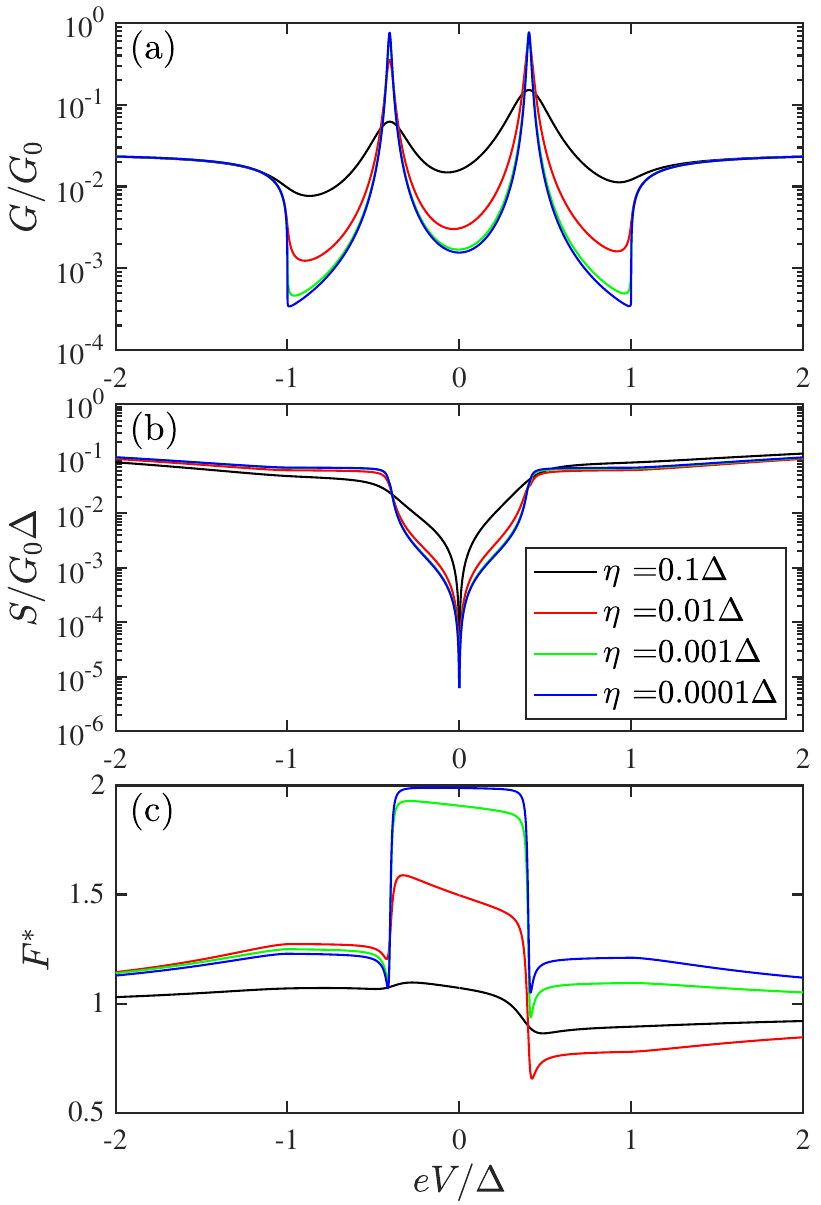}
\caption{Differential conductance (a), shot noise (b), and Fano factor (c) as a function of the voltage for the
case of a single impurity coupled to a normal tip and SC substrate for different values of the Dynes' parameter 
$\eta$ of the SC electrode, as indicated in the legend of panel (b). The parameters used are $\Gamma_{\mathrm{S}} 
= 100\Delta,\ J = 80\Delta,\ U = 60\Delta, \Gamma_{\mathrm{t}} = \Delta$, and $T=0$. With these parameters the 
junction has a normal-state conductance of $0.026G_0$ and the corresponding YSR energy is $\epsilon_{\rm YSR} = 
0.41\Delta$. }
\label{fig-NS1}
\end{figure}

Let us now illustrate the results. We begin by analyzing the impact of the broadening or lifetime of the YSR states on the
different transport properties. For this purpose, we choose the following system parameters: $\Gamma_{\mathrm{S}} = 
100\Delta,\ J = 80\Delta,\ U = 60\Delta, \Gamma_{\mathrm{t}} = \Delta$, and assume zero temperature. With these parameters 
the junction has a total normal-state conductance of $G_{\rm N} = 0.025G_0$, where $G_0 =2e^2/h$. This corresponds 
to the tunnel regime in which the STM usually operates, and the corresponding YSR energy given by Eq.~(\ref{eq-YSR1}) 
is $\epsilon_{\rm YSR} = 0.41\Delta$. In Fig.~\ref{fig-NS1} we show the bias dependence of the differential conductance, 
shot noise and Fano factor for these parameters and for different values of the Dynes' parameter $\eta$ of the substrate. 
The most salient feature in the conductance is the appearance of a peak (for both positive and negative voltages) exactly 
at the energy of the YSR states. The broadening of these peaks increases as $\eta$ increases, as expected, and the
height goes from being independent of the bias polarity for very small $\eta$ to exhibit a very clear asymmetry when
$\eta$ is relatively large. In the case of the shot noise, see Fig.~\ref{fig-NS1}(b), the presence of the YSR states
results in an abrupt increase of the noise at the energy of these states, while the value of $\eta$ determines the 
rounding of the noise step. Finally, the impact of the YSR lifetime is most notable in the Fano factor, see 
Fig.~\ref{fig-NS1}(c). In this case, for very long lifetimes, $F^{\ast}$ exhibits a plateau for voltages below the
YSR energy ($e|V| < \epsilon_{\rm YSR}$), while it adopts values very close to 1 for higher voltages. As the broadening
of the YSR states increases (or their lifetime decreases), the Fano factor at low voltages progressively diminishes.
Moreover, the dependence on the bias polarity also becomes more apparent. Notice also that values of $F^{\ast} < 1$
become possible, in particular, inside the gap. Another thing that it worth mentioning is the absence of pronounced 
features at $eV=\pm \Delta$ in all transport characteristics, contrary to what happens in the absence of
YSR states. This is due to the fact that, due to the conservation of the number of states, the appearance of
in-gap states is accompanied by the disappearance of the BCS gap edge singularities. This fact will become important 
in the next section when we compare with recent experimental results.

\begin{figure}[t]
\includegraphics[width=\columnwidth,clip]{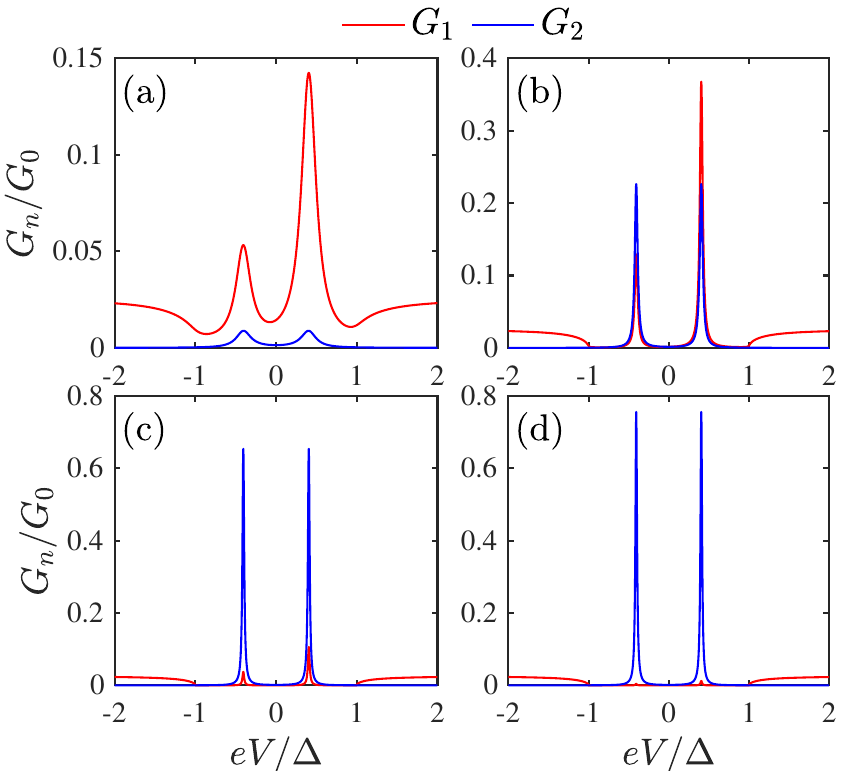}
\caption{Charge-resolved differential conductance as a function of the voltage for the cases considered in 
Fig.~\ref{fig-NS1}. Every panel corresponds to a value of the Dynes' parameter $\eta$ of the SC electrode: 
(a) $\eta = 0.1\Delta$, (b) $\eta = 0.01\Delta$, (c) $\eta = 0.001\Delta$, and (d) $\eta = 0.0001\Delta$. The 
red lines correspond to the contribution of single-quasiparticle tunneling ($G_1$) and the blue ones to the 
contribution of the Andreev reflection ($G_2$).}
\label{fig-NS2}
\end{figure}

The previous results can be easily understood thanks to the unique insight that FCS offers us by identifying 
every individual tunneling process that contributes to the transport. In this regard, we show in
Fig.~\ref{fig-NS2} the results for the charge-resolved contributions to the differential conductance corresponding
to the example of Fig.~\ref{fig-NS1}. In other words, we present in Fig.~\ref{fig-NS2} the individual contributions
to the conductance due to single-quasiparticle tunneling ($G_1$) and to the Andreev reflection ($G_2$). These
processes are schematically represented in Fig.~\ref{fig-NS-processes}. The first thing to notice is that while 
the contribution of single-quasiparticle tunneling depends on the bias polarity, it is not the case for the 
Andreev reflection. Thus, any asymmetry in the transport characteristics must be produced by the contribution 
of the single-quasiparticle tunneling. The main message of this figure is that the Dynes parameter $\eta$ 
determines the relative contribution of both tunneling processes: in the limit of large $\eta$ the 
single-quasiparticle tunneling dominates the transport characteristics, while the Andreev reflection takes 
over in the opposite limit of very long-lived YSR states. This naturally explains the fact that the Fano 
factor is reduced upon increasing the broadening, which is simply due to the fact that in this case the 
transport is dominated by single-quasiparticle tunneling, see Fig.~\ref{fig-NS-processes}(a). This also explains
the doubling of the Fano factor at low bias ($e|V| < \epsilon_{\rm YSR}$) in the limit of small $\eta$ because
in this case the Andreev reflection (transferring two electron charges) dominates the transport,
see Fig.~\ref{fig-NS-processes}(b). A less trivial issue is to understand the origin of the abrupt jump of 
the Fano factor when $eV = \pm \epsilon_{\rm YSR}$ and why in the limit of $\eta \to 0$ the Fano factor gets much 
smaller than 2 in the voltage range $\epsilon_{\rm YSR} < e|V| < \Delta$ where the Andreev reflection completely 
dominates the transport. These interesting issues deserve a detailed explanation, as the one we are about to 
provide in the following paragraphs.  

To clarify these issues we make use of the following analytical approximation for the probabilities of the two tunneling 
processes. Assuming zero temperature and focusing on energies close to the YSR energy, the probabilities for 
single-quasiparticle tunneling and the Andreev reflection for positive bias in Eq.~(\ref{V>}) can be approximated 
as Lorentzians of the form  
\begin{eqnarray} \label{patysr-P1}
    P^{\rm YSR}_1(E) & \approx & \frac{P^{\mathrm{max}}_1}{1+(E-\epsilon_{\mathrm{YSR}})^2/W^2} , \\
    P^{\rm YSR}_2(E) &\approx  & \frac{P^{\mathrm{max}}_2}{1+(E-\epsilon_{\mathrm{YSR}})^2/W^2}, \label{patysr-P2}
\end{eqnarray}
where $P^{\mathrm{max}}_{1/2}$ are the energy-independent maxima of these two probabilities that is reached at 
$E=\epsilon_{\mathrm{YSR}}$ and $W$ is the broadening of the YSR state in our model. If we assume electron-hole
symmetry ($U=0$) for simplicity, the factors $P^{\mathrm{max}}_{1/2}$ are given by
\begin{equation}\label{pmax}
    P_1^{\mathrm{max}} = \frac{2Z}{(1+Z)^2} \;\; \; \mbox{and} \;\;\;  P_2^{\mathrm{max}} = \frac{1}{(1+Z)^2} ,
\end{equation}
where $Z = \frac{\eta J}{ \Gamma_{\mathrm{t}} \Delta} \frac{(\Gamma_{\mathrm{S}}^2+J^2)^2}{4\Gamma_{\mathrm{S}}^2J^2}$. 
Notice that in the limiting case $\eta \to 0$, it holds that $Z = 0$ and then, $P_2^{\mathrm{max}}=1$ and 
$P_1^{\mathrm{max}} = 0$, i.e., there is no single-quasiparticle tunneling as expected. On the other hand, 
in the limit $\Gamma_{\mathrm{L}} \ll \Gamma_{\mathrm{R}},J$ and $\eta \ll \Delta$, the broadening of the YSR 
states is given by 
\begin{equation}\label{eq-width}
    W = \eta + \frac{\Gamma_{\mathrm{t}}}{J}\frac{4J^2\Gamma_{\mathrm{S}}^2}
                              {(\Gamma_{\mathrm{S}}^2+J^2)^2}\Delta .
\end{equation}
With these approximate expressions, one can compute the current and shot noise in the voltage range 
$\epsilon_{\rm YSR} < e|V| < \Delta$. Considering first the ideal case of $\eta=0$, where only the Andreev reflection 
contributes to the in-gap transport, we obtain
\begin{eqnarray}
I(V) & = & \frac{2e}{h} \int_{-\infty}^\infty dE \, P^{\rm YSR}_2(E) = \frac{2e}{h} \pi W , \\
S(V) & = & \frac{8e^2}{h} \int_{-\infty}^\infty dE \, P^{\rm YSR}_2(E) (1 - P^{\rm YSR}_2(E)) \nonumber \\
& = &  \frac{4 e^2}{h} \pi W .
\end{eqnarray}
\begin{figure}[t]
\includegraphics[width=1\columnwidth,clip]{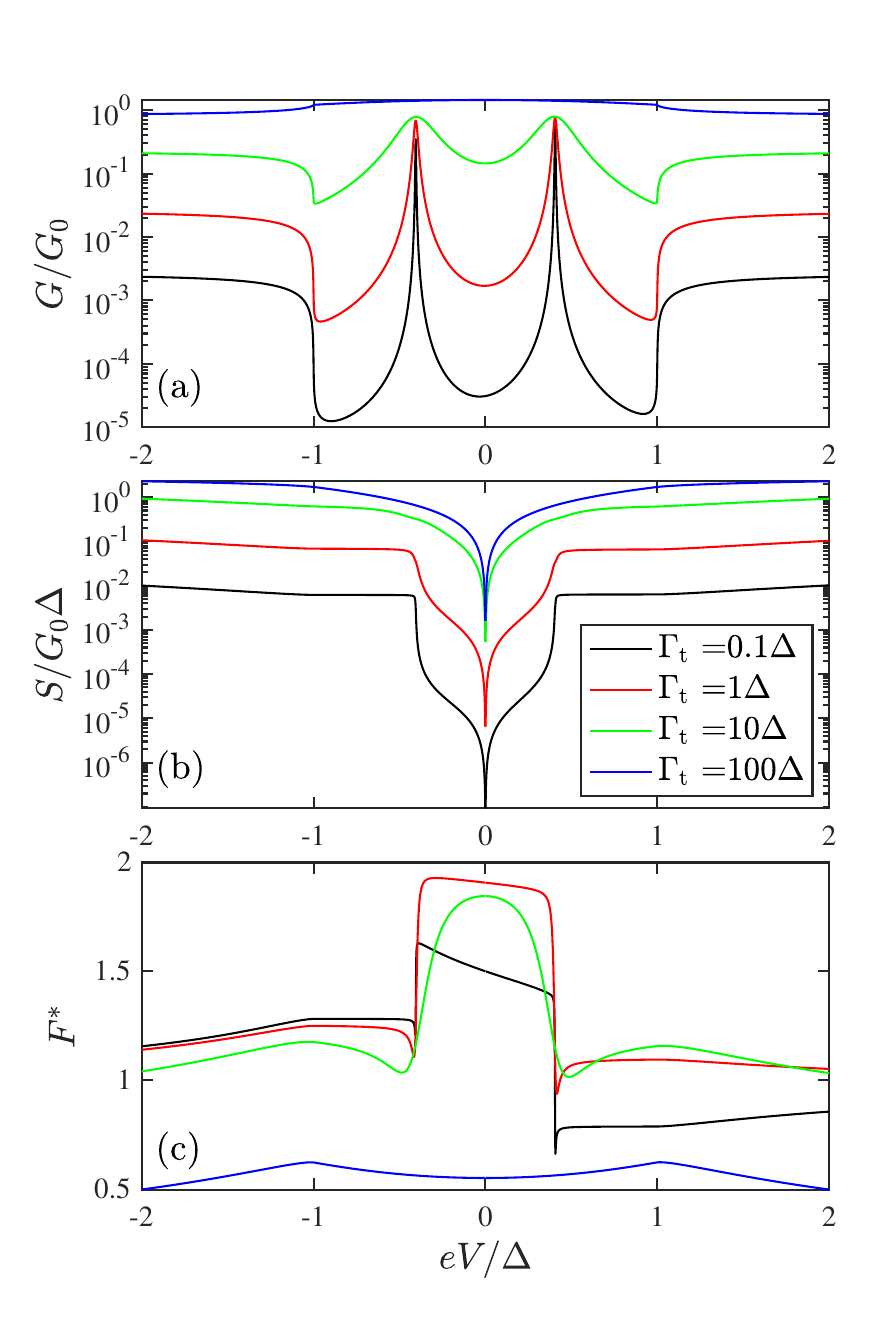}
\caption{Differential conductance (a), shot noise (b), and Fano factor (c) as a function of the voltage for the
case of a single impurity coupled to a normal tip and SC substrate and for different values of the tip tunneling rate 
$\Gamma_{\rm t}$, as indicated in the legend of panel (b). The parameters used are $\Gamma_{\mathrm{S}} = 100\Delta,
\ J = 80\Delta,\ U = 60\Delta, \eta = 0.001 \Delta$, and $T=0$.}
\label{fig-NS3}
\end{figure}

\noindent 
Thus, the corresponding Fano factor is $F^{\ast} = S/(2eI) = 1$
in this voltage range, while it is easy to show that $F^{\ast} = 2$ for $e|V| < \epsilon_{\rm YSR}$ (as long as 
$\eta=0$). Thus, the abrupt reduction of the Fano factor at $e|V| = \epsilon_{\rm YSR}$ is a signature of the fact 
that the Andreev reflection becomes resonant because of the presence of the bound state, see 
Fig.~\ref{fig-NS-processes}(c). The occurrence of this resonant Andreev reflection can reduce the Fano factor all 
the way down to 1 inside the gap, see Fig.~\ref{fig-NS1}(c). 
This is due to the fact that the Andreev reflection can have a probability as high as 1 at the energy of the 
YSR states, which leads to a reduction of the Fano factor from 2 to 1 when crossing the bound state. In other 
words, as the Andreev reflection is resonant at the YSR energy, the transport is not longer in the 
Poissonian limit (with independent tunneling events) and the Fano factor cannot longer be interpreted as an 
effective charge. The fact that we observe in Fig.~\ref{fig-NS1}(c) that the Fano can be larger than 1 in
the voltage range $\epsilon_{\rm YSR} < e|V| < \Delta$ is because $U \neq 0$ in this example. In this case,
it can be shown that the probability of the Andreev reflection does not reach unity, see Fig.~S2(d) in the 
Supplemental Material \cite{SM}, and the Fano factor reduction upon crossing the YSR state is not complete, 
i.e., $F^{\ast} > 1$. Actually, all of this is analogous to what happens in other resonant situations like 
in the case of a normal conductive 
double barrier structure. In that case, the Fano factor becomes $1/2$ for a symmetric situation 
(two identical barriers), and it gets close to 1 for a very asymmetric case \cite{Blanter2000,Chen1991,Buettiker1991}. 
It is also worth mentioning that this crossover of the Fano factor, related to an Andreev reflection from 2 to 1
when crossing a bound state, has been reported theoretically in mesoscopic normal-superconducting structures
\cite{Fauchere1998}. To conclude this discussion, let us mention that one can also show using Eqs.~(\ref{patysr-P1}) 
and (\ref{patysr-P2}) that, as long as $U=0$, even in the case of a finite $\eta$, the zero-temperature Fano factor 
is equal to 1 in the voltage range $\epsilon_{\rm YSR} < e|V| < \Delta$.

\begin{figure}[t]
\includegraphics[width=\columnwidth,clip]{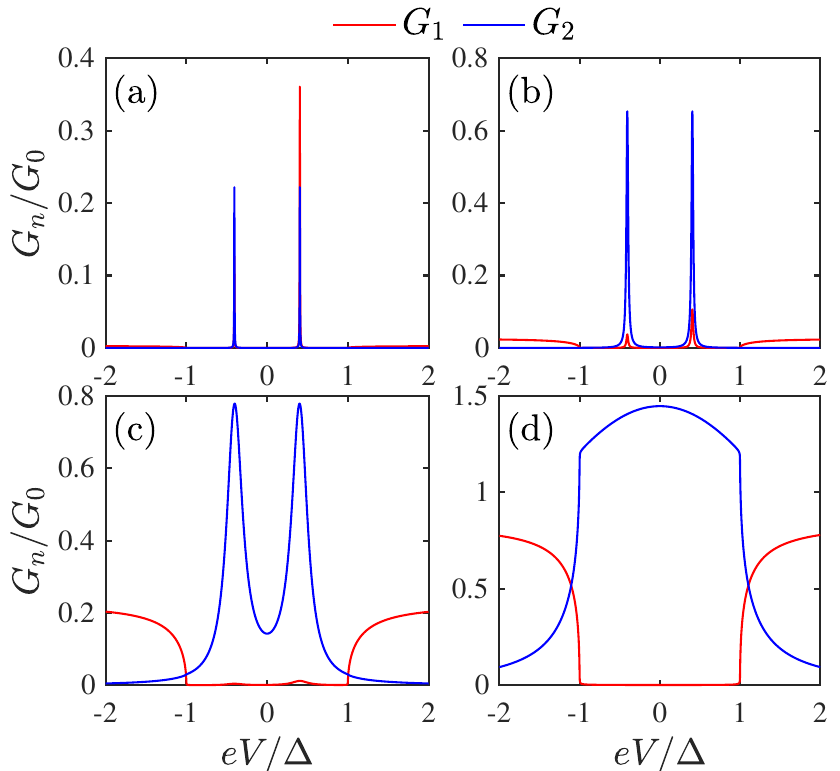}
\caption{Charge-resolved differential conductance as a function of the voltage for the cases considered in 
Fig.~\ref{fig-NS3}. Every panel corresponds to a given value of the tip tunneling rate $\Gamma_{\rm t}$: 
(a) $\Gamma_{\rm t} = 0.1\Delta$, (b) $\Gamma_{\rm t} = 1\Delta$, (c) $\Gamma_{\rm t} = 10\Delta$, and (d) 
$\Gamma_{\rm t} = 100\Delta$. The red lines correspond to the contribution of single-quasiparticle tunneling 
($G_1$) and the blue ones to the contribution of the Andreev reflection ($G_2$).}
\label{fig-NS4}
\end{figure}

Let us now analyze the evolution of the transport characteristics in the crossover between the tunneling regime and
a highly transparent situation. For this purpose, we show in Fig.~\ref{fig-NS3} the bias dependence of the 
differential conductance, shot noise and Fano factor for different values of the tunneling rate $\Gamma_{\rm t}$
and $\Gamma_{\mathrm{S}} =  100\Delta,\ J = 80\Delta,\ U = 60\Delta, \eta = 0.001\Delta$, and $T=0$. 
In this case the conductance evolves from exhibiting peaks at the YSR energies (see curve for $\Gamma_{\rm t} = 0.1\Delta$, 
which corresponds to $G_{\rm N} = 0.0025G_0$), to display a plateau inside the gap in which the conductance is
close to $2G_0$ (see curve for $\Gamma_{\rm t} = 100\Delta$, which corresponds to $G_{\rm N} = 0.83G_0$). 
This latter case essentially corresponds to a fully transparent standard (spin-degenerate) NS point contact.
Notice that in this case the Fano factor has a complex evolution, namely it first increases at low bias upon increasing
the transparency of the contact and then becomes sub-Poissonian ($F^{\ast} < 1$) in the whole voltage range due 
to the reduction caused by the Pauli exclusion principle.

Again, these results can be easily rationalized making use of the charge-resolved contributions to the total
differential conductance, which are displayed in Fig.~\ref{fig-NS4} for this example. These results show that
at the lowest transparency in this example the single-quasiparticle tunneling and the Andreev reflection give
similar contributions. However, as the transmission increases the Andreev reflection completely dominates
the transport inside the gap. For this reason, the evolution of the Fano factor can be solely understood from
the evolution of the probability of the Andreev reflection. In particular, the sub-Poissonian Fano factor inside
the gap for the highest transmission is simply due to the fact that the Andreev reflection reaches almost perfect 
transparency. Outside the gap, the Factor factors remains smaller than 1 due to the competition between the 
two tunneling processes that both give finite contributions. Finally, notice again that the contribution of
the Andreev reflection is independent of the bias polarity and, thus, any asymmetry in the transport characteristics
must be due to the contribution of single-quasiparticle tunneling.

\section{Single-impurity YSR junctions: Comparison with shot noise measurements} \label{sec-comp}

Very recently, Thupakula and coworkers reported the first measurements of the shot noise through a magnetic
impurity on a superconducting substrate featuring YSR states \cite{Thupakula2022}. To be precise, these
researchers employed shot-noise scanning tunneling microscopy to measure the nonequilibrium current 
fluctuations through a magnetic impurity deposited on 2H-NbSe$_2$ at temperatures around $0.7$ K, way 
below the critical temperature of this superconductor. The STM tip was made of a normal metal and the main 
observation was the appearance of a Fano factor above 1. This fact was interpreted as the evidence of the
contribution of an Andreev reflection to the charge transport. Moreover, those authors presented a theoretical
analysis that suggested that the broadening of the YSR state was of the order of $1$ $\mu$eV, which was clearly
below the thermal energy ($k_{\rm B}T$) in this experiment demonstrating that shot noise can probe energy 
scales that are not accessible in conventional tunneling spectroscopy measurements. The goal of this section 
is to provide a thorough analysis of those experimental results in the light of the theory presented in 
Sec.~\ref{Sec:NS}.

\begin{figure}[t]
\includegraphics[width=1\columnwidth,clip]{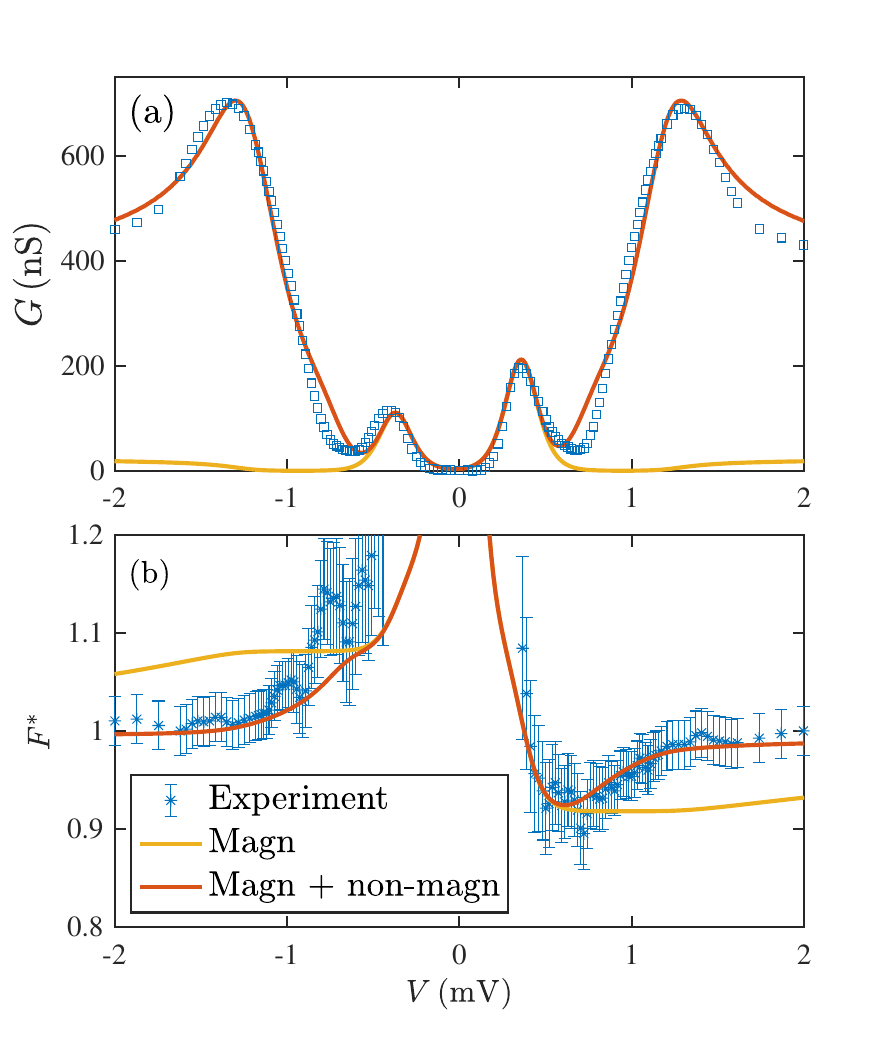}
\caption{Conductance (a) and Fano factor (b) as a function of voltage. The symbols correspond to the experimental
data of Ref.~\cite{Thupakula2022} that were obtained with a normal tip at $T=0.7$ K and a magnetic impurity deposited
on a 2H-NbSe$_2$ superconducting surface. The orange lines in both panels correspond to a fit with our theory using 
a two-channel model, as described in the main text. The yellow lines in both panels correspond to the theory results
considering only the contribution of the magnetic channel (see explanation in the text).}
\label{fig-NS-exp}
\end{figure}

In Fig.~\ref{fig-NS-exp} we show an example of the experimental results reported in Ref.~\cite{Thupakula2022},
which corresponds to the conductance and Fano factor measured in the tail of the YSR states of a magnetic
impurity. The most notable feature in the conductance is the appearance of two peaks inside the gap, 
which correspond to the YSR states. Notice also the asymmetry in the height of the peaks for positive and
negative bias. With respect to the Fano factor, the main observation is the appearance of values above 1
for negative bias, but inside the gap region. This signature was taken as the evidence of the occurrence of
Andreev reflections. Notice also that the Fano factor is asymmetric and for positive bias it remains
below 1 for most voltages inside the gap. The Fano factor at very low bias was not reported simply due to
the difficulty of measuring the very low currents in this voltage range.

These experimental results were analyzed in Ref.~\cite{Thupakula2022} in the light of a model for the YSR state
similar in spirit to ours, but just focusing on voltages inside the gap. To be precise, these authors 
used an approximation similar to that summarized in our Eqs.~(\ref{patysr-P1}) and (\ref{patysr-P2}). 
Although such a simplified model qualitatively captures the physics of the YSR states, it is obvious that
it cannot provide an overall correct picture of the experimental results. Actually, the major problem is
that no model based on a single magnetic channel can describe the simultaneous appearance in the conductance
of YSR peaks and pronounced coherent peaks as those in Fig.~\ref{fig-NS-exp}(a). As we explained above,
these models do not predict the appearance of the coherent peaks simply due to the conservation of the 
number of states. So, as a way out to provide an overall consistent fit of the experimental results, we propose 
that there are at least two channels or pathways for the current: one due to the magnetic impurity and
another non-magnetic channel, which probably results from the direct tunneling of the STM tip to the 
superconducting substrate. This is actually a solution that has been proposed before to explain the 
reported conductance spectra, see for instance Refs.~\cite{Huang2020a,Karan2022}. There is another 
subtlety that we need to take into account, namely 2H-NbSe$_2$ is not a regular BCS superconductor. In 
fact, it has been shown by several groups that 2H-NbSe2 can be described as a two-band superconductor 
\cite{Noat2015,Dvir2018,Senkpiel2019}. So, we propose here to explain the experimental results assuming
that the transport takes place via two independent channels, one magnetic that is described by our YSR
model of the previous sections and a non-magnetic channel that proceeds from the normal tip to the
2H-NbSe$_2$ substrate that we describe as a two-band superconductor. Moreover, we shall assume that this non-magnetic
channel can be described in the tunneling regime, i.e., taking only into account the quasiparticle tunneling
at the lowest order in transmission. We proceed now to describe the details of such a two-channel model.

First, for the non-magnetic channel we describe the superconductivity in 2H-NbSe$_2$ using the two-band model
described in Ref.~\cite{Dvir2018}. In this model one considers that the SC order parameter has two energy-dependent 
components $\Delta_1(E)$ and $\Delta_2(E)$ that can be computed by solving the following self-consistent set of 
equations
\begin{align} \label{Delta1}
    \Delta_1(E) &= \Delta_1^{\mathrm{BCS}} -\Gamma_{12} \frac{\Delta_1(E)-\Delta_2(E)}{\sqrt{\Delta_2^2(E)-E^2}} \\
    \Delta_2(E) &= \Delta_2^{\mathrm{BCS}} -\Gamma_{21} \frac{\Delta_2(E)-\Delta_1(E)}{\sqrt{\Delta_1^2(E)-E^2}}, \label{Delta2}
\end{align}
where $\Delta_{1/2}^{\mathrm{BCS}}$ describes the bare SC gap of the separate bands. The density of states for 
the two bands can be computed as follows
\begin{equation}
    \rho_i (E) = \rho_i(E_{\rm F}) \int d\theta \, \Re \left( \frac{|E|}{\sqrt{(1+\alpha \cos(\theta))
    \Delta^2_i(E)-E^2}}\right),
\end{equation}
where the DOS at the Fermi energy is adjusted to fit the experimental data and $\alpha$ is a measure of the
band anisotropy. The total DOS then follows
\begin{equation} \label{rho_total}
    \rho_{\rm S}(E) = \rho_1(E) +\rho_2(E).
\end{equation}
Using the procedure explained in Ref.~\cite{Senkpiel2019}, we solved the algebraic system of Eqs.~(\ref{Delta1}) 
and (\ref{Delta2}) and found that the best set of parameters is given by $\Delta_1^{\mathrm{BCS}} = 1.23$ meV, 
$\Gamma_{12} = 0.27$ meV, $\rho_1(E_{\rm F}) = 1$, $\Delta_2^{\mathrm{BCS}} = 0.29$ meV, $\Gamma_{21} = 1.25$ meV, 
$\rho_2(E_{\rm F}) = 0.18$ and $\alpha = 0.2$. As mentioned above, we assume that the non-magnetic channel 
operates in the tunnel regime such that its transport properties are solely determined by single-quasiparticle
tunneling, whose probabilities can be computed adapting Eqs.~(\ref{eq-P1}) and (\ref{eq-P2}) to a non-magnetic 
situation in the tunneling regime, i.e., 
\begin{align}\label{twochannel1}
    P^{(\mathrm{nm})}_1  &= |t|^2 \rho_{\rm S} \left[ f_{1}(1-f_{0})+ f_0 (1-f_{-1}) \right] , \\
    P^{(\mathrm{nm})}_{-1} & = |t|^2 \rho_{\rm S} \left[ f_0 (1-f_{1}) + f_{-1} (1-f_{0}) \right],
\end{align}
where $\rho_{\rm S}$ is given by Eq.~(\ref{rho_total}). We have made use of the fact that all the transmission
coefficients are the same (for electrons and holes and for spin up and spin down) and equal to $|t|^2$. This 
transmission coefficient was adjusted to fit the experimental results for the conductance and we obtained a 
value of $|t|^2 = 0.0044$. For the second, magnetic channel we simply used the theory of Sec.~\ref{Sec:NS} and 
adjusted the different parameters to describe as well as possible the in-gap conductance due to the YSR states. 
In particular, our best fit was produced with the following set of parameters: $\Delta = 1.23$ meV, 
$\Gamma_{\mathrm{S}} = 123$ meV, $\Gamma_{\mathrm{t}} = 0.014$ meV, $J = 99.32$ meV, $U = 49.2$ meV, 
$\eta = 0.0011$ meV, and we used the temperature of the experiment $T = 0.7$ K.

The results of the best fit with our two-channel theory are shown in Fig.~\ref{fig-NS-exp} alongside with the 
experimental results for both the conductance and Fano factor. As a reference, we also include the corresponding 
theory results obtained taking only into account the contribution of the magnetic channel. As one can see in 
panel (a), the theory captures very well the salient features of the differential conductance, namely the YSR peaks 
inside the gap and the coherent peaks related to the double-gap structure. Notice that the magnetic channel by itself 
reproduces well the YSR peaks, but it is unable to properly describe the conductance close to the gap edges and 
outside the largest gap. 

Concerning the Fano factor, which was simply calculated from the parameters extracted from the fit of the 
conductance, there is an excellent agreement for positive voltages, see Fig.~\ref{fig-NS-exp}(b). The Fano 
Factor is super-Poissonian for voltages below the YSR energy and becomes sub-Poissonian for voltages higher than 
the bound state energy. For negative voltages there seems to be fine structure that we are not able to perfectly 
reproduce, but overall the agreement is quite satisfactory. In particular, we are able to reproduce the fact that 
the Fano factor is always bigger than 1 inside the gap and it tends to 1 for higher voltages. Here, it becomes even 
more apparent that a two-channel model is necessary. The contribution of the magnetic channel alone does not 
reproduce very well the structure of the Fano factor and it supports our hypothesis on the need of an additional 
contribution. Let us also say that at very low bias (not shown here) the Fano factor becomes very large simply
because the current fluctuations are dominated by a finite thermal noise, while the current tends to zero.

Something that is very important to emphasize is the fact that, as we showed in Sec.~\ref{Sec:NS}, the Fano 
factor is very sensitive to the broadening or lifetime of the YSR states. In our fit we obtained a value of 
$\eta = 1.1$ $\mu$eV for the Dynes' parameter in the superconducting substrate. Using Eq.~(\ref{eq-width}) and 
the rest of the parameter values extracted from the fit, we obtain that $W \approx \eta = 1.1$ $\mu$eV, which 
is much smaller than the thermal energy in this case ($k_{\rm B}T \approx 60$ $\mu$eV). Thus, as stated in
Ref.~\cite{Thupakula2022}, the analysis of the noise and the corresponding Fano factor allows us to have 
access to energy and time scales that are usually out of the scope of conventional conductance measurements 
due to thermal broadening.

\begin{figure*}[!t]
\includegraphics[width=0.95\textwidth,clip]{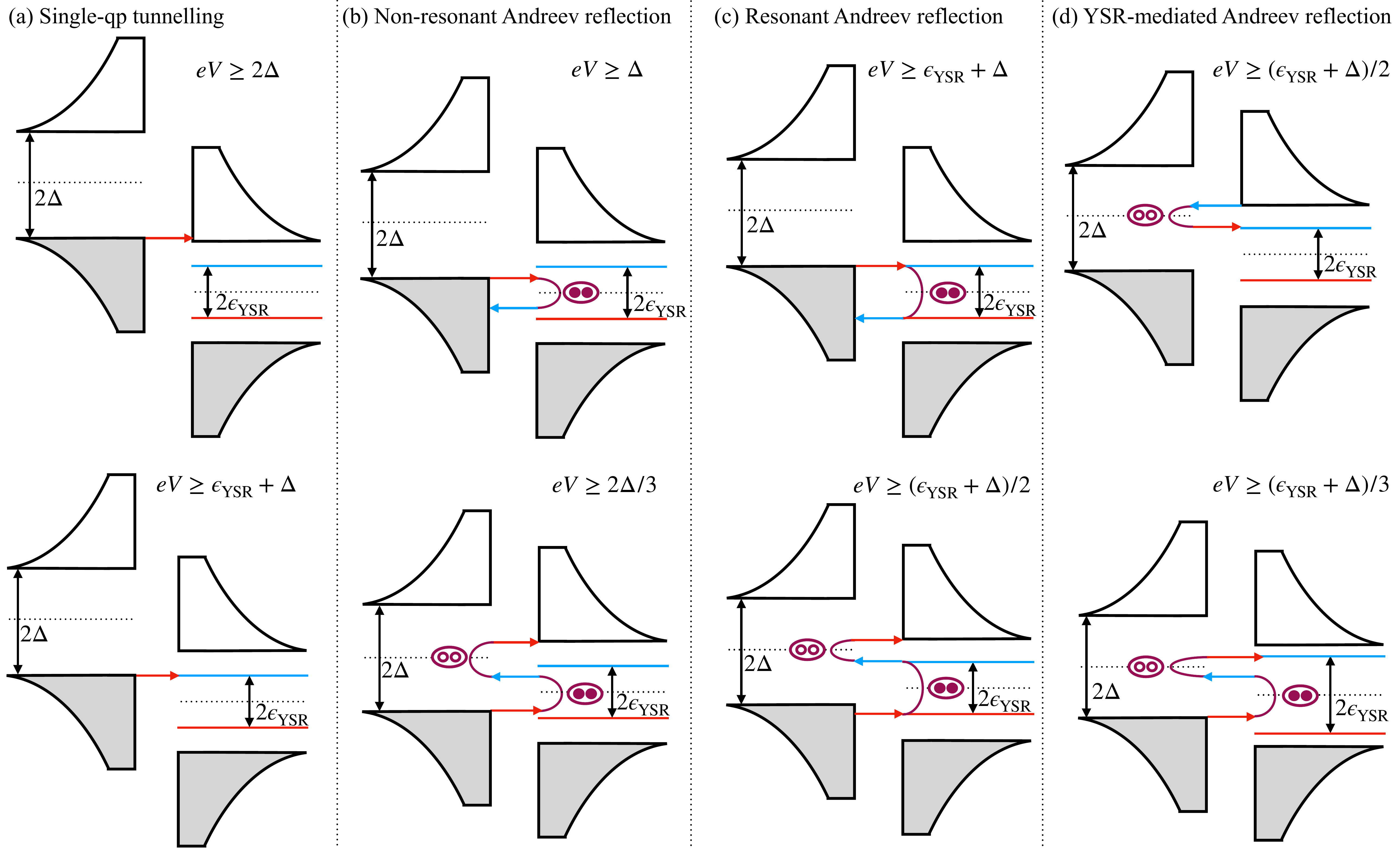}
\caption{Tunneling processes through an impurity coupled to two superconducting leads. Here, the left electrode is 
a superconducting tip and the right one is the impurity coupled to the superconducting substrate featuring YSR states. 
The diagrams follow the same convention as in Fig.~\ref{fig-NS-processes}. (a) Single-quasiparticle tunneling, both 
non-resonant (upper) and resonant one (lower). (b) Standard non-resonant (multiple) Andreev reflections.
(c) Resonant (multiple) Andreev reflections in which at least an Andreev reflection at the energy of the
YSR states takes place. (d) YSR-mediated Andreev reflections, which are Andreev reflections that involve the 
tunneling from or into a YSR state.}
\label{fig-MARs}
\end{figure*}

\section{Single-impurity YSR junctions: Superconducting tip} \label{Sec:SS}

We now analyze the case in which a magnetic impurity is coupled to two superconducting leads, which for 
simplicity we assume to have the same gap $\Delta$. In this case, the novelty with respect to the previous case 
is the possibility of having MARs. As a reference for our discussions below, we illustrate the relevant tunneling 
processes in this case in Fig.~\ref{fig-MARs}.

Technically speaking, this case is considerably more complicated than the NS one because the GFs of the terminals 
are not longer diagonal in energy space, which is due to the ac Josephson effect. This means in practice that 
we have to treat the problem in the Floquet language, as explained in Appendix~\ref{App:GFs}. For this reason, 
it is not possible to provide a complete analytical solution for the CGF and we have to resort to numerics. 
The technical details are presented in Appendix~\ref{App:GFs},\ref{App:action} and here we shall focus on the discussion 
of the physical results.
 
As discussed in Sec.~\ref{Sec:action}, in this case the CGF reads
\begin{equation}\label{Action_SS}
    \mathcal{A}_{t_0}(\chi) = \frac{t_0}{2h}\sum_{\sigma = \Uparrow , \Downarrow} \int^{eV}_{-eV} dE \ln 
    \left[ \sum_{n=-\infty}^{\infty} P_n^\sigma(E,V) e^{\imath n\chi} \right] ,
\end{equation}
where $P_n^\sigma(E,V)$ corresponds to the probability of transferring $n$ charges across the junction 
for spin $\sigma$ and $E$ is the Floquet energy. Notice that the main difference with respect to 
Eq.~(\ref{Action_NS}) is the fact that $n$ now runs from $-\infty$ to $\infty$ due to the occurrence of 
MARs and we integrate over a finite energy range. Again, from the knowledge of the probabilities $P_n^\sigma(E,V)$ 
we can easily compute the current and the noise, which are given by the standard formulas of a multinomial distribution
\begin{eqnarray} \label{current-SS}
    I(V) & = & \frac{e}{2h}\sum_{\sigma =\Uparrow,\Downarrow} \int^{eV}_{-eV} dE \sum^{\infty}_{n=-\infty} 
    n P_n^\sigma(E,V), \\
    S(V) & = & \frac{e^2}{h} \sum_{\sigma = \Uparrow,\Downarrow} \int^{eV}_{-eV}  dE
    \left\{ \sum^{\infty}_{n=-\infty} n^2 P_n^\sigma(E,V) - \right. \\ 
    & & \hspace{2.8cm} \left. \left(\sum^{\infty}_{n=-\infty} n P_n^\sigma(E,V)\right)^2 \right\} . \nonumber
\end{eqnarray}
\begin{figure}[t]
\includegraphics[width=1\columnwidth,clip]{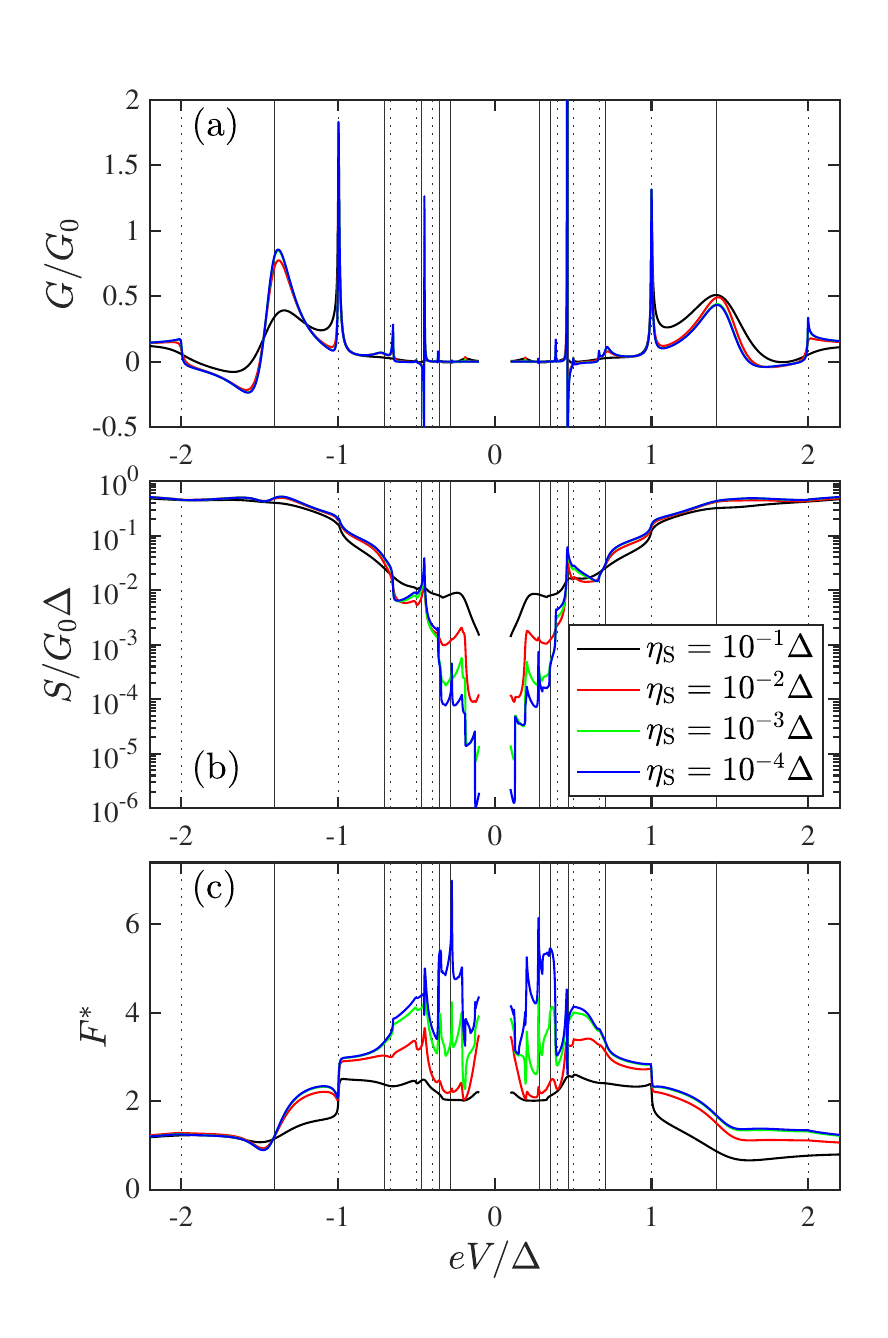}
\caption{Differential conductance (a), shot noise (b), and Fano factor (c) as a function of the voltage for the
case of a single impurity coupled to two superconducting electrodes and for different values of the Dynes' parameter 
$\eta$ of the SC electrode, as indicated in the legend of panel (b). The parameters used are $\Gamma_{\mathrm{S}} 
= 100\Delta,\ J = 80\Delta,\ U = 60\Delta, \Gamma_{\mathrm{t}} = 5\Delta$, $\eta_{\rm t} = 0.001\Delta$ and $T=0$. 
With these parameters the junction has a normal-state conductance of $0.12G_0$ and the corresponding YSR energy is
$\epsilon_{\rm YSR} = 0.41\Delta$. The vertical lines indicate the position of the voltages $eV = \pm 2\Delta/n$
(dotted lines) and $eV = \pm (\Delta + \epsilon_{\rm YSR})/n$ (solid lines) with $n=1,2,\dots$}
\label{fig-SS1}
\end{figure}

We begin by analyzing the impact of the broadening or lifetime of the YSR states on the different transport properties. 
For this purpose, we choose the following system parameters: $\Gamma_{\mathrm{S}} = 100\Delta,\ J = 80\Delta,
\ U = 60\Delta, \Gamma_{\mathrm{t}} = 5\Delta$, $\eta_{\rm t} = 0.001\Delta$ and assume zero temperature. With 
these parameters the junction has a normal-state conductance of $G_{\rm N} = 0.12G_0$ and the corresponding YSR 
energy given by Eq.~(\ref{eq-YSR1}) is $\epsilon_{\rm YSR} = 0.41\Delta$. Figure~\ref{fig-SS1} displays the conductance, 
shot noise, and Fano factor for these parameters and different values of the Dynes parameter of the substrate 
$\eta_{\rm S}$. Notice that the conductance exhibits a very rich subgap structure due to the occurrence
of MARs. In any case, the most visible conductance peaks appear at $eV = \pm (\Delta + \epsilon_{\mathrm{YSR}})$, 
which as we show below are due to both single-quasiparticle tunneling and the lowest-order resonant Andreev 
reflection. Notice also that the conductance depends on the bias polarity because we are dealing with a situation
with electron-hole asymmetry ($U=60\Delta$). The additional conductance peaks appear at $eV = \pm 2\Delta/n$, 
which are due to conventional (non-resonant) Andreev reflections, and at $eV = \pm (\Delta + \epsilon_{\mathrm{YSR}})/n$,
whose origin will be discussed below. Overall, the role of the Dynes' parameter is to determine the width and the 
height of all these conductance peaks, as expected. On the other hand, the shot noise exhibits steps at the 
voltages at which the conductance peaks appear and these steps are progressively more pronounced as $\eta_{\rm S}$ 
decreases, see Fig.~\ref{fig-SS1}(b). Again, this parameter has a strong impact in the Fano factor, see 
Fig.~\ref{fig-SS1}(c), which now exhibits super-Poissonian values well above 2 as $\eta_{\rm S}$ decreases. This is 
obviously a signature of the occurrence of MARs. It is also important to notice that the Fano factor does not 
exhibit integer values inside the gap, as it occurs in the absence of in-gap states \cite{Cuevas1999,Cron2001}, 
which suggests that no tunneling process completely dominates the transport at any subgap voltage.
Let us also say that we do not report the results at very low bias because the current becomes exceedingly small 
and it is not measurable in practice.

\begin{figure}[t]
\includegraphics[width=\columnwidth,clip]{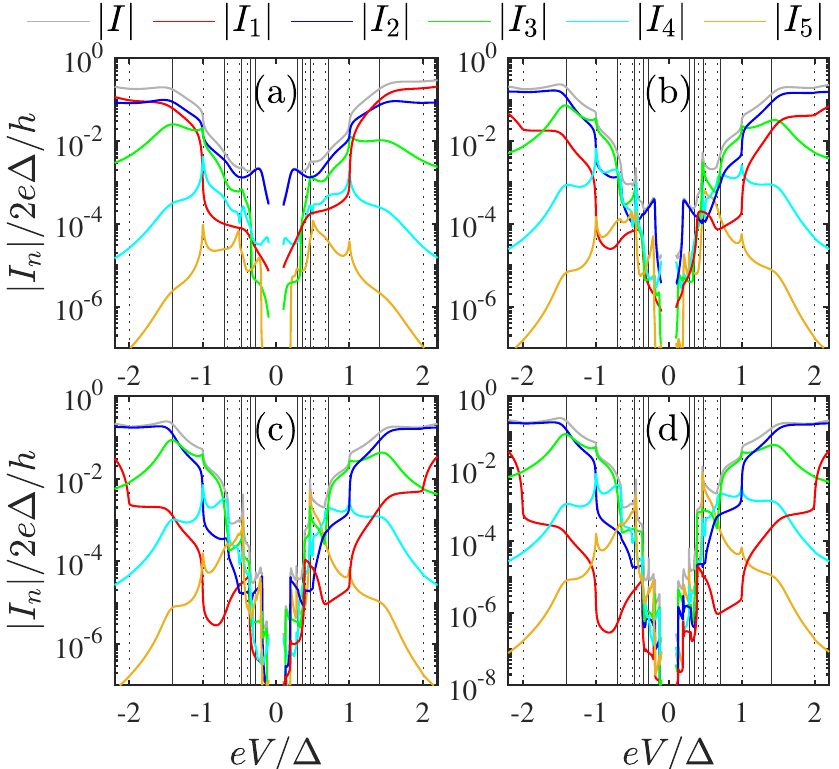}
\caption{Charge-resolved currents as a function of the voltage for the cases considered in Fig.~\ref{fig-SS1}.
Every panel corresponds to a given value of the Dynes' parameter of the substrate: 
(a) $\eta_{\rm S} = 0.1\Delta$, (b) $\eta_{\rm S} = 0.01\Delta$, (c) $\eta_{\rm S} = 0.001\Delta$, and 
(d) $\eta_{\rm S} = 0.0001\Delta$. Notice that the absolute value of the current is plotted for clarity.
The vertical lines follow the same convention as in Fig.~\ref{fig-SS1}.}
\label{fig-SS2}
\end{figure}

Again, we can rationalize these results with the analysis of the individual contributions of the different tunneling 
processes classified according to the charge they transfer. In Fig.~\ref{fig-SS2}, the charge-resolved currents
are shown for different values of $\eta_{\rm S}$ (we plot the absolute value of those currents for clarity). 
The corresponding charge-resolved conductances are shown in the Supplemental Material (Fig.~S7) \cite{SM}. The light 
grey line indicates the total current, whereas the others correspond to the contributions $I_n$ due to the 
processes transferring $n$ electron charges. Notice that for the smallest value of the Dynes' parameter, 
$\eta_{\rm S} = 0.1\Delta$ in panel (a), the single-quasiparticle tunneling and Andreev reflection dominate the 
transport and the MAR contributions with $n>2$ are negligible for all voltages. However, the resonant Andreev reflection
illustrated in the upper graph of Fig.~\ref{fig-MARs}(c) gives a sizeable contribution to the conductance peak
at $eV = \pm (\Delta + \epsilon_{\mathrm{YSR}}) = \pm 1.41\Delta$. Decreasing $\eta_{\rm S}$ results in the 
suppression of the quasiparticle current ($I_1$) because the SC DOS in the gap region is lowered. For 
$\eta_{\rm S} = 0.01\Delta$ in panel (b), we see that the lowest-order Andreev reflection takes over at 
voltages $\Delta < e|V| < 2\Delta$ and it is responsible for the conductance peaks at $eV = \pm \Delta$ 
(non-resonant Andreev reflection, see Fig.~\ref{fig-MARs}(b)) and at $eV = \pm (\Delta + \epsilon_{\mathrm{YSR}})
= \pm 1.41\Delta$ (resonant Andreev reflection, see Fig.~\ref{fig-MARs}(c)). Notice, on the other hand, that 
the third-order MAR starts to play a fundamental role in the subgap transport and it is responsible for the 
conductance peak at $eV = \pm (\Delta + \epsilon_{\mathrm{YSR}})/2 = \pm 0.705\Delta$ and $eV = \pm (\Delta + 
\epsilon_{\mathrm{YSR}})/3 = \pm 0.47\Delta$, this latter one exhibiting moreover a pronounced negative 
differential conductance. The conductance peak at $eV = \pm (\Delta + \epsilon_{\mathrm{YSR}})/2 = \pm 0.705\Delta$
is extremely interesting because it has been observed in a number of experiments and its origin has been
attributed to a second-order Andreev reflection that starts or ends in a YSR state (transferring 2 charges)
\cite{Randeria2016,Farinacci2018,Huang2021,Trahms2023}. This process, which we term YSR-mediated Andreev reflection, 
is illustrated in the upper diagram of Fig.~\ref{fig-MARs}(d) and, it has in fact a threshold voltage 
equal to $eV = \pm (\Delta + \epsilon_{\mathrm{YSR}})/2$. However, this interpretation is incorrect 
in our case, as it is evidenced by the charge-resolved currents and by the fact that the Fano factor is
clearly above 2 in this voltage region. Such a conductance peak originates indeed from a MAR of order 3 that 
becomes resonant precisely at $eV = \pm (\Delta + \epsilon_{\mathrm{YSR}})/2$, as we illustrate in the lower 
diagram of Fig.~\ref{fig-MARs}(c). It is easy to show that such a resonant MAR requires the YSR energy to 
fulfill $\epsilon_{\rm YSR} \ge \Delta/3$, which is satisfied in this example. Another fact that
confirms our interpretation is the absence of negative differential conductance at that bias, which would
be expected from a YSR-mediated Andreev reflection, but not from a resonant MAR. This discussion illustrates 
again the unique insight of the FCS approach, without which it would be hard to draw the right conclusion in this case. 

Coming back to the features at $eV = \pm (\Delta + \epsilon_{\mathrm{YSR}})/3 = \pm 0.47\Delta$, they also 
originate from a third-order MAR, but in this case it is a MAR process that involves the tunneling into an 
YSR state (i.e., a YSR-mediated MAR), which explains the negative differential conductance. This process is 
illustrated in the lower graph of Fig.~\ref{fig-MARs}(d). For $\eta_{\rm S} = 0.001\Delta$ in panel (c), 
the current steps are very abrupt leading to very high conductance peaks. In this case, the conductance 
peaks at $eV = \pm (\Delta+\epsilon_{\mathrm{YSR}})/3$ can now be mainly attributed to the fifth-order 
YSR-mediated MAR. For the smallest $\eta_{\rm S} = 0.0001\Delta$ in panel (d), the total current exhibits pronounced 
negative differential conductance whenever the tunneling into a YSR state is involved. An important observation 
in Fig.~\ref{fig-SS1} is that, contrary to the case of a normal tip, the contribution of the different Andreev 
reflections does depend on the bias polarity. Thus, asymmetries in the the current-voltage characteristics cannot 
be solely attributed to the contribution of single-quasiparticle tunneling. 

\begin{figure}[t]
\includegraphics[width=1\columnwidth,clip]{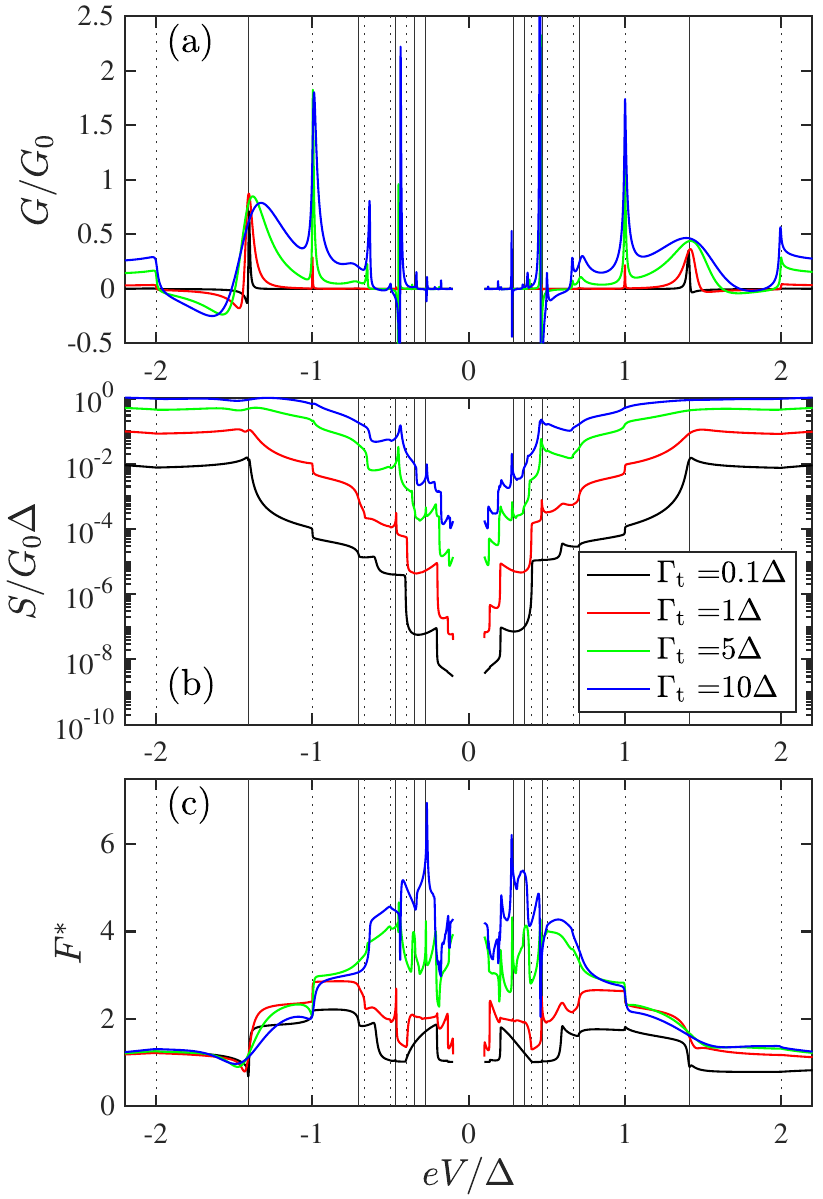}
\caption{Differential conductance (a), shot noise (b), and Fano factor (c) as a function of the voltage 
for the case of a single impurity coupled to two superconducting electrodes and for different values of 
the tunneling rate $\Gamma_{\rm t}$, as indicated in the legend of panel (b). The parameters used are 
$\Gamma_{\mathrm{S}} = 100\Delta,\ J = 80\Delta,\ U = 60\Delta, \eta_{\rm S} = 0.001\Delta, \eta_{\rm t} 
= 0.001\Delta$ and $T=0$. The vertical lines follow the same convention as in Fig.~\ref{fig-SS1}.}
\label{fig-SS3}
\end{figure}

Let us explore now the impact of the junction transmission by changing the tip tunneling rate $\Gamma_{\rm t}$,
while maintaining constant the other parameters: $\Gamma_{\mathrm{S}} = 100\Delta,\ J = 80\Delta,\ U = 60\Delta, 
\eta_{\rm S} = 0.001\Delta, \eta_{\rm t} = 0.001\Delta$, and $T=0$. In Fig.~\ref{fig-SS3}, the conductance, 
shot noise, and Fano factor are shown as a function of the voltage and for different tip tunneling rates 
$\Gamma_{\mathrm{t}}$. Considering the conductance, for the smallest value of $\Gamma_{\mathrm{t}} = 0.1\Delta$
($G_{\rm N} = 0.0025G_0$), the first YSR resonance at $eV = \pm (\Delta + \epsilon_{\mathrm{YSR}})$ dominates 
the subgap structure. Increasing $\Gamma_{\mathrm{t}}$ firstly broadens the conductance peaks, and secondly shifts 
the position of the conductance peaks slightly. This is caused by the renormalization of the YSR energies due 
to the finite coupling to the tip. For $\Gamma_{\mathrm{t}} = \Delta$ ($G_{\rm N} = 0.025G_0$), the first YSR 
resonance at $eV = \pm (\Delta + \epsilon_{\mathrm{YSR}})$ becomes increasingly pronounced and additional subgap 
structure starts to appear, namely at the Andreev reflection onsets at $eV = \pm \Delta$ but also at the onset of
the YSR-mediated and resonant Andreev reflections $eV = \pm (\Delta + \epsilon_{\mathrm{YSR}})/n$ (with $n>1$). 
For $\Gamma_{\mathrm{t}} = 5\Delta$ ($G_{\rm N} = 0.12G_0$), the conductance peaks are increasingly broadened 
and, in particular, the peaks at $eV = \pm (\Delta+\epsilon_{\mathrm{YSR}})/3$ become clearly visible. 
For $\Gamma_{\mathrm{L}} = 10\Delta$ ($G_{\rm N} = 0.22G_0$), the YSR peaks at $eV = \pm (\Delta + 
\epsilon_{\mathrm{YSR}})$ are almost completely washed out. The peaks at $eV = \pm \Delta$ are clearly visible and
correspond to the onset of the regular Andreev reflection. At $eV = \pm (\Delta + \epsilon_{\mathrm{YSR}})/3$, 
we see high conductance peaks that correspond to YSR-mediated Andreev reflections. If one continued increasing
the coupling (not shown here), the system would resemble a highly transparent superconducting point contact 
\cite{Cuevas1996}, and the YSR resonances are no longer resolvable. On the other hand, the shot noise exhibits 
steps at voltages corresponding to onsets of the different types of Andreev processes. Concerning the Fano factor, 
for large values of $\Gamma_{\mathrm{t}}= 5,10\Delta$ it exhibits very high super-Poissonian values due to the 
occurrence of MARs. In addition, we see the characteristic drop-off of the Fano factor from roughly 2 to almost 
1 at the first YSR resonance at $eV = \pm (\Delta + \epsilon_{\mathrm{YSR}})$, which has the same origin as in 
the NS case just shifted by the gap energy $\Delta$. Decreasing the tip coupling decreases the Fano factor 
because the MAR contributions are suppressed. Notice that for small $\Gamma_{\mathrm{t}} = 0.1\Delta$, the 
Fano Factor never reaches over 2 indicating that MARs do not contribute to the transport. In addition to the 
characteristic drop-off at $eV = \pm (\Delta + \epsilon_{\mathrm{YSR}})$, there are other signatures in the 
Fano factor which correspond to the MAR onsets at $eV = \pm (2\Delta/n)$ and to the onset of both resonant 
MARs and YSR-mediated Andreev reflections at $eV = \pm (\Delta + \epsilon_{\mathrm{YSR}})/n$.

\begin{figure}[t]
\includegraphics[width=\columnwidth,clip]{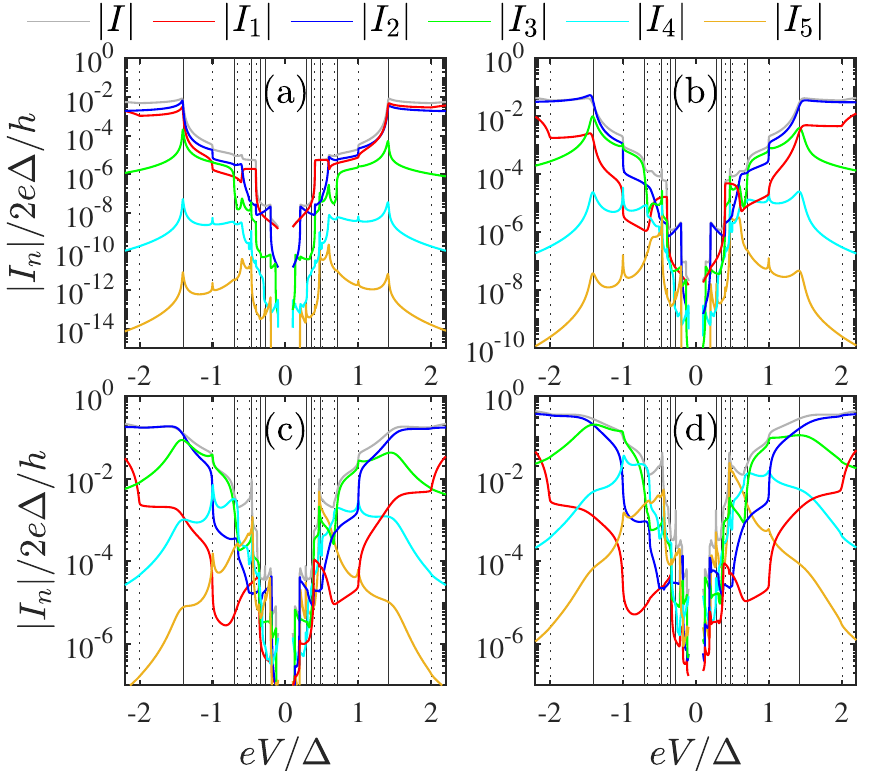}
\caption{Charge-resolved currents as a function of the voltage for the cases considered in Fig.~\ref{fig-SS3}.
Every panel corresponds to a given value of the tip tunneling rate: (a) $\Gamma_{\rm t} = 0.1\Delta$, 
(b) $\Gamma_{\rm t} = \Delta$, (c) $\Gamma_{\rm t} = 5\Delta$, and (a) $\Gamma_{\rm t} = 10\Delta$.
Notice that the absolute value of the current is plotted for clarity. The vertical lines follow the same 
convention as in Fig.~\ref{fig-SS1}.}
\label{fig-SS4}
\end{figure}

Again, we can pinpoint the origin of every feature in the transport characteristics by considering the 
charge-resolved currents, as we illustrate in Fig.~\ref{fig-SS4} (see Fig.~S8 for the corresponding 
charge-resolved conductances \cite{SM}). Notice that for the smallest coupling 
$\Gamma_{\mathrm{t}} = 0.1\Delta$ in panel (a) the single-quasiparticle tunneling and the lowest-order Andreev 
reflection dominate the transport and give similar contributions at $eV = \pm (\Delta + \epsilon_{\mathrm{YSR}})$.
However, as the transmission increases, the quasiparticle tunneling becomes progressively more irrelevant in the subgap
transport. Thus, for instance, for $\Gamma_{\mathrm{t}} = \Delta$, the structure at $eV = \pm (\Delta + 
\epsilon_{\mathrm{YSR}})$ is solely due to the resonant Andreev reflection. For $\Gamma_{\mathrm{t}} = 5\Delta$ 
in panel (c), the subgap structure becomes much more pronounced with current steps not only at the standard
MAR onsets $eV = \pm 2\Delta/n$, but also at voltages $eV = \pm (\Delta + \epsilon_{\mathrm{YSR}})/2$ and 
$eV = \pm (\Delta + \epsilon_{\mathrm{YSR}})/3$ which, as explained above, are due to resonant MARs and
to YSR-mediated Andreev reflections, respectively. Thus, for instance, the  structure at $eV = \pm (\Delta + 
\epsilon_{\mathrm{YSR}})/2$ is mainly due to the third-order MAR, whereas several YSR-mediated MARs contribute 
to the structure at $eV = \pm (\Delta + \epsilon_{\mathrm{YSR}})/3$. The steps in the charge-resolved
currents are smoothed out when $\Gamma_{\mathrm{t}}$ is further increased towards $\Gamma_{\mathrm{t}} = 10\Delta$ 
in panel (d). In this case the first YSR resonance becomes so broad that it can barely be identified as a 
resonance, whereas the higher-order YSR resonances are still very sharp and become increasingly easy to detect.

To conclude this section, let us say that one can provide some analytical insight in the tunnel regime when
$\Gamma_{\mathrm{t}} \ll \Gamma_{\mathrm{S}}$. In this limit, and with the help of Ref.~\cite{Villas2020}, 
one can derive perturbative expressions for the current contributions of single-quasiparticle tunneling ($I_1$)
and the lowest-order Andreev reflection ($I_2$). These expressions are given by
\begin{eqnarray}
    I^{\rm tunnel}_{1}(V) & \approx & \frac{4e\pi^2}{h} \Gamma_{\mathrm{t}} \int^{\infty}_{-\infty} \rho_{\rm t}(E-eV)
    \rho_{\mathrm{imp}}(E) \times \nonumber \\ & & \left[ f(E-eV) - f(E) \right] dE, \label{eq-I1-tunnel} \\
    I^{\rm tunnel}_{2}(V) & \approx & \frac{8e\pi^2}{h} \Gamma_{\mathrm{t}}^2 \int^{\infty}_{-\infty} \rho_{\rm t}(E-eV) 
    \rho_{\rm t}(E+eV) \times \nonumber \\ 
    & & \hspace*{-0.5cm} |F(E)|^2 \left[ f(E-eV) - f(E+eV) \right] dE. \label{eq-I2-tunnel} 
\end{eqnarray}
Here, $\rho_{\rm t}(E)$ is the BCS DOS of the superconducting tip, $\rho_{\mathrm{imp}}(E)$ is the total DOS of
the impurity coupled to the superconducting substrate, and $F(E)$ is the corresponding anomalous Green's
function of the impurity coupled to the substrate. The impurity DOS is given by
\begin{eqnarray}
\rho_{\mathrm{imp}}(E) & = & \frac{1}{\pi} \Im \left[ \frac{E \Gamma_{\rm S} +(U+E-J)\sqrt{\Delta^2-E^2}}{D(E)} +
\right. \nonumber \\ & & \left. \frac{-E \Gamma_{\rm S} + (U-E-J)\sqrt{\Delta^2-E^2}}{D(-E)} \right] ,
\end{eqnarray}
where
\begin{eqnarray}
D(E) & = & 2\Gamma_{\rm S} E(E-J) + \nonumber \\ & & 
\left[ (E-J)^2 - U^2 -\Gamma_{\rm S}^2 \right] \sqrt{\Delta^2-E^2},	
\end{eqnarray}
and the impurity anomalous Green's function is given by
\begin{equation}
	F(E) = \frac{\Gamma_{\rm S} \Delta}{D(E)} .
\end{equation}

\begin{figure}[t]
\includegraphics[width=0.9\columnwidth,clip]{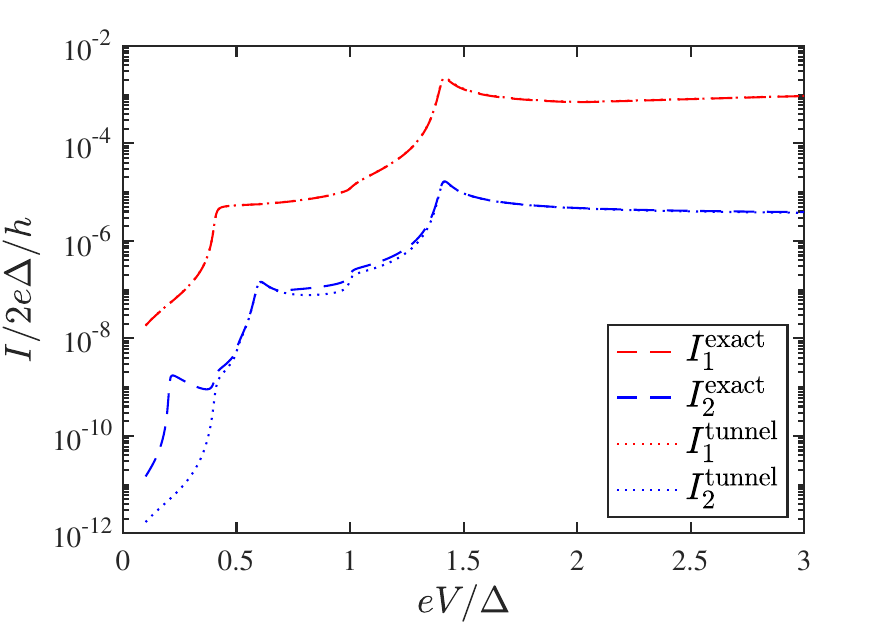}
\caption{Comparison between the analytical results of Eqs.~(\ref{eq-I1-tunnel}) and (\ref{eq-I2-tunnel}) for
the current contributions from single-quasiparticle tunneling and Andreev reflection and the numerically 
exact results obtained from the CGF. The different parameters of the model are: $\Gamma_{\mathrm{S}} = 
100\Delta$, $J = 80\Delta$, $U = 60\Delta$, $\eta_{\rm t/S} = 0.01\Delta$, $\Gamma_{\mathrm{t}} = 0.01\Delta$, 
and $T=0$.}
\label{fig-comp-tunnel}
\end{figure}

By construction, these perturbative current formulas are only valid in the tunnel regime, but they also require 
the broadening of the YSR states to be large in comparison with the tip tunneling rate $\Gamma_{\rm t}$. To test these
approximate formulas, we present in Fig.~\ref{fig-comp-tunnel} a comparison with the exact numerical results for the
case in which $\Gamma_{\mathrm{S}} = 100\Delta$, $J = 80\Delta$, $U = 60\Delta$, $\eta_{\rm t,S} = 0.01\Delta$, 
$\Gamma_{\mathrm{t}} = 0.01\Delta$, and $T=0$. Notice that the analytical formulas nicely reproduce the numerical
results, except for the Andreev reflection contribution at very low bias when higher-order terms in the 
transmission are expected to play a role.

\section{Double-impurity junctions: Tunneling between YSR states} \label{Sec:2imp}

As mentioned in Sec.~\ref{sec:intro}, recently it has been experimentally demonstrated that a superconducting STM tip 
can be functionalized with a magnetic impurity that then features YSR states \cite{Huang2020b}. More importantly, it
was shown that this YSR-STM can be used to probe other magnetic impurities deposited on a superconducting substrate 
that also features YSR states. In this way, these experiments realized for the first time the tunneling between 
individual superconducting bound states at the atomic scale, which is the ultimate limit for quantum transport. 
In particular, the current-voltage characteristics in these double-impurity junctions were shown to exhibit huge
current peaks inside the gap (with an extremely pronounced negative differential conductance). These current 
peaks have been interpreted as the evidence of tunneling between YSR states (both direct at low temperatures
and thermally excited at high temperatures) \cite{Huang2020b,Villas2021,Huang2021}. In this section we want to 
analyze this unique situation from the FCS point of view and, in particular, provide very concrete predictions 
for the shot noise and Fano factor in these junctions. In turn, this problem gives us the opportunity to show
how the FCS approach can be used in a case in which the scattering matrix is not diagonal in spin space.

\subsection{Model and scattering matrix}

In the spirit of the Keldysh action of Eq.~(\ref{action}) we now need a scattering matrix describing these double
impurity junctions. For this purpose, we follow the model put forward in Ref.~\cite{Villas2021}, which has been
very successful describing the experimental observations for the current-voltage characteristics 
\cite{Huang2020b,Huang2021}. We briefly describe this model and then proceed to the determination of the 
corresponding normal-state scattering matrix.

\begin{figure}[t]
\includegraphics[width=0.8\columnwidth,clip]{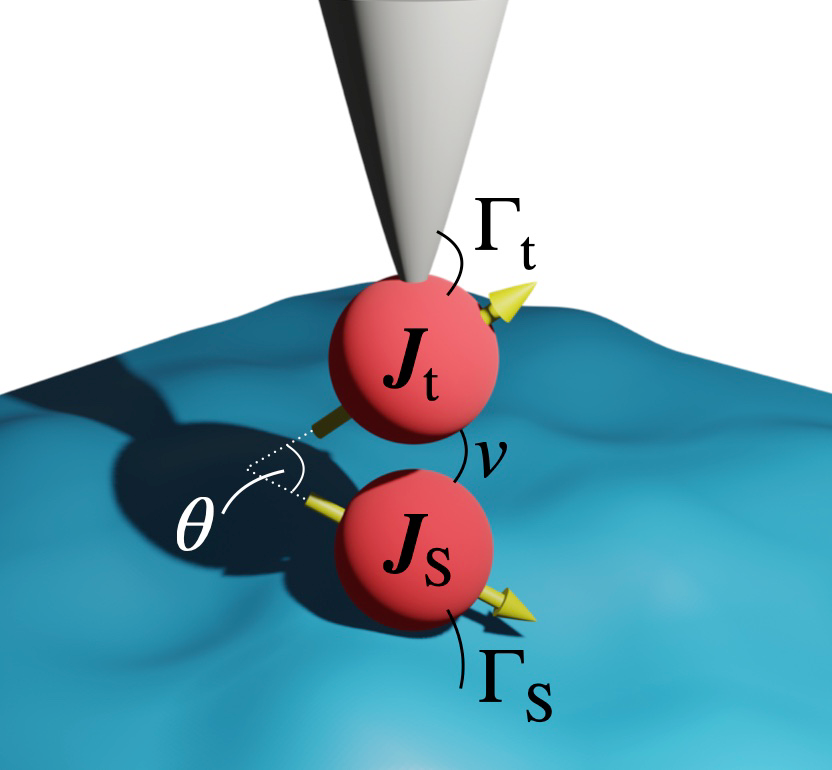}
\caption{Schematic representation of our two-impurity model. Two magnetic impurities are respectively coupled 
to a superconducting substrate and to an STM tip that is also superconducting. The tunneling rates 
$\Gamma_{\rm t}$ and $\Gamma_{\rm S}$ measure the strength of the coupling of the impurity to the tip and 
substrate, respectively, and $v$ is the hopping matrix element describing the tunnel coupling between the 
impurities. These impurities have magnetizations $\boldsymbol{J}_\mathrm{t}$ and $\boldsymbol{J}_\mathrm{S}$ 
whose relative orientation is described by the angle $\theta$.}
\label{fig-model-2imp}
\end{figure}

The double-impurity model of Ref.~\cite{Villas2021} is schematically illustrated in Fig.~\ref{fig-model-2imp}. In this
model, and inspired by the experiments of Ref.~\cite{Huang2020b}, every impurity is strongly coupled to a given 
superconducting lead (tip and substrate) and both impurities are coupled via a tunnel coupling. The Hamiltonian 
describing this model is given by \cite{Villas2021}
\begin{eqnarray}
H & = &  H_{\mathrm{lead,t}} + H_{\mathrm{imp,t}} + H_{\mathrm{hopping,t}} + \nonumber \\ 
& & H_{\mathrm{lead,S}} + H_{\mathrm{imp,S}} + H_{\mathrm{hopping,S}} + V, \label{eq-H-2imp}
\end{eqnarray}
where $H_{\mathrm{lead},j}$ describes the Hamiltonian of lead $j={\rm t,S}$, $H_{\mathrm{imp},j}$ describes 
the respective impurity which is coupled to lead $j$ via the Hamiltonian $H_{\mathrm{hopping,j}}$ and $V$ 
describes the coupling between the two impurities. We choose a global $z$-quantization axis in which to 
describe the creation and annihilation operators
\begin{align}\label{spinors1}
    \bar{c}_{\bm{k},j}^\dagger &= \left( c_{\bm{k}\uparrow,j}^\dagger, c_{-\bm{k}\downarrow,j}, 
    c_{\bm{k}\downarrow,j}^\dagger, - c_{-\bm{k}\uparrow,j}\right) \\ \label{spinors2}    
    \bar{d}^\dagger_j &= \left( d_{\uparrow,j}^\dagger, d_{\downarrow,j}, 
    d_{\downarrow,j}^\dagger, -d_{\uparrow,j} \right), 
\end{align}
where $\bar{c}_{\bm{k},j}^\dagger$ corresponds to lead $j$ and $\bar{d}^\dagger_j$ to impurity $j$. 
We can express the different terms of the Hamiltonian in their respective basis as follows
\begin{align}
    H_{\mathrm{lead},j} &= \frac{1}{2} \sum_{\bm{k}} \bar{c}_{\bm{k},j}^\dagger \bar{H}_{\bm{k},j} \bar{c}_{\bm{k},j} \\
    H_{\mathrm{imp},j} &= \frac{1}{2} \bar{d}_j^\dagger \bar{H}_{\mathrm{imp}} \bar{d}_j \\
    H_{\mathrm{hopping},j} &= \frac{1}{2} \sum_{\bm{k}}\left(\bar{c}_{\bm{k},j}^\dagger \bar{V}_{j} 
    \bar{d}_j+ \bar{d}_j^\dagger \bar{V}_{j}^\dagger \bar{c}_{\bm{k},j} \right) ,
\end{align}
where the respective matrices in spin-Nambu space take the form
\begin{align} \label{eq-Hs-2imp-1}
\bar{H}_{\bm{k},j} &= \sigma_0 \otimes \left( \xi_{\bm{k},j} \tau_3+\Delta_j e^{\imath \varphi_j \tau_3}\tau_1 \right) \\
\bar{H}_{\mathrm{imp},j} &= U_j  \sigma_0 \otimes \tau_3 + \bm{J}_j \cdot (\bm{\sigma} \otimes \tau_0) \\
\bar{V}_{j} & = t_j \sigma_0 \otimes \tau_3, \label{eq-Hs-2imp-3}
\end{align}
with the on-site energies $U_j$ and the exchange energies $\bm{J}_j$. The tunneling term between the two 
magnetic impurities reads
\begin{equation}
    V = \frac{1}{2} \bar{d}^\dagger_{\mathrm{t}} \bar{V}_{\mathrm{tS}} \bar{d}_{\mathrm{S}} + \frac{1}{2} 
    \bar{d}^\dagger_{\mathrm{S}} \bar{V}_{\mathrm{St}} \bar{d}_{\mathrm{t}},
\end{equation}
with the coupling matrix
\begin{equation} \label{eq-hopp-2imp}
    \bar{V}_{\mathrm{St}} = v (\sigma_0 \otimes \tau_3) = \bar{V}_{\mathrm{tS}}^\dagger ,
\end{equation}
where $v$ is the tunneling coupling between the impurities. To simplify things, it is convenient to 
transfer the dependence on $\theta_j$ and $\varphi_j$ to the coupling term $V$ in Eq.~(\ref{eq-H-2imp})
and work with Hamiltonians describing the subsystems in which the corresponding spin points along its 
quantization axis. For this purpose, we introduce the combined unitary transformation $\bar{R}_j = 
e^{\imath \theta_j\sigma_2/2} \otimes e^{-\imath \varphi_j \tau_3/2}$, where $\theta_j$ is the angle 
formed between the exchange energy $\bm{J}_j$ and the global $z$-quantization axis. Upon performing this 
unitary transformation, the Hamiltonian matrices of Eqs.~(\ref{eq-Hs-2imp-1})-(\ref{eq-Hs-2imp-3}) become
now
\begin{align}
\bar{\mathcal{H}}_{\bm{k},j} &= \sigma_0 \otimes \left( \xi_{\bm{k},j} \tau_3 + \Delta_j \tau_1 \right) \\
\bar{\mathcal{H}}_{\mathrm{imp},j} &= U_j  \sigma_0 \otimes \tau_3 +J_j(\sigma_3 \otimes \tau_0) \\
\bar{\mathcal{V}}_j & = t_j \sigma_0 \otimes \tau_3 ,
\end{align}
while the coupling matrices in Eq.~(\ref{eq-hopp-2imp}) adopt now the form
\begin{align}
    \bar{\mathcal{V}}_{\mathrm{tS}} &= v \left( e^{-\imath \theta \sigma_2/2} \otimes \tau_3
    e^{-\imath \varphi_0 \tau_3/2} \right) \\
    \bar{\mathcal{V}}_{\mathrm{St}} &= v \left( e^{\imath \theta \sigma_2/2} \otimes \tau_3 
    e^{\imath \varphi_0 \tau_3/2} \right).
\end{align}
Notice that these coupling matrices effectively describe a spin-active interface in which there are spin-flip 
processes whose probabilities depend on the relative orientation of the impurity spins described by $\theta$.

To obtain the scattering matrix, we shall make use again of the Fischer-Lee relations. For this purpose, we 
first notice that the central region is now twice as big as in the single-impurity case, and the 
corresponding Hamiltonian describing this region reads
\begin{equation}
    \mathcal{H}_{\mathrm{C}} = \begin{pmatrix}
    \bar{\mathcal{H}}_{\mathrm{imp,t}} & \bar{\mathcal{V}}_{\mathrm{tS}} \\ \bar{\mathcal{V}}_{\mathrm{St}} & 
    \bar{\mathcal{H}}_{\mathrm{imp,S}}
    \end{pmatrix} 
\end{equation}
in the rotated basis $(\mathcal{d}_{\mathrm{t}}^\dagger, \mathcal{d}_{\mathrm{S}}^\dagger) \equiv 
(\bar{R}_{\mathrm{t}}\bar{d}_{\mathrm{t}}^\dagger,\bar{R}_{\mathrm{S}}\bar{d}_{\mathrm{S}}^\dagger)$. 
As in previous cases, only the electron-components of the matrix representation are needed for 
the computation of the scattering matrix, namely the first and third component of the spinor basis 
in Eq.~(\ref{spinors2}). Explicitly, that part of the Hamiltonian of the central system reads
\begin{widetext}
\begin{equation}
    \mathcal{H}_{\mathrm{C},e} = \begin{pmatrix} U_{\mathrm{t}}+eV+J_{\mathrm{t}} & 0 & v \cos(\theta/2) 
    & -v \sin (\theta/2) \\
    0 & U_{\mathrm{t}}+eV-J_{\mathrm{t}} & v \sin(\theta/2) & v \cos(\theta/2) \\
    v \cos(\theta/2) & v \sin(\theta/2) & U_{\mathrm{S}}+J_{\mathrm{S}} & 0 \\
    -v \sin(\theta/2) & v \cos(\theta/2) & 0 & U_{\mathrm{S}}-J_{\mathrm{S}},   \end{pmatrix}
\end{equation}
\end{widetext}
in the basis $(\mathcal{d}_{\uparrow,\mathrm{t}}^\dagger,\mathcal{d}_{\downarrow,\mathrm{t}}^\dagger,
\mathcal{d}_{\uparrow,\mathrm{S}}^\dagger,\mathcal{d}_{\downarrow,\mathrm{S}}^\dagger)$. Notice that
we have included the bias voltage $V$ in the tip subsystem as a shift of the corresponding impurity level.
This way we describe a situation in which the voltage entirely drops between the two impurities, as it happens
in the STM experiments. On the other hand, the couplings can be expressed as $H_{\mathrm{hopping},j} = 
\frac{1}{2} \sum_{\bm{k}} (\mathcal{c}_{\bm{k}\uparrow,\mathrm{j}}^\dagger, 
\mathcal{c}_{\bm{k}\downarrow,\mathrm{j}}^\dagger) \mathcal{V}_{j,\rm e} 
(\mathcal{d}_{\uparrow,\mathrm{t}},\mathcal{d}_{\downarrow,\mathrm{t}},
\mathcal{d}_{\uparrow,\mathrm{S}},\mathcal{d}_{\downarrow,\mathrm{S}})^{\rm T}+\mathrm{c.c.}$, where the matrices
$\mathcal{V}_{j,\rm e}$ are given by
\begin{align}
    \mathcal{V}_{\mathrm{t},e} &= \begin{pmatrix}
    t_{\mathrm{t}} & 0 & 0 & 0 \\
    0 & t_{\mathrm{t}} & 0 & 0
    \end{pmatrix} \\
    \mathcal{V}_{\mathrm{S},e} &= \begin{pmatrix}
    0 & 0 & t_{\mathrm{S}} & 0 \\
    0 & 0 & 0 & t_{\mathrm{S}} 
    \end{pmatrix}.
\end{align}
With these couplings, the self-energies of the tip and substrate can be readily computed as we are only 
interested in the normal state scattering matrix, and thus the lead GFs are normal metal ones. In particular, 
the (normal-state) electron part of the GFs are given by $g_{\mathrm{N,e}}^{\rm r/a}  = \mp \imath \sigma_0$. 
Thus, the self-energies read
\begin{align}
    \Sigma_{\rm t,e}^{\rm r/a} &= \mathcal{V}_{\mathrm{t,e}}^\dagger g^{\rm r/a}_{\mathrm{N,e}} 
    \mathcal{V}_{\mathrm{t,e}} \\
    \Sigma_{\rm S,e}^{\rm r/a} &= \mathcal{V}_{\mathrm{S,e}}^\dagger g^{\rm r/a}_{\mathrm{N,e}} 
    \mathcal{V}_{\mathrm{S,e}} , 
\end{align}
or more explicitly
\begin{align}
    \Sigma_{\rm t,e}^{\rm r/a} = \mp \imath \begin{pmatrix}
    \Gamma_{\rm t} & 0 & 0 & 0 \\
    0 & \Gamma_{\rm t} & 0 & 0 \\
    0 & 0 & 0 & 0 \\
    0 & 0 & 0 & 0
    \end{pmatrix} \\
        \Sigma_{\rm S,e}^{\rm r/a} = \mp \imath \begin{pmatrix}
   0 & 0 & 0 & 0 \\
    0 & 0 & 0 & 0 \\
    0 & 0 &  \Gamma_{\rm S} & 0 \\
    0 & 0 & 0 &  \Gamma_{\rm S}
    \end{pmatrix}, 
\end{align}
so they are $4\times 4$ matrices in impurity-spin space and they are not of full rank. The dressed central 
GFs can then be computed from the Dyson equation
\begin{equation}
    G_{\mathrm{C},e}^{\mathrm{r/a}} = \left[ (E\pm \imath \eta_{\mathrm{imp}})1 - 
    \mathcal{H}_{\mathrm{C},e} -\Sigma_{\rm t,e}^{\rm r/a} - \Sigma_{\rm S,e}^{\rm r/a} \right]^{-1},
\end{equation}
with the regularization factor $\eta_{\mathrm{imp}}$ which can again be chosen as zero, as in the single 
impurity case. To compute the components, we also need the scattering rate matrices $\bm{\Gamma}_{\rm t} 
\equiv \Im (\Sigma_{\mathrm{t,e}}^{\rm a})$ and $\bm{\Gamma}_{\rm S}\equiv \Im(\Sigma_{\mathrm{S,e}}^{\rm a})$. 
Finally, the scattering matrix then follows from the Fisher-Lee relation using the $4 \times 4$ dressed 
central GFs, namely
\begin{align}\label{FisherLee}
s_{\rm e}(E,V) &= \begin{pmatrix}
    \mathfrak{r}_{\rm{e}} & \mathfrak{t}_{\rm{e}}^\prime \\ \mathfrak{t}_{\rm{e}} & \mathfrak{r}_{\rm{e}}^\prime
\end{pmatrix} \\
&= \rho_3\sigma_0 -2\imath \left(\bm{\Gamma}_{\rm t}^{1/2}+\imath \bm{\Gamma}_{\rm S}^{1/2} \right) 
G_{\rm C,e}^{\rm r} \left(\bm{\Gamma}_{\rm t}^{1/2}+\imath \bm{\Gamma}_{\rm S}^{1/2} \right), \nonumber 
\end{align}
where $\rho_3$ is the third Pauli matrix in lead space. For the hole part of the scattering matrix, it holds that
\begin{equation}
    s_{\rm h}(E,V) = \sigma_2 s_{\rm e}(-E,V)^{\rm T} \sigma_2,
\end{equation}
and the total scattering matrix reads
\begin{equation}
    s(E,V) = \mathrm{diag}(s_{\rm e}(E,V),s_{\rm h}(E,V)).
\end{equation}

\subsection{Results}

The calculation of the Keldysh action in this double-impurity case is very similar to that of 
a single impurity discussed in Sec.~\ref{Sec:SS}. Again, we have to treat the problem in the Floquet 
language and resort to numerics to compute the charge-resolved probabilities. On a conceptual level, 
the main differences are that the scattering matrix has a different energy dependence and it is non-diagonal
in spin space. Therefore, it can be shown that the CGF in this case reads 
\begin{equation}\label{Action_SYYS}
    \mathcal{A}_{t_0}(\chi) = \frac{t_0}{2h}\int^{eV}_{-eV} dE \ln 
    \left[ \sum_{n=-\infty}^{\infty} P_n(E,V) e^{\imath n\chi} \right] ,
\end{equation}
where $P_n(E,V)$ corresponds to the probability of transferring $n$ charges across the junction and 
$E$ is the Floquet energy. Notice the absence of the spin index, when compared to Eq.~(\ref{Action_SS}). 
We can then compute the current and the noise, which are given by the standard formulas of a multinomial 
distribution
\begin{eqnarray} \label{current-2imp}
    I(V) & = & \frac{e}{2h} \int^{eV}_{-eV} dE \sum^{\infty}_{n=-\infty} 
    n P_n(E,V), \\
    S(V) & = & \frac{e^2}{h}  \int^{eV}_{-eV}  dE
    \left\{ \sum^{\infty}_{n=-\infty} n^2 P_n(E,V) - \right. \\ 
    & & \hspace{2.8cm} \left. \left(\sum^{\infty}_{n=-\infty} n P_n(E,V)\right)^2 \right\} . \nonumber
\end{eqnarray}

\begin{figure*}[t]
\includegraphics[width=0.95\textwidth,clip]{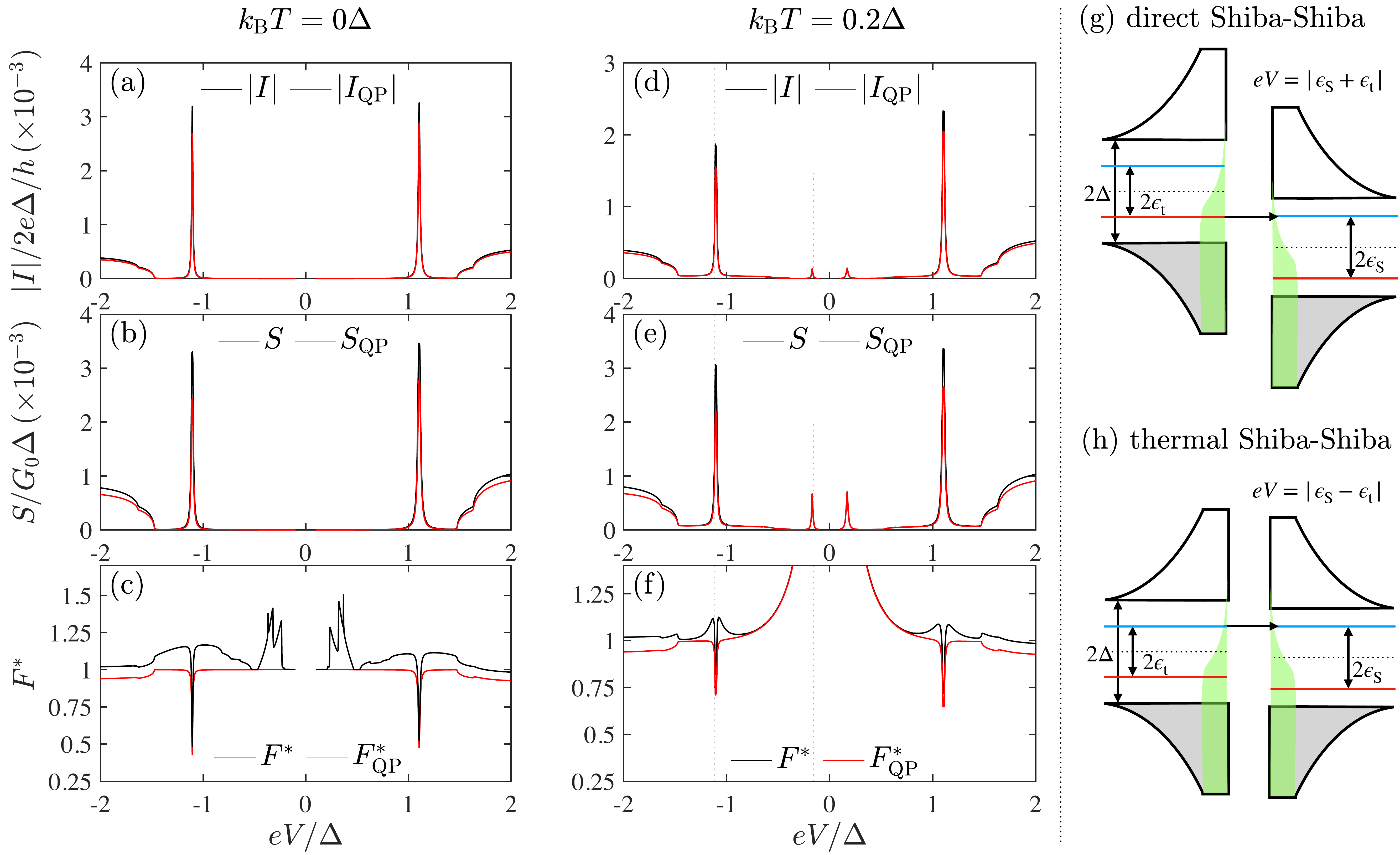}
\caption{Zero-temperature current (a), shot noise (b), and Fano factor (c) as a function of the voltage for the 
case of two impurities coupled to their respective superconducting electrodes at zero temperature. The two 
superconductors are assumed to be identical and have a gap equal to $\Delta$. The model parameters used in this
example are: $\Gamma_{\rm t} = \Gamma_{\rm S} = 100\Delta$, $J_{\rm t} = 60\Delta$, $U_{\rm t}= 0$, $J_{\rm S} = 60\Delta$, 
$U_{\rm S} = 60\Delta$, $\eta_{\rm t} = \eta_{\rm S} = 0.001\Delta$, $v = \Delta$, and $\theta = \pi/2$. With 
these parameters the junction has a normal state conductance of $4 \times 10^{-4}G_0$ and the YSR energies are 
$\epsilon_{\rm t} = 0.48\Delta$ and $\epsilon_{\rm S} = 0.64\Delta$. In the different panels, the black lines
correspond to the exact results (including all possible contributions) and the red lines to the contribution of 
single-quasiparticle tunneling. The dotted vertical lines indicate the position of the voltages 
$eV = \pm |\epsilon_{\rm S}+\epsilon_{\rm t}|$ of the direct Shiba-Shiba tunneling. (d-f) The same as in panels (a-c), 
but for a temperature $k_{\rm B}T = 0.2\Delta$ with additional dotted vertical lines indicating the position 
of the voltages $eV = \pm |\epsilon_{\rm S}-\epsilon_{\rm t}|$ of the thermal Shiba-Shiba tunneling. 
(g,h) Schematic representation of the tunneling processes between the two YSR states. Here, the left 
electrode is a magnetic impurity coupled to a SC tip and the right one is another impurity coupled to a
superconducting substrate both featuring YSR states. The green shaded areas represent the corresponding
Fermi functions of both electrodes. The diagrams follow the same convention as in Fig.~\ref{fig-NS-processes}. 
Panel (g) corresponds to the direct Shiba-Shiba tunneling enabled by a finite spin-mixing angle $\theta$ and
panel (h) to the thermal Shiba-Shiba tunneling enabled by a finite temperature and a spin-mixing angle $\theta \neq \pi$. }
\label{fig-2imp1}
\end{figure*}

To make a connection with the experiments of Refs.~\cite{Huang2020b,Huang2021}, we shall mainly focus here on 
the analysis of the results in the regime of weak coupling between the impurities in which the transport properties 
are mainly determined by the tunneling of quasiparticles. Moreover, we shall assume that the two superconducting
electrodes are identical and the corresponding gap is given by $\Delta$. To illustrate the results in the 
weak-coupling regime, we consider an example in which the different model parameters are given by: 
$\Gamma_{\rm S,t} = 100 \Delta$, $J_{\rm t} = 60\Delta$, $U_{\rm t}= 0$, $J_{\rm S} = 60\Delta$, 
$U_{\rm S} = 60\Delta$. With these parameters, the YSR energies are $\epsilon_{\rm t} = 0.48\Delta$ for the SC 
tip and $\epsilon_{\rm S} = 0.64\Delta$ for the substrate. The Dynes' parameter is chosen to be the same for both 
SCs $\eta_{\rm t,S} = 0.001\Delta$. Moreover, we choose the impurity coupling $v = \Delta$ such that the normal 
state conductance is relatively low ($\approx 4 \times 10^{-4}G_0$) and the quasiparticle tunneling dominates the 
transport. Finally, the spin-mixing angle is chosen as $\theta = \pi/2$, which has been shown to reproduce the 
experimental results \cite{Huang2021}. Actually, this value was interpreted as a way to describe the average 
transport properties in a situation in which the two spins are freely rotating, see Ref.~\cite{Huang2021} for
details. The results for the current, shot noise, and Fano factor as a function of the bias voltage are shown 
in Fig.~\ref{fig-2imp1} for two cases: zero temperature ($k_{\rm B}T=0$) and $k_{\rm B}T = 0.2\Delta$. Let us
first discuss the zero-temperature results. The main salient feature in the current is the appearance of very
pronounced peaks (with a huge negative differential conductance) inside the gap region at voltages given by 
$eV = \pm |\epsilon_{\rm S} + \epsilon_{\rm t}| \approx \pm 1.12\Delta$, see Fig.~\ref{fig-2imp1}(a), which
reproduces the main observation in Refs.~\cite{Huang2020b,Huang2021}. As explained in
Refs.~\cite{Huang2020b,Villas2021,Huang2021}, these current peaks are due to the resonant quasiparticle 
tunneling between the lower and upper YSR states of the impurities in the tip and substrate, as we 
illustrate in Fig.~\ref{fig-2imp1}(g). Such a tunneling process, which we shall refer to as \emph{direct
Shiba-Shiba tunneling}, can occur at any temperature and it only requires the impurity spins to be nonparallel
(i.e., $\theta \neq 0$), otherwise it would be forbidden due to the full spin polarization of the YSR states. Notice 
that in Fig.~\ref{fig-2imp1}(a) we compare the exact result taking into account any possible contribution 
(also from Andreev reflections) and the result only including single-quasiparticle tunneling ($n = \pm 1$). 
Such a comparison shows that the current is dominated in this case by quasiparticle tunneling. With respect 
to the shot noise, see Fig.~\ref{fig-2imp1}(b), it exhibits the same voltage dependence as the current and it 
shows two pronounced peaks at $eV = \pm |\epsilon_{\rm S} + \epsilon_{\rm t}|$ as a result of the direct 
Shiba-Shiba tunneling. Again, the noise is dominated by the contribution of single-quasiparticle tunneling. 
Turning now to the Fano factor, see Fig.~\ref{fig-2imp1}(c), the main feature is the strong reduction at the
voltages of the direct Shiba-Shiba tunneling. In this particular case, the Fano factor reaches a value around 
$0.5$ that actually depends on the bias polarity. Notice that the Fano factor reduction is also dominated by 
single-quasiparticle tunneling. Obviously, the origin of this pronounced reduction of the Fano factor must be 
due to the resonant character of this tunneling process between the two individual YSR states. This will be 
explained below. Notice, on the other hand, that the Fano factor is close to 1 away from the resonant voltage and it 
reaches values above 1 inside the gap due to the contribution of Andreev reflections.

Turning now to the high temperature case ($k_{\rm B}T = 0.2\Delta$), the main novelty in the current is the
presence of two additional current peaks inside the gap that appear at $eV = \pm |\epsilon_{\rm t} - 
\epsilon_{\rm S}| \approx 0.16\Delta$, see Fig.~\ref{fig-2imp1}(d). These current peaks, which were in fact 
experimentally observed \cite{Huang2020b,Huang2021}, are due to the thermally activated tunneling between the 
two upper (or excited) YSR states, as we illustrate in Fig.~\ref{fig-2imp1}(h). As explained in 
Refs.~\cite{Huang2020b,Villas2021,Huang2021}, this thermally activated tunneling process, which we shall refer 
to as \emph{thermal Shiba-Shiba tunneling}, can occur when the temperature is high enough to have a 
partial occupation of the excited YSR states and it requires that the impurity spins are not antiparallel 
(i.e., $\theta \neq \pi$), otherwise it would be forbidden due to the full spin polarization of the YSR states.
Again, we note that the current in this case is also dominated by single-quasiparticle tunneling, see 
Fig.~\ref{fig-2imp1}(d). The thermal Shiba-Shiba tunneling is also visible in the shot noise in the 
form of two peaks at $eV = \pm |\epsilon_{\rm t} - \epsilon_{\rm S}|$. In the case of the Fano factor, the
values at the voltages $eV = \pm |\epsilon_{\rm t} + \epsilon_{\rm S}|$ associated to direct Shiba-Shiba
tunneling have increased as compared to the zero-temperature case, while there is no visible feature at the
biases of the thermal Shiba-Shiba tunneling because in this example it is masked by the rapid increase of the 
thermal noise that dominates the low-bias regime in the Fano factor at finite temperatures.

\begin{figure}[t]
\includegraphics[width=\columnwidth,clip]{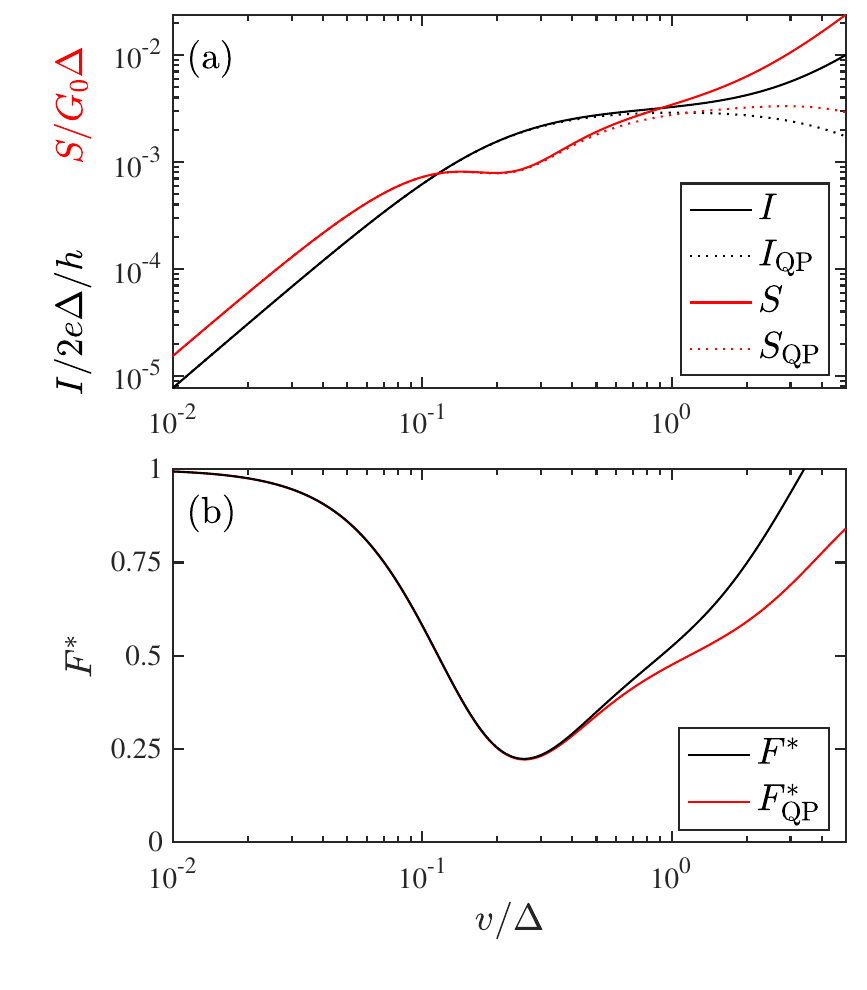}
\caption{(a) Zero-temperature current and shot noise at the direct Shiba-Shiba energy $eV = |\epsilon_{\rm S} + 
\epsilon_{\rm t}|$ as a function of the coupling $v$. The rest of the parameters of the model correspond to those 
of Fig.~\ref{fig-2imp1}. The solid lines correspond to the exact results, while the dotted lines were computed
taking only into account the contribution of the tunneling of single quasiparticles. (b) The corresponding
Fano factor $F^{\ast}$. The black line corresponds to the exact results and the red line to the contribution 
of single-quasiparticle tunneling.}
\label{fig-2imp2}
\end{figure}

From the discussion of the example of Fig.~\ref{fig-2imp1} it is obvious that the most important feature related
to the tunneling between two YSR states is the Fano factor reduction at the resonant bias of the direct Shiba-Shiba
tunneling. To understand its origin and magnitude we first need to analyze more systematically how this reduction
depends on the different parameters of the model. In particular, the Fano factor related to the direct Shiba-Shiba
tunneling is expected to depend on the relative value between the effective tunneling rate (related to the hopping element
$v$) and the broadening (or lifetime) of the bound states. In fact, this issue is very much related to another
interesting observation reported in Ref.~\cite{Huang2020b}, namely the fact that the height of the direct 
Shiba-Shiba current peaks (and their area) undergoes a crossover between a linear regime at very low transmission
(or normal state conductance) and a sublinear regime at higher transmission when the STM tip with its impurity 
was brought closer to the impurity on the substrate. To understand the impact of the junction transmission, we 
considered the zero-temperature example of Fig.~\ref{fig-2imp1} and computed the value of the current, noise
and Fano factor at the direct Shiba-Shiba voltage $eV = |\epsilon_{\rm S}+\epsilon_{\rm t}|$ as a function of 
the hopping matrix element $v$ describing the coupling between the impurities, while the rest of the parameters are 
kept constant. The results are displayed in Fig.~\ref{fig-2imp2}, where we show both the exact results and those 
computed taking only into account the contribution of single-quasiparticle processes. As it can be seen in panel (a),
the peak height indeed exhibits the type of crossover observed experimentally. For small values of $v$, i.e.,
deep into the tunnel regime, the current peak height increases linearly with the transmission (which in this regime
is proportional to $v^2$) and it crosses over to a sublinear regime at higher couplings. Moreover, this crossover
is mainly about the quasiparticle current and only at high enough couplings ($v \gtrsim \Delta$) we start to see 
additional contributions due to Andreev reflections. Interestingly, the shot noise exhibits a similar crossover,
but with an important difference, namely the fact that it enters the sublinear regime more quickly than the current
and even exhibits a local minimum at a certain value of $v$. This peculiar behavior is clearly reflected in the 
evolution of Fano factor, see Fig.~\ref{fig-2imp2}(b). The Fano factor for very small couplings is equal to 1,
as expected from standard (non-resonant) tunneling, it is progressively reduced as the coupling increases reaching
a minimum value of around ($\sim 0.22$), and it increases again monotonically for even higher couplings. Notice
also that the quasiparticle contribution dominates the Fano factor up to coupling values clearly beyond that at 
which the minimum takes place.

\begin{figure*}[t]
\includegraphics[width=\textwidth,clip]{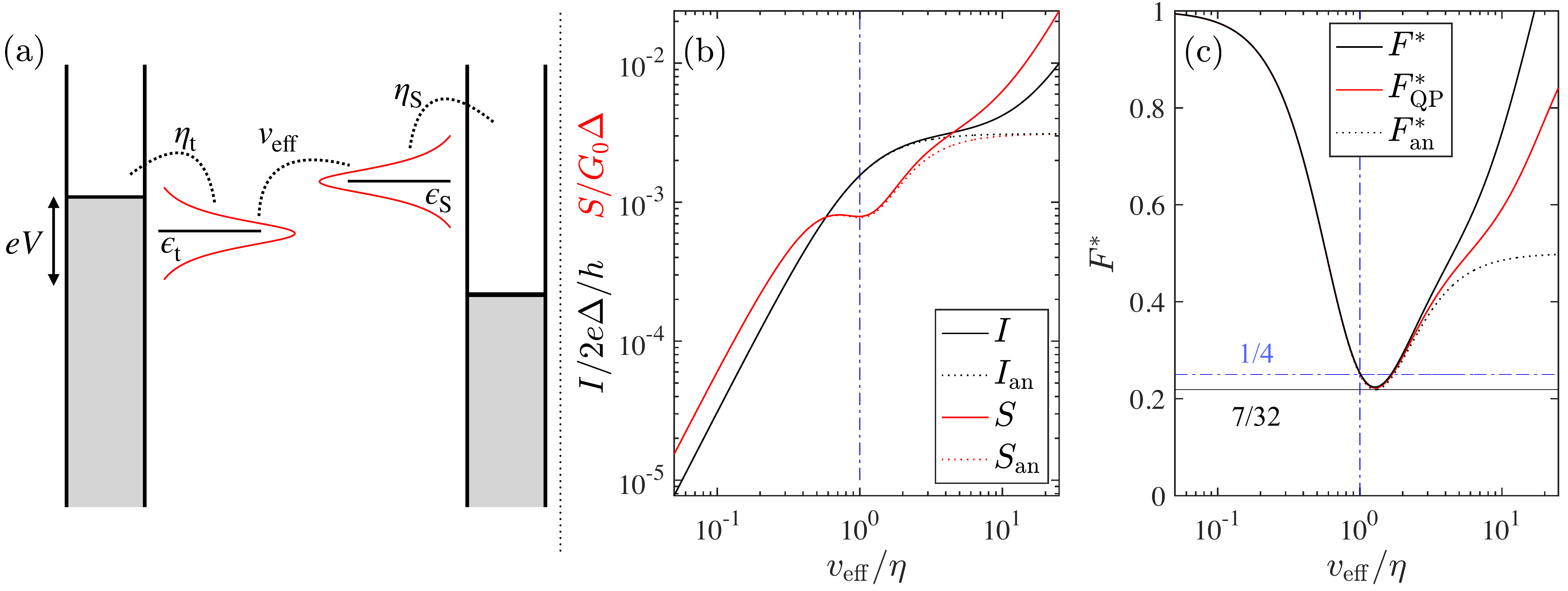}
\caption{(a) Schematic representation of the model used to describe the tunneling between two bound states.
(b) Zero-temperature current and shot noise computed with the model of panel (a) (dotted lines) as a function
of the ratio $v_{\rm eff}/\eta$. The solid lines correspond to the exact results taken from Fig.~\ref{fig-2imp2} 
and using the expression of $v_{\rm eff}$ of Eq.~(\ref{eq-veff}). The vertical line indicates the case 
$v_{\rm eff} = \eta$. (c) The corresponding Fano factor computed with the model (dotted line) and the 
results of Fig.~\ref{fig-2imp2}(b) (both exact and single-quasiparticle contributions). The solid horizontal 
line indicates a Fano factor equal to $7/32$, while the the horizontal dashed line corresponds to the Fano factor 
equal to $1/4$.}
\label{fig-2imp3}
\end{figure*}

To understand the origin of this crossover, the existence and magnitude of the Fano factor minimum, and whether this
behavior is universal, we put forward the following toy model. We consider two energy levels $\epsilon_{\rm t}$
and $\epsilon_{\rm S}$ that are coupled to their respective electron reservoirs. Due to this coupling, the 
quantum levels acquire a broadening, which we assume to be equal for both levels and given 
by $\eta_{\rm S} = \eta_{\rm t} = \eta$. Additionally, these two levels are coupled via a tunnel coupling 
$v_{\rm eff}$. This model is illustrated in Fig.~\ref{fig-2imp3}(a). Notice that we do not even need to invoke the
existence of superconductivity. The energy and bias dependent electron transmission in this toy model is given by
\cite{Cuevas2017}
\begin{equation}
    T(E,V) = \frac{4 \pi^2 v_{\rm eff}^2 \rho_{\rm t}(E-eV) \rho_{\rm S}(E)}
    {|1 - v_{\rm eff}^2 g_{\rm t}(E-eV) g_{\rm S}(E)|^2} ,
\end{equation}
where $g_{\rm t,S}(E)$ are advanced GFs related to both energy levels and given by
\begin{equation}
    g_{\rm t,S}(E) = \frac{1}{E - \epsilon_{\rm t,S} - \imath \eta}, 
\end{equation}
and $\rho_{\rm t/S}(E) = (1/\pi) \Im (g_{\rm L/R})$ are the DOS associated to those two levels. Without loss
of generality, we choose $\epsilon_{\rm t}<0$, $\epsilon_{\rm S}>0$, and a positive bias. 
Since we want to emulate the transport properties at the direct Shiba-Shiba energy, we shall set from now on 
$eV = |\epsilon_{\rm t}| + |\epsilon_{\rm S}|$. At this bias, the transmission adopts the form
\begin{align}
    T(E) & = \frac{4 v_{\rm eff}^2 \eta^2}{[(E-\epsilon_{\rm S}-v_{\rm eff})^2+\eta^2]
    [(E-\epsilon_{\rm S}+v_{\rm eff})^2+\eta^2]} \nonumber \\
    & =  \frac{4\gamma^2}{\left[ \left( \frac{E -\epsilon_{\rm S}}{\eta} -\gamma \right)^2 + 1 \right] 
                          \left[ \left( \frac{E - \epsilon_{\rm S}}{\eta} + \gamma \right) ^2 + 1 \right]} ,
\end{align}
where $\gamma = v_{\rm eff}/\eta$. Within this model, the corresponding zero-temperature current and noise are
given by
\begin{eqnarray}
    I & = & \frac{2e}{h} \int_0^{eV} T(E) \, dE , \\
    S & = & \frac{4e^2}{h} \int_0^{eV} T(E) \left[ 1 - T(E) \right] dE .
\end{eqnarray}
Assuming that $\eta \ll |\epsilon_{\rm t,S}|$ and $v_{\rm eff}< |\epsilon_{\rm t}|+|\epsilon_{\rm S}|$, one can 
compute
analytically the previous integrals to obtain 
\begin{eqnarray} \label{i_v}
    I & = & \left(\frac{2e}{h}\right) 2 \pi \eta \frac{\gamma^2}{\gamma^2+1} , \\ \label{shot_v}
    S & = & \left(\frac{4e^2}{h}\right) 2 \pi \eta \left[ \frac{\gamma^2}{\gamma^2+1} - 
\frac{\gamma^4(\gamma^2+5)}{2(\gamma^2+1)^3} \right] .
\end{eqnarray}
The corresponding Fano factor is then given by
\begin{equation}\label{fano_v}
    F^*= \frac{S}{2eI} = \frac{\gamma^4-\gamma^2+2}{2(\gamma^2+1)^2}.
\end{equation}
These expressions nicely summarize the type of crossover that we discussed above and show that it is simply controlled
by the parameter $\gamma$, i.e., ratio between the tunnel coupling and the level width. Before comparing with the actual
results of the two impurity system, let us analyze the Fano factor in different limiting cases. First, in the weak
coupling regime $\gamma \ll 1$, the Fano factor becomes 1, which is the expected result for non-resonant tunneling
situations. Second, in the high coupling regime $\gamma \gg 1$, the Fano factor tends to $1/2$, which is the generic
expected result for two very narrow energy levels. We shall see that this regime cannot be easily achieved with the 
YSR states because for very large $v$ the Fano factor is altered by the Andreev reflections. More interestingly, the
Fano factor can be much smaller than $1/2$. Thus, for instance, for $\gamma=1$ the Fano factor is exactly $1/4$. In
particular, the shot noise exhibits a local minimum at this value of $\gamma$, which we propose as a new method to extract
the intrinsic lifetime of the states. However, the Fano factor of $1/4$ does not mark the lowest Fano factor. It is 
easy to show that the minimum Fano factor value is achieved when $\gamma = \sqrt{5/3} > 1$, which results in a Fano 
factor equal to $7/32$. This is a unique value that, to our knowledge, is not found in any other generic situation.

To establish a quantitative comparison between the toy model and the direct tunneling between YSR states, 
we still need to identify the effective tunneling coupling in the latter situation. In the presence of YSR states 
the effective coupling must take into account both the spin-mixing angle $\theta$ and the coherent factors that
determine the height of the DOS associated to the YSR states. These coherent factors are given in our model by
\begin{align}
    u_{j}^2 &= \frac{2J_{j} \Gamma_{j}^2 \Delta_j}{\left[ \Gamma_{j}^2+(J_{j}+U_{j})^2 \right]^{3/2}
    \left[\Gamma_{j}^2+(J_{j}-U_{j})^2 \right]^{1/2}} , \\
    v_{j}^2 &= \frac{2J_{j} \Gamma_{j}^2 \Delta_j}{\left[ \Gamma_{j}^2+(J_{j}-U_{j})^2 \right]^{3/2} 
    \left[\Gamma_{j}^2+(J_{j}+U_{j})^2 \right]^{1/2}} .
\end{align}
with $j = \rm t,S$. Thus, the effective tunnel coupling $v_{\rm eff}$ describing the direct Shiba-Shiba tunneling
can be expressed in terms of the bare $v$ hopping matrix elements as follows 
\begin{equation} \label{eq-veff}
    v_{\rm eff}= v \sin(\theta/2) \times \begin{cases} u_{\rm t} v_{\rm S} & \mbox{for} \; V>0 \\ 
u_{\rm S} v_{\rm t} & \mbox{for} \; V<0 \end{cases}.
\end{equation}

Finally, we are in position to establish the desired comparison, which is shown in Fig.~\ref{fig-2imp3}(b,c). There
we present both the analytical results obtained with the toy model for the current, noise, and Fano factor at the
resonant bias, as well as the exact results shown in Fig.~\ref{fig-2imp2}(a,b) where the tunnel coupling has been 
renormalized according to Eq.~(\ref{eq-veff}). The results are presented as a function of the ratio $v_{\rm eff}/
\eta$. Notice that the toy model is able to quantitative reproduce the results of our two-impurity model over a broad 
range of values of that ratio. In particular, the toy model nicely reproduces the crossover in all transport 
properties up to coupling values in which Andreev reflections start playing a role. Interestingly, we now
can see that the local minimum in the shot noise exactly corresponds to the case $v_{\rm eff}= \eta$ when the
effective coupling is equal to the broadening of the YSR states. Moreover, the minimum of the Fano factor is shown
to reach the value $7/32$ predicted by the toy model, and the Fano factor becomes $1/4$ exactly when the 
shot noise is locally minimal. On the other hand, notice that the prediction of the toy model of a Fano factor equal
to $1/2$ when $v_{\rm eff} \gg \eta$ is not reached in the direct Shiba-Shiba tunneling case due to the onset of 
the Andreev reflections, as one can see in Fig.~\ref{fig-2imp3}(c). Finally, the most important conclusion of the
impressive agreement with the toy model is the universality of the results when considered as a function of the
the ratio $v_{\rm eff}/\eta$, as long as the transport is dominated by the tunneling of single quasiparticles.
This universality is illustrated in Ref.~\cite{SM} where we present the results for this crossover for 
different sets of values of the model parameters, see Fig.~S15.

It is important to stress that our FCS approach is valid for arbitrary junction transparency and can also describe
situations in which the transport is dominated by MARs. Actually, this is a very interesting subject since, as it
has been theoretically shown in Ref.~\cite{Villas2021} for the current, the transport properties are expected 
to exhibit an extremely rich subgap structure due to the occurrence of a variety of different types of MARs, some
of which have no analog in the single-impurity case. However, due to the richness of that physics, and to keep
the length of this manuscript under control, we shall postpone the analysis of that regime to a forthcoming
publication. In any case, we have included some examples of the high transparency regime in Ref.~\cite{SM} to 
illustrate once more the power of our FCS approach, see Figs.~S10-S14.

\section{Discussion and conclusions} \label{sec:conclusions}

As explained in Sec.~\ref{sec:intro}, the concept of FCS has already been used to understand the transport
properties of several basic superconducting junctions. However, the Keldysh action
described in Sec.~\ref{Sec:action} \cite{Snyman2008,Nazarov2015}, which has remained fairly unnoticed, puts 
this concept at a whole new level because it allows us to describe the coherent transport in situations
where the scattering matrix of the system may depend on energy, spin, and even involve an arbitrary number of
superconducting terminals. Here, we have shown how this action can be used in practice by combining it
with model Hamiltonians, and we have illustrated its power in the several examples concerning the spin-dependent 
transport in hybrid superconducting systems that exhibit YSR states. Such a combination opens up the possibility 
to shed new light on the coherent transport in numerous superconducting nanostructures. Thus, for instance, 
among the natural extensions of this work we can mention the study of Majorana physics in magnetic atomic chains
from the FCS point of view \cite{Nadj-Perge2014,Ruby2015b,Kezilebieke2018,Ruby2017,Ruby2018,Kamlapure2021,Schneider2022}.
In fact, it has been recently suggested that shot-noise tomography in
hybrid magnetic-superconducting wire systems could be used to distinguish Majorana modes from other trivial
fermionic states \cite{Perrin2021}. Another context in which one could straightforwardly apply our FCS approach 
is that of superconductor-semiconductor nanowire junctions in which one can realize single and double quantum dot 
systems that exhibit YSR states, see Refs.~\cite{Grove-Rasmussen2018,Estrada2018,Steffensen2022} and references 
therein. An interesting situation in which it would be very interesting to apply the FCS approach is that of 
microwave-irradiated junctions \cite{Cuevas2002,Chauvin2006,Kot2020}. It has been recently shown that one can 
gain some new insight into the interplay between YSR states and Andreev transport by studying the single-impurity 
systems discussed in this work with the assistance of a microwave field \cite{Peters2020,Gonzalez2020,Siebrecht2023}. 
Probably the most exciting possibility of the approach put forward in this work is the analysis of multiterminal
superconducting systems. As mentioned above, the Keldysh action described here is valid for arbitrary number of 
terminals. In this sense, we are very much interested in extending this work to study different aspects of 
superconducting multiterminal systems that are currently attracting a lot of attention such as the
generation of Cooper quartets \cite{Freyn2011,Jonckheere2013,Pfeffer2014,Cohen2018,Huang2022c}, the study
of the Josephson effects \cite{Draelos2019,Graziano2020,Pankratova2020,Arnault2021}, or the possibility 
to engineer Andreev bound states with interesting topological properties \cite{vanHeck2014,Yokoyama2015,Riwar2016,
Eriksson2017,Meyer2017,Xie2017,Xie2019,Repin2019,Gavensky2019,Houzet2019,Klees2020,Coraiola2023}. 
Last but not least, nothing prevents us from using this FCS approach in the case of junctions involving 
unconventional superconductors.

It is worth remarking that our whole analysis of the FCS in these impurity systems has been done
making use of mean-field models. In that regard, it would be interesting, albeit very challenging, to investigate
the role of electronic correlations or spin fluctuations in the transport properties discussed in this work, most 
notably in the noise. This clearly goes beyond the scope of this work and it cannot be done with the action of 
Eq.~(\ref{action}) used here as starting point. The success of these mean-field models is indeed quite remarkable,
as we have shown in Sec.~\ref{sec-comp} and has been demonstrated in numerous publications \cite{Ruby2015,Huang2020a,
Huang2020b,Huang2021,Karan2022,Peters2020,Gonzalez2020,Villas2021,Siebrecht2023}. So, it will be interesting to see
if future shot noise experiments reveal the presence of significant correlation effects.

To summarize the present work, let us say that we have shown here that the FCS approach can be extended to 
describe the spin-dependent transport in systems involving magnetic impurities coupled to superconducting
leads. In particular, with the analysis of different situations we have illustrated how the concept of FCS provides
an unprecedented insight into the interplay between YSR states and electronic transport. Among the lessons and 
predictions put forward here, we can highlight the following ones. First, in the case of single-impurity junctions 
with only one superconducting electrode, we have shown that the whole subgap transport can be understood as a competition 
between single-quasiparticle tunneling and a resonant Andreev reflection. Such a competition is especially reflected
in the shot noise and Fano factor, which allow us to access energy scales that are out of the scope of conventional
conductance measurements, most notably the YSR lifetimes. In particular, we have illustrated this fact with the
analysis of very recent experiments that have reported for the first time shot noise measurements in these hybrid
atomic-scale systems \cite{Thupakula2022}. Second, in the case of single-impurity junctions with two superconducting 
leads, we have shown how the FCS concept allows us to unambiguously identify the contribution of all the tunneling processes,
including multiple Andreev reflections. In particular, this has helped us to correct common misinterpretations on the
origin of certain features in the subgap structure of the conductance. In particular, we have discussed the unique
signatures of the so-called resonant multiple Andreev reflections (multiple versions of the known lowest-order 
resonant Andreev reflection), which should enable their experimental identification, something that to our knowledge
has not been reported thus far. Moreover, we have presented extensive predictions on how the occurrence of these 
resonant Andreev reflections, and YSR-mediated Andreev reflections, are reflected in the shot noise and Fano factor.
Finally, in connection to recent experiments on the tunneling between two individual YSR states in two-impurity systems
\cite{Huang2020b,Huang2021}, we have shown that such a tunneling leads to an unambiguous signature in the Fano factor
in the form of a strong reduction at the resonant bias voltage at which the two YSR states are aligned. 
In particular, we have demonstrated that the direct Shiba-Shiba tunneling at low temperatures can exhibit a Fano factor 
as small as $7/32$, which results from the resonant quasiparticle tunneling between two YSR states. This constitutes a
novel result in quantum transport that has not been realized in any other system. So, in short, we think that these 
examples and predictions will motivate other theoretical groups to use the concept of FCS to provide a new point of 
view on the superconducting transport in numerous situations. Moreover, our work clearly shows the importance of going
beyond conductance measurements to truly understand the Andreev transport in the presence of superconducting bound 
states. In this sense, we are convinced that this work will trigger off the realization of new shot noise measurements 
in these atomic-scale superconducting systems exhibiting YSR states.

\begin{acknowledgments}

We would like to thank Freek Massee, Alexandra Palacio-Morales, and Marco Aprili for numerous discussions and for 
providing us the experimental data of Ref.~\cite{Thupakula2022}. We also thank Raffael Klees, Christian R.\ Ast,
Pascal Simon, and Matthias H\"ubler for helpful discussions. D.C.O.\ and W.B.\ acknowledge support by the Deutsche 
Forschungsgemeinschaft (DFG; German Research Foundation) via SFB 1432 (Project No. 425217212). J.C.C.\ thanks the 
Spanish Ministry of Science and Innovation (Grant PID2020-114880GB-I00) for financial support and the DFG and SFB 
1432 for sponsoring his stay at the University of Konstanz as a Mercator Fellow.

\end{acknowledgments}

\appendix

\section{Keldysh Green function formalism} \label{App:GFs}

One ingredient for the Keldysh action in Eq.~(\ref{action}) are the Green's functions (GFs) of the 
leads, which are described in the framework of the Keldysh formalism \cite{Keldysh1965}. We provide
a more in-depth discussion of these functions in this Appendix. The retarded and advanced GFs of a 
BCS superconductor (SC) in spin-Nambu space are given by
\begin{align} \label{GFspinnambu}
    \bar{g}^{\rm r/a}(E) & = \sigma_0 \otimes \frac{-\imath}{\sqrt{\Delta^2-(E\pm \imath \eta)^2}} \begin{pmatrix} 
    (E\pm \imath \eta) & \Delta e^{\imath \phi} \\ -\Delta e^{-\imath \phi} & -(E\pm \imath \eta) 
    \end{pmatrix} \nonumber \\
    & = \sigma_0 \otimes \begin{pmatrix} g^{\rm r/a} & f^{\rm r/a} \\ {\mathfrak{f}^{\rm r/a}} & -g^{\rm r/a} 
    \end{pmatrix}, 
\end{align}
where we have used the spinor basis $(\Psi_\uparrow^\dagger, \Psi_\downarrow,\Psi_\downarrow^\dagger,-\Psi_\dagger)$ 
and the Pauli matrices $\sigma_i$ in spin space, $\sigma_0$ being the identity. Note that we already 
eliminated the $k$-dependence by summing it out. The bar ($\ \bar{}\ $) indicates that a quantity is expressed 
in spin-Nambu space. The SC is described by a gap $\Delta>0$ and the phase of the SC condensate $\phi$. 
The regularization parameter $\eta$ should in principle tend to zero in these expressions, but it can be kept
constant to describe in simple terms inelastic mechanisms that tend to broaden the electronic states (Dynes'
parameter). Note that we are using here GFs subjected to the normalization condition $\bar{g}^{\rm r/a} 
\bar{g}^{\rm r/a} = \bar{1}$, where $\bar{1}$ is the identity in spin-Nambu space. Notice also that the GFs 
are block-diagonal in spin space
\begin{equation}
    \bar{g}^{\rm r/a}(E) = \begin{pmatrix} g^{\rm r/a}_\Uparrow(E) & 0 \\ 0 & g^{\rm r/a}_\Downarrow(E) \end{pmatrix},
\end{equation}
where $g^{\rm r/a}_\Uparrow(E)$ is the GF in the spinor basis $\Psi_\Uparrow = (\Psi_\uparrow^\dagger, 
\Psi_\downarrow)$, including spin-up electrons and spin-down holes, whereas $g^{r/a}_\Downarrow(E)$ is the 
GF in the spinor basis $\Psi_\Downarrow = (\Psi_\downarrow^\dagger, -\Psi_\uparrow)$, describing spin-down 
electrons and spin-up holes. We shall assume in this work that $g_\Uparrow = g_\Downarrow$ (no net 
spin polarization in the electrodes). 

In the case of a normal metal ($\Delta = 0$), the GFs reduce to
\begin{equation}
    \bar{g}^{\rm r/a}_{\mathrm{N}}(E) = \pm \sigma_0 \otimes \tau_3 , 
\end{equation}
where we have used the Pauli matrices $\tau_i$ in Nambu space. 

To deal with nonequilibrium situations, we consider the Keldysh GFs of a SC given by
\begin{equation}
    \check{g}(E) = \begin{pmatrix} \bar{g}^{--} & \bar{g}^{-+} \\ \bar{g}^{+-} & \bar{g}^{++} \end{pmatrix} = \frac{1}{2}
    \begin{pmatrix}
        \bar{g}^{\rm a}+\bar{g}^{\rm r}+\bar{g}^{\rm k} & \bar{g}^{\rm a}-\bar{g}^{\rm r}+\bar{g}^{\rm k} \\
        \bar{g}^{\rm a}-\bar{g}^{\rm r}-\bar{g}^{\rm k} & \bar{g}^{\rm a}+\bar{g}^{\rm r}-\bar{g}^{\rm k} 
    \end{pmatrix},
\end{equation}
with the retarded and advanced GF in spin-Nambu basis from Eq.~(\ref{GFspinnambu}) and the Keldysh component of 
the GF $\bar{g}^{\rm k} = (\bar{g}^{\rm r}-\bar{g}^{\rm a})(1-2f)$ with the Fermi function 
of the superconductor $f(E) = 1/(e^{E/k_{\mathrm{B}}T}+1)$ with the temperature $T$ and the Boltzman 
constant $k_{\mathrm{B}}$. The index $(\ \check{} \ )$ indicates that the GF is expressed in 
Keldysh-spin-Nambu space. The Keldysh GF can be brought into the time-domain by a Fourier transformation as follows:
\begin{equation}
    \check{g}(t_{\mathrm{rel}}) = \int dE \, \check{g}(E) e^{\imath Et_{\mathrm{rel}}} ,
\end{equation}
where $t_{\mathrm{rel}} = t-t^\prime$ is the relative time. If a reservoir is voltage biased, the voltage can be 
gauged onto the GF. Namely, for a constant bias voltage $V$, the SC phase is given by $\phi(t) = 2eVt/\hbar$ 
(where the dc phase $\phi_0$ is already included in the GF in Eq.~(\ref{GFspinnambu})). The voltage dependent GF 
is then obtained via
\begin{equation}
    \check{g}(t,t^\prime) = e^{\imath \phi(t) (\nu_0\otimes \sigma_0 \otimes \tau_3)/2}
    \check{g}(t_{\mathrm{rel}}) e^{-\imath \phi(t^\prime) (\nu_0\otimes \sigma_0 \otimes \tau_3)/2},
\end{equation}
where the Pauli matrices in Keldysh $\nu_i$, spin $\sigma_i$ and Nambu $\tau_i$ space and their respective 
identities $\nu_0$, $\sigma_0$ and $\tau_0$ are used. Similar to the voltage, the counting field $\chi$ can 
be gauged onto the SC GF
\begin{equation}\label{A8}
    \check{g}(\chi,t,t^\prime) = e^{-\imath \chi(\nu_3\otimes \sigma_0 \otimes \tau_3)/2}\check{g}(t,t^\prime ) 
    e^{\imath \chi(\nu_3\otimes \sigma_0 \otimes \tau_3)/2}.
\end{equation}
Since we only consider two-terminal settings in this work, we only need a single counting field (the other
one can be gauged away). Additionally, the non-diagonal entries in Nambu space result in a dependence of the GF 
on the average time $t_{\rm av} = (t+t^\prime)/2$. We can thus express the GF as a function of relative time 
$t_{\rm rel}$ and average time $t_{\rm av}= (t+t^\prime)/2$, namely $\check{g}(t_{\rm rel},t_{\rm av})$.
For a normal metal, the off-diagonal entries in Nambu space are zero and the GFs only depend on the 
relative time $t_{\mathrm{rel}}$. Then, the GFs can be trivially Fourier transformed back with respect to 
$t_{\mathrm{rel}}$ to the energy space. However, because of the occurrence of $t_{\mathrm{av}}$, this is 
not possible in the SC case. Hence, we introduce the Wigner representation of the GFs following 
Ref.~\cite{Floquet2008}. The $n$th moment of the Keldysh GF is defined by
\begin{equation}
    \check{g}^n(E) \equiv \int_{\mathbb{R}} dt_{\mathrm{rel}} \frac{1}{\tau} 
    \int_{-\tau/2}^{\tau/2}dt_{\mathrm{av}} e^{\imath E t_{\mathrm{rel}}+\imath n 
    \Omega t_{\mathrm{av}}} \check{g}(t_{\mathrm{rel}},t_{\mathrm{av}}),
\end{equation}
where $\Omega = 2eV$ is the fundamental energy (periodicity of the average time argument) and 
$\tau = 2\pi/\Omega$ is the corresponding period. This way, we restrict the Floquet energy to the first 
Floquet Brioullin zone, namely $E \in [-\Omega/2,\Omega/2]=[-eV,eV]$. From the Wigner representation of 
the GF, one can compute the components of the so-called Floquet matrix $\mathcal{G}$ which are given by
\begin{equation}\label{A10}
    (\mathcal{G})^{mn}(E) \equiv \check{g}^{mn}(E) \equiv \check{g}^{m-n}\left(E+(m+n)eV\right),
\end{equation}
which explicitly written out for the case of the voltage-biased SC GF in Keldysh space reads (we omit the 
spin structure as the GF is identity in spin space)
\begin{widetext}
\begin{equation}\label{SCvoltage}
         \check{g}^{mn} = \begin{pmatrix}
    g^{--}_{2m+1} \delta_{m,n} &  
    e^{-\imath \chi} f^{--}_{2m+1} \delta_{m+1,n} &
    e^{-\imath \chi} g^{-+}_{2m+1} \delta_{m,n}&
    f^{-+}_{2m+1} \delta_{m+1,n} \\
    e^{\imath \chi} \mathfrak{f}^{--}_{2m-1} \delta_{m,n+1} &
     -g^{--}_{2m-1} \delta_{m,n} & 
    \mathfrak{f}^{-+}_{2m-1} \delta_{m,n+1} &
     -e^{\imath \chi} g^{-+}_{2m-1} \delta_{m,n}\\
    e^{\imath \chi} g^{+-}_{2m+1}\delta_{m,n} &
    f^{+-}_{2m+1} \delta_{m+1,n} &
   g^{++}_{2m+1} \delta_{m,n} &  
   e^{\imath \chi} f^{++}_{2m+1} \delta_{m+1,n} \\
    \mathfrak{f}^{+-}_{2m-1} \delta_{m,n+1} &
    -e^{-\imath \chi} g^{+-}_{2m-1} \delta_{m,n} &
    e^{-\imath \chi} \mathfrak{f}^{++}_{2m-1} \delta_{m,n+1} & 
    -g^{++}_{2m-1} \delta_{m,n} 
    \end{pmatrix}.
\end{equation}
\end{widetext}
where we have used the notation $g_{n}\equiv g(E+neV)$. Hence, the GF is an infinitely large matrix in Floquet space
with each entry being an $8\times 8$ Keldysh-spin-Nambu GF. In the case of a constant voltage, there are off-diagonal
contributions in the Floquet space, characterized by factors of $\delta_{m+1,n}$ and $\delta_{m,n+1}$. 
In the case when the SC is not voltage-biased, there are no off-diagonal entries in Floquet space and we obtain
\begin{widetext}
\begin{equation}\label{Eq_floquet_SC_V0}
     \check{g}^{mn}_{\mathrm{SC}}(\chi,\omega) = \begin{pmatrix}
    g^{--}_{2m} &  
    e^{-\imath \chi} f^{--}_{2m}&
    e^{-\imath \chi} g^{-+}_{2m} &
    f^{-+}_{2m} \\
    e^{\imath \chi} \mathfrak{f}^{--}_{2m} &
     -g^{--}_{2m} & 
     \mathfrak{f}^{-+}_{2m} &
    -e^{\imath \chi}g^{-+}_{2m} \\
    e^{\imath \chi} g^{+-}_{2m} &
    f^{+-}_{2m} &
    g^{++}_{2m} &  e^{\imath \chi} f^{++}_{m} \\
    \mathfrak{f}^{+-}_{2m} &
    -e^{-\imath \chi}g^{+-}_{2m}&
    e^{-\imath \chi} \mathfrak{f}^{++}_{2m} & 
    -g^{++}_{m} 
    \end{pmatrix}  \delta_{m,n} 
\end{equation}
\end{widetext}
so the Floquet GF is diagonal in Floquet space. In the case of a normal metal, $g_{\rm N}^{\rm r/a} = \pm 1$ 
and $f^{\rm r/a} = \mathfrak{f}^{\rm r/a} = 0$ and the voltage-biased normal metal Floquet GF 
follows easily by taking these limits in Eq.~(\ref{SCvoltage}). In particular, the voltage-biased normal 
metal Floquet GF is diagonal in Floquet space. Let us remark that in the case of a normal metal and one SC, 
the voltage can be gauged onto the normal metal. Hence, both of the Floquet GFs are fully diagonal in Floquet 
space and the problem can be treated by just integrating over the whole energy range.

The other ingredient for the action in Eq.~(\ref{action}) is the normal state scattering matrix. The structure 
of the scattering matrix is explained in detail in Sec.~\ref{sec:NS-scatt-matrix} for the single-impurity and 
in Sec.~\ref{Sec:2imp} for the double-impurity case. The scattering matrix is a function of the energy $E$. 
Thus, upon a Fourier transformation, the scattering matrix only depends on relative time $\tilde{s} = 
\tilde{s}(t_{\rm rel})$ where ($ \ \tilde \ \ $) signifies that the quantity is expressed in lead-Keldysh-spin-Nambu 
space. Thus, its Wigner representation is trivial and its Floquet matrix $\breve{\mathcal{S}}$ is diagonal in Floquet
space. In particular, the element $(n,m)$ of the Floquet matrix is given by
\begin{equation}
    (\breve{\mathcal{S}})^{mn} \equiv \tilde{s}^{mn}(E) = \tilde{s}(E+2meV) \delta_{m,n},
\end{equation}
with the energy-dependent scattering matrix in lead-Keldysh-spin-Nambu space.

\section{More on the Keldysh action} \label{App:action}

In the following, we want to go through the derivation of the Keldysh action in the Floquet representation in detail. 
We start by the Keldysh action
\begin{equation} \label{B1}
    \mathcal{A}(\chi) = \frac{1}{2}\mathrm{Tr} \ln \left[ \underbrace{ \frac{\hat{1} + \hat{G}(\chi)}{2} + 
    \hat{S}\frac{\hat{1}-\hat{G}(\chi)}{2}}_{\hat{Q}(\chi)} \right] - \frac{1}{2}\mathrm{Tr} \ln \hat{Q}(0),
\end{equation}
where $\hat{S}$ is the normal-state scattering matrix and $\hat{G}(\chi) = 
\mathrm{diag}(G_{\rm L}(\chi),G_{\rm R})$ is a block-diagonal matrix containing the lead GFs $G_{\rm L/R}$.
In this formalism, the reservoir GFs are expressed as infinitely large matrices in time-space. Namely, the 
element $(t,t^\prime)$ of the GF $G_{\rm L/R}(\chi)$ is the Keldysh GF $\check{g}_{\rm L/R}(\chi,t,t^\prime)$ 
from Eq.~(\ref{A8}). In addition, the scattering matrix is expressed in the same way as a matrix $\hat{S}$ 
with its element $(t,t^\prime)$ being the scattering matrix $\tilde{s}(t,t^\prime)$. Note that by expressing 
the GFs and scattering matrix as infinitely large matrices in time space, the trace and logarithm in Eq.~(\ref{B1}) 
imply convolutions over intermediate arguments. In particular, it holds
\begin{equation}\label{lemma2}
 (G_{\rm L}  G_{\rm R})_{t,t^\prime} = (\check{g}_{\rm L} \otimes \check{g}_{\rm R})(t,t^\prime),
\end{equation}
where $G_{\rm L}G_{\rm R}$ describes a matrix multiplication in time-Keldysh-spin-Nambu space between two 
GFs $G_{\rm L}$ and $G_{\rm R}$ in time-Keldysh-spin-Nambu space. The subindex ($_{t,t^\prime}$) refers to 
extracting the element $(t,t^\prime)$. As a reminder, the Floquet matrix representation of the GFs $G_{\rm L}$ 
is $\mathcal{G}_{\rm L}(E)$ with its element $(n,m)$ being the Keldysh matrix $\check{g}_{\rm L}^{nm}(E)$ from 
Eq.~(\ref{SCvoltage}) and equivalently for the GF $G_{\rm R}$ and its Floquet matrix $\mathcal{G}_{\rm R}$. 
An important mathematical relation of the Floquet matrix representation in Eq.~(\ref{A10}) is that upon 
mapping the GFs in time space to the Floquet space, the algebraic structure of a generalized convolution is 
preserved \cite{Floquet2008}
\begin{align} \label{lemma1}
   & (\check{g}_{\rm L}\otimes \check{g}_{\rm R})(t,t^\prime)  =  (G_{\rm L}  G_{\rm R})_{t,t^\prime} \nonumber \\
  \Leftrightarrow &  \sum_{l=-\infty}^{\infty} \check{g}_{\rm L}^{ml}(E) \check{g}_{\rm R}^{ln}(E) = 
  (\mathcal{G}_{\rm L}(E) \mathcal{G}_{\rm R}(E))^{mn},
\end{align}
where $\check{g}_{\rm L}^{ml}$ and $\check{g}_{\rm R}^{ln}$ are the components of the Floquet matrices of the
Keldysh functions $\check{g}_{\rm L}$ and $\check{g}_{\rm R}$ respectively. It is seen that a multiplication 
of two Keldysh functions in time-Keldysh-spin-Nambu-space can just be translated into a multiplication of Floquet 
matrices in Floquet-Keldysh-spin-Nambu-space. Thus we can rewrite the action in Eq.~(\ref{B1}) using Floquet matrices
\begin{align}\label{B4}
    \mathcal{A}(\chi) =& \frac{1}{2}\mathrm{Tr} \ln \left[ \underbrace{\frac{\breve{1} + \breve{\cal G}(\chi,E)}{2} + 
    \breve{\cal S}(E)\frac{\breve{1}-\breve{\cal G}(\chi,E)}{2}}_{\breve{\cal Q}(\chi,E)}\right] \nonumber \\
    &- \frac{1}{2}\mathrm{Tr} \ln \breve{\cal Q}(0,E),
\end{align}
where the GF $\breve{\mathcal{G}}(\chi,E) = \mathrm{diag}(\mathcal{G}_{\rm L}(E),\mathcal{G}_{\rm R}(\chi,E))$ is 
block-diagonal in lead space with $\mathcal{G}_{\rm L/R}(\chi,E)$ the Floquet matrices of the GFs $G_{\rm L/R}$ 
and the Floquet matrix $\breve{\mathcal{S}}(E)$ of the scattering matrix $\hat{S}$. Hence, the Keldysh action is 
form-invariant under the transformation. The infinitely large uncountable time space is mapped onto an infinitely 
large uncountable energy argument $E$ and an infinitely large countable Floquet index. Hence, the trace is mapped 
from time space to a trace over the Floquet index and an integral over the Floquet energy $E \in [-eV,eV]$
\begin{widetext}
\begin{equation} \label{B7}
    \mathcal{A}(\chi) = \frac{1}{2}\int_{-eV}^{eV} dE \left( \ln \det \left[ \frac{\breve{1} + \breve{\cal G}(\chi,E)}{2} + 
    \breve{\cal S}(E)\frac{\breve{1}-\breve{\cal G}(\chi,E)}{2}\right]-\ln \det \breve{\mathcal{Q}}(0,E) \right) .
\end{equation}
\end{widetext}
Practically, the Floquet matrices cannot be expressed as infinitely large matrices. Thus, when computing 
the determinant in Eq.~(\ref{B7}) numerically, one introduces a cutoff in the Floquet index large enough for the 
cumulants to converge.

In the case of single-impurity junctions, the matrix $\breve{\mathcal{Q}}$ is block-diagonal in spin space, thus 
$\det \breve{\mathcal{Q}} = \det \breve{\mathcal{Q}}^\Uparrow \det \breve{\mathcal{Q}}^\Downarrow$ and the charge- 
and spin-resolved tunneling probabilities are computed using the inverse Fourier transformation
\begin{equation}
    P_n^\sigma(E,V) = \frac{1}{2\pi} \int_0^{2\pi}d\chi \left( \frac{\det \breve{\mathcal{Q}}^\sigma(\chi,E)}
    {\det \breve{\mathcal{Q}}^\sigma(0,E)} \right) e^{-\imath n \chi}
\end{equation}
for $\sigma = \Uparrow, \Downarrow$. In the two-impurity case, the matrix $\breve{\mathcal{Q}}$ is not block-diagonal 
in spin space and the charge-resolved tunneling probabilities can be obtained as follows
\begin{equation}
    P_n(E,V) = \frac{1}{2\pi} \int_0^{2\pi}d\chi \left( \frac{\det \breve{\mathcal{Q}}(\chi,E)}
    {\det \breve{\mathcal{Q}}(0,E)} \right) e^{-\imath n \chi}.
\end{equation}
%


\end{document}